\documentclass[fleqn,usenatbib]{mnras}

\usepackage{newtxtext,newtxmath}
\usepackage{xspace}
\usepackage{bm}
\usepackage{xcolor}
\usepackage{hyperref}
\usepackage{comment}

\usepackage[T1]{fontenc}
\usepackage{booktabs}
\usepackage{cleveref}
\usepackage{tabularx}
\usepackage{siunitx}
\usepackage{textcomp}
\usepackage{orcidlink}
\usepackage{adjustbox}  
\crefname{figure}{Figure}{Figures}
\crefname{table}{Table}{Tables}
\crefname{equation}{Equation}{Equations}
\crefname{section}{Section}{Sections}
\usepackage{enumitem}

\DeclareRobustCommand{\VAN}[3]{#2}
\let\VANthebibliography\thebibliography
\def\thebibliography{\DeclareRobustCommand{\VAN}[3]{##3}\VANthebibliography}

\usepackage{graphicx}	
\usepackage{amsmath}	

\newcommand{\illustrisTNG}{\mbox{\textsc{IllustrisTNG}}\xspace}

\newcommand{\LCDM}{$\Lambda$CDM\xspace}

\newcommand{\Msun}{{\rm M_\odot}}
\newcommand{\msun}{{\,\rm M_\odot}}

\newcommand{\kpc}{{\rm kpc}}
\newcommand{\Mpc}{{\rm Mpc}}
\newcommand{\Gyr}{{\rm Gyr}}
\newcommand{\kms}{{\rm km\, s^{-1}}}
\newcommand{\cpm}{{\rm cm^{2}\, g^{-1}}}

\newcommand{\arepo}{{\scshape arepo}\xspace}
\newcommand{\arepoOne}{{\scshape arepo-1}\xspace}
\newcommand{\arepoTwo}{{\scshape arepo-2}\xspace}

\DeclareSIUnit\erg{erg}

\newcommand\orcid[1]{\href{https://orcid.org/#1}{\adjustbox{trim={-.15\width} 0 {-.15\width} 0,clip}{\includegraphics[height=9pt]{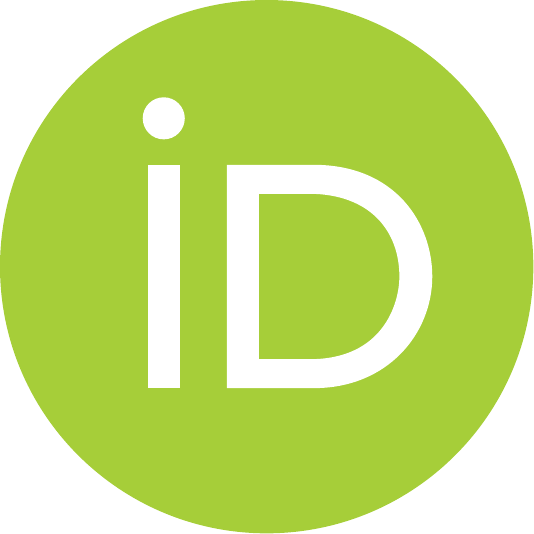}}}}
\newcommand{\appropto}{\mathrel{\vcenter{
  \offinterlineskip\halign{\hfil$##$\cr
    \propto\cr\noalign{\kern2pt}\sim\cr\noalign{\kern-2pt}}}}}

\title[SIDM in \arepo]{A Novel Implementation of Self-Interacting Dark Matter in \arepo}

\author[Zier et al.]{
Oliver Zier\orcid{0000-0003-1811-8915},$^{1}$\thanks{E-mail: \href{mailto:oliver.zier@cfa.harvard.edu}{oliver.zier@cfa.harvard.edu}}
Xuejian Shen\orcid{0000-0002-6196-823X},$^{2, 3}$
Vinh Tran\orcid{0009-0003-6068-6921},$^{2, 3}$
Martin Rosenlyst\orcid{0000-0003-1245-7748},$^{2, 3}$
Mark Vogelsberger\orcid{0000-0001-8593-7692},$^{2, 3}$\newauthor
Rongrong Liu\orcid{0000-0003-0685-3525}$^1$
\\
$^{1}$ Center for Astrophysics $|$ Harvard $\&$ Smithsonian, 60 Garden Street, Cambridge, MA 02138, USA\\
$^{2}$ Massachusetts Institute of Technology, Kavli Institute for Astrophysics and Space Research, 77 Massachusetts Avenue, Cambridge, MA 02139, USA\\
$^{3}$ Department of Physics, Massachusetts Institute of Technology, 77 Massachusetts Avenue, Cambridge, MA 02139, USA\\%
}

\date{Accepted XXX. Received YYY; in original form ZZZ}

\pubyear{\the\year{}}

\begin{document}
\label{firstpage}
\pagerange{\pageref{firstpage}--\pageref{lastpage}}
\maketitle

\begin{abstract}
Self-interacting dark matter (SIDM) influences halo structure through collisional heat transport and offers potential solutions for a range of small-scale puzzles in structure formation. SIDM creates thermalized cores in low-mass haloes, which may account for the observed cored dwarf galaxies. In the meantime, during the late-time gravothermal core collapse, SIDM can produce dense low-mass DM haloes and substructures that have been detected through perturbations to cold stellar streams and strong gravitational lenses. In this work, we present a new Monte-Carlo SIDM implementation in the moving-mesh code \arepoTwo, designed for efficiency, scalability, and extensibility.
The central feature of the implementation is a dedicated DM-only neighbour-search tree that decouples the scattering solver from gravity. This preserves compatibility with the hierarchical time integration used by \arepoTwo while leaving the optimized gravity solver unconstrained. A pairwise communication scheme between MPI tasks allows tracking multiple scattering events in a single timestep while conserving momentum and energy and maintaining parallel consistency by construction. This is complemented by a per-pair timestep criterion that significantly reduces unnecessary timestep restrictions. The implementation natively supports velocity-dependent cross-sections and inelastic interactions, while a compact interface is designed for additional SIDM physics to be implemented without knowledge of the parallelization layer.
We validate the implementation for isotropic, elastic scattering using a suite of idealized and cosmological tests. We assess performance and scalability in isolated core-collapse simulations and in cosmological boxes, both DM-only and with baryons. Except during the late stages of gravothermal collapse, SIDM simulations incur only modest overhead relative to the corresponding CDM runs and are substantially faster than the previous SIDM implementation in \arepoOne.
\end{abstract}




\section{Introduction}

Collisionless cold dark matter (CDM) is a key ingredient of the concordance $\Lambda$CDM cosmological model. Dark matter (DM) accounts for the majority of the matter content of the Universe \citep{Planck2018Parameters}, but has so far been detected only through its gravitational impact. Nevertheless, it plays a central role in cosmic structure formation \citep{White1978,Blumenthal1984,Davis1985}, seeding the cosmic web and haloes within which baryons cool and condense to form galaxies. Despite the empirical success of $\Lambda$CDM, the microscopic nature of DM remains unknown \citep[e.g.,][]{BertoneHooper2018HistoryDM}. On galactic and sub-galactic scales, gravity-only $\Lambda$CDM simulations predict that hierarchical structure formation produces virialized haloes with approximately universal, cuspy density profiles \citep{navarro1997universal} and a large population of bound substructures (subhaloes) spanning many decades in mass \citep[e.g.,][]{Springel2008Aquarius,Diemand2008ViaLacteaII}.
For typical CDM candidates, such as weakly interacting massive particles (WIMPs), the primordial power-spectrum cut-off (due to free streaming) occurs at very small scales, so the hierarchical structures of CDM extend down to Earth-mass microhaloes \citep[e.g.,][]{Diemand2005EarthMassMicrohalos,wang2020universal,zheng2024abundance}. 

Historically, comparing these predictions with the observed nearby dwarf galaxies revealed apparent discrepancies, most prominently the ``cusp--core'' \citep[e.g.,][]{Flores1994,Moore1994CuspCore} and ``missing satellites'' problems \citep[e.g.,][]{Klypin1999MissingSatellites,Moore1999}. Later work pointed to further potential challenges, such as the ``too big to fail'' problem \citep[e.g.,][]{BoylanKolchin2011TBTF,BoylanKolchin2012,Tollerud2014} and the diversity of dwarf-galaxy rotation curves \citep[e.g.,][]{Oman2015Diversity,Kaplinghat2019}. The underlying observational inferences of these tensions and their interpretation remain actively debated \citep[see, e.g.,][]{BullockBoylanKolchin2017SmallScaleReview}. The ``missing satellites'' problem, for instance, has largely eased following the discovery of new faint Milky Way satellites \citep{Bechtol2015,Drlica2015} once survey incompleteness and baryonic disruption are accounted for \citep{Kim2018}. Additional uncertainties arise from the baryonic physics in galaxy formation that can substantially reshape both the visible galaxy properties and the structure of DM haloes, which may alleviate many of the tensions above \citep[e.g.,][]{Governato2010Cores,PontzenGovernato2012CuspCore,Sawala2016APOSTLE,Wetzel2016,GarrisonKimmel2019}. 

Nevertheless, these observational findings have driven a stream of work exploring DM models alternative to the standard collisionless CDM model \citep[e.g.,][]{Hogan2000,SpergelSteinhardt2000SIDM,Dalcanton2001,Buckley2018}. Additional motivation for the field to consider these models comes from the increasingly tight constraints on classical CDM particle candidates, such as WIMPs, despite decades of efforts in direct detection experiments and collider searches \citep[e.g.,][]{Bertone2005,Bertone2010,Aprile2018,Meng2021,Aalbers2023}. Among alternatives to collisionless CDM, one theoretically motivated model is self-interacting dark matter (SIDM), in which DM can interact non-gravitationally \citep[e.g.,][]{Carlson1992, deLaix1995, Firmani2000, SpergelSteinhardt2000SIDM}. It is well motivated by hidden dark sectors as extensions to the Standard Model~\citep[e.g.,][]{Ackerman2009,ArkaniHamed2009,Feng2009,LoebWeiner2011YukawaSIDM,CyrRacine2013,TulinYuZurek2013SIDM,Cline2014,Boddy2014}. In the simplest case of elastic scattering with velocity-independent cross-section ($\sigma/m$), SIDM thermalizes and produces approximately isothermal cores in the centres of haloes \citep[e.g.,][]{Dave2001SIDM,Colin2002,rocha2013cosmological,Elbert2015SIDMCores}. For the typical DM particle velocities in dwarf galaxies, cross-sections in the range $\sigma/m \sim 0.1\text{--}10~{\rm cm^2\,g^{-1}}$ can reproduce the observed cores in dwarf galaxies \citep[e.g.,][]{KaplinghatTulinYu2016ParticleColliders,TulinYu2018SIDMReview}. These simple models with constant $\sigma/m$, however, are constrained on cluster scales by halo density profiles, shapes, and the dynamics of merging systems \citep[e.g.,][]{MiraldaEscude2002Shapes,Randall2008BulletCluster,Peter2013SIDMHaloShapes,Harvey2015ClusterMergers,Shen2022esidm}, with recent bounds tightened by joint strong-lensing and stellar-kinematic modelling of individual systems \citep{ODonnell2026} and by the dynamics of double radio relic clusters \citep{Jee2026}. This has motivated velocity-dependent models in which $\sigma/m$ is enhanced at the low relative velocities of dwarfs but small at cluster velocities. Such a velocity dependence arises naturally for interactions mediated by light force carriers with Yukawa-like potentials, linking halo phenomenology to microscopic dark-sector physics \citep[e.g.,][]{BuckleyFox2010LightMediators,LoebWeiner2011YukawaSIDM,TulinYuZurek2013SIDM}.

Recent observational findings have driven a new wave of interest in SIDM models, one that points in the opposite direction from the early explorations described above. Several recent studies of strong gravitational lensing systems reported signatures of dark substructures that are significantly more concentrated than predicted in the standard CDM model, on both cluster \citep{Meneghetti2020,Ragagnin2022,Meneghetti2023,Natarajan2026} and galaxy scales \citep{Minor2021,Ballard2024,Despali2025,Enzi2025,Kong2025}.
In addition, observations of perturbed cold stellar streams in the Milky Way provide a complementary laboratory to study structures of low-mass subhaloes. Several recent analyses inferred perturbers whose central densities exceed typical $\Lambda$CDM expectations at the same mass \citep{PriceWhelanBonaca2018GD1,Bonaca2019,Boer2020,Zhang2024GD1SIDM}.
A less-explored SIDM parameter regime offers a natural mechanism to produce these extremely dense substructures. 

Because self-gravitating systems have a negative heat capacity, an SIDM halo eventually undergoes gravothermal core collapse as heat is conducted outward through self-interactions. This is similar to the gravothermal collapse first studied in the context of globular clusters \citep{Lynden1968}. For models with cross-section $\sigma /m \gtrsim 10\text{--}100\,\cpm$, the collapse can produce central densities in dwarf haloes a few orders of magnitude higher than in collisionless CDM within the age of the Universe \citep[e.g.,][]{Balberg2002,KodaShapiro2011Gravothermal,Turner2021}, offering a coherent explanation for the dense substructures inferred from observations, with further implications for the diversity of dwarf-galaxy rotation curves and for dense Milky Way satellites \citep{Yu2025}. Another potential consequence of this class of models is core collapse in massive, concentrated haloes at high redshifts, which could produce DM-seeded supermassive black holes that could explain several key features of the faint active galactic nuclei revealed by JWST, known as ``Little Red Dots'' \citep{Shen2025,Shen2026,Jiang2026,Roberts2026}, as well as massive quasars \citep{Xiao2021,Feng2021,Shen2025}.

Turning this evidence into quantitative arguments or constraints on SIDM is challenging, especially for models that result in core collapse. In principle, one would directly solve the Boltzmann equation with an appropriate collision operator for DM scatterings, but this is computationally prohibitive over the dynamic range relevant for cosmic structure formation. Practical approaches often rely on effective descriptions. The most widely used class embeds Monte-Carlo scattering algorithms in traditional $N$-body simulation codes, stochastically scattering particle pairs with probabilities calibrated to reproduce the desired collisional dynamics in the continuum limit \citep[e.g.,][]{Yoshida2000,Burkert2000,Dave2001SIDM,vogelsberger2012subhaloes,rocha2013cosmological,Robertson2017SIDM}. However, predictions for halo structure deep in the gravothermal collapse phase require extremely accurate and efficient algorithms along with careful choices of numerical parameters \citep[e.g.,][]{palubski2024numerical,Mace2026}. One particular difficulty for Monte-Carlo SIDM schemes is the treatment of multiple scattering events within a single timestep, which become common during core collapse. Violations of energy conservation can occur for massively parallel codes, where specialized communication schemes are required \citep{Robertson2017SIDM,Fischer2021,Fischer2024}.
 
A reliable interpretation of observations also requires understanding the interplay between baryonic physics and SIDM, and this coupling can be decisive in certain systems, such as local dwarf galaxies and stellar streams. This in turn calls for hydrodynamical simulations that treat SIDM and galaxy formation in a single, self-consistent framework \citep[e.g.,][]{Despali2019InterplaySIDMBaryons}.
Recent efforts have therefore embedded Monte-Carlo SIDM modules in full galaxy-formation simulations: EAGLE-SIDM \citep{Robertson2018,Robertson2021,Forouhar2022}, BAHAMAS-SIDM \citep{Robertson2019BAHAMASSIDM}, TANGO-SIDM \citep{Correa2022,Correa2025} and the \textsc{AIDA}-TNG suite \citep{Despali2025AIDATNG}, as well as high-resolution zoom-in simulations of individual dwarf and Milky Way-mass galaxies, such as SIDM in the FIRE model \citep{robles2017sidm,Sameie2021,Vargya2022} and the Lyra simulations \citep{Gutcke2025}.

Both \textsc{AIDA}-TNG and Lyra use the moving-mesh code \arepo with the SIDM implementation of \citet{vogelsberger2012subhaloes,vogelsberger2019evaporating}, which uses the gravity tree for a neighbour search required for SIDM.
This tight coupling prevents several recent optimizations of the gravity solver from being used and complicates further extensions of the SIDM module.
The aim of this paper is to introduce a new SIDM solver in \arepoTwo that (i) exposes an easily extensible interface for generic two-body interactions, (ii) employs a parallelization scheme that natively supports multiple scattering events, and (iii) decouples the SIDM module from gravity by means of a dedicated neighbour-search tree, allowing an efficient SIDM treatment alongside the optimized gravity solver.

The paper is organized as follows.
\cref{sec:sidm_arepo2} reviews the physical description of SIDM scattering and the landscape of numerical methods used to model it.
\cref{sec:arepo2} gives an overview of \arepoTwo, focusing on its gravity solver.
In \cref{sec:sidm_impl} we introduce our new SIDM module.
In \cref{sec:codeVerification} we verify the implementation against analytic test problems and against the legacy implementation for the gravothermal collapse of an isolated halo and a cosmological zoom-in simulation.
\cref{sec:performance} presents the computational performance and scalability of the module, both for isolated core collapse and for cosmological box simulations with and without baryons, comparing against a pure \LCDM\ run and the \arepoOne\ implementation.
\cref{sec:conclusions} summarizes our results.

\section{SIDM theory and numerical methods}
\label{sec:sidm_arepo2}

In this section, we briefly review the theory of collisional particle systems and the corresponding numerical methods, covering the Monte-Carlo method implemented in \arepoTwo and alternative approaches from the literature. 

\subsection{Kinetic theory of SIDM}

A numerical scheme for SIDM scattering is most naturally derived from the Boltzmann equation,
\begin{equation}
    \dfrac{\mathrm{D}f(\mathbf{x}, \mathbf{v}, t)}{\mathrm{D}t}
    = \dfrac{\partial f}{\partial t}
    + \mathbf{v}\!\cdot\!\nabla_{\mathbf{x}} f
    - \nabla_{\mathbf{x}}\Phi \!\cdot\! \nabla_{\mathbf{v}} f
    = \mathcal{C}[f, \sigma],
    \label{eq:CBE}
\end{equation}
where $\mathrm{D}/\mathrm{D}t$ is the Lagrangian derivative, $f(\mathbf{x}, \mathbf{v}, t)$ is the DM phase-space distribution function, $\Phi$ is the gravitational potential, and $\mathcal{C}[f,\sigma]$ is the collision operator. The left-hand side describes collisionless phase-space advection under gravity, while the right-hand side encodes the non-gravitational scattering of DM. For SIDM, $\mathcal{C}$ is set by the underlying microphysics through the differential cross-section ${\rm d}\sigma/{\rm d}\Omega$, which may depend on the relative velocity of the scattering particles, the scattering angle $\theta$ (the deflection angle of DM particles in the centre-of-momentum frame), and, in more general models, on internal states or inelastic channels. For a system consisting of equal-mass DM particles, one can replace $f(\mathbf{x}, \mathbf{v}, t)$ with the phase-space mass distribution $\rho(\mathbf{x}, \mathbf{v}, t) \equiv f(\mathbf{x}, \mathbf{v}, t)\,m$ ($m$ is the mass of the DM particles), so $\sigma/m$, rather than $\sigma$, determines the evolution of the system.

Several integrated cross-sections are particularly useful for numerical and phenomenological modelling~\citep[e.g.,][]{TulinYu2018SIDMReview}. We briefly describe them here. The first is the total cross-section,
\begin{equation}
    \sigma = \int {\rm d}\Omega\, \dfrac{{\rm d}\sigma}{{\rm d}\Omega},
\end{equation}
which controls the overall DM scattering rate and sets, for example, the mean free path and the average interaction time at fixed local density and relative velocity. However, the total cross-section is often a poor measure of the dynamical impact of scattering. For example, for certain types of interactions, ${\rm d}\sigma/{\rm d}\Omega$ can peak or even exhibit a singularity at $\theta=0$, inflating $\sigma$ even though the corresponding small-angle scatterings contribute little to the heat conduction rates that govern SIDM halo evolution. It is therefore common to introduce the momentum-transfer cross-section~\citep[e.g.,][]{Feng2010,TulinYuZurek2013SIDM},
\begin{equation}
    \sigma_{\rm T} = \int {\rm d}\Omega\, \dfrac{{\rm d}\sigma}{{\rm d}\Omega}\,(1-\cos{\theta}),
    \label{eq:transfer_cross}
\end{equation}
which down-weights forward scattering by weighting each collision by its longitudinal momentum transfer. For identical DM particles, where the $\theta=0$ and $\theta=\pi$ cases are physically indistinguishable, a modified definition has been proposed~\citep[e.g.,][]{Kahlhoefer2014},
\begin{equation}
    \tilde{\sigma}_{\rm T} = \int {\rm d}\Omega\, \dfrac{{\rm d}\sigma}{{\rm d}\Omega}\,(1-|\cos{\theta}|). \label{eq:mod_transfer_cross}
\end{equation}
Another widely used definition is the viscosity cross-section~\citep[e.g.,][]{TulinYuZurek2013SIDM, Cline2014, Boddy2016}
\begin{equation}
    \sigma_{\rm V} = \int {\rm d}\Omega\, \dfrac{{\rm d}\sigma}{{\rm d}\Omega}\,(1-\cos^{2}{\theta}),
\end{equation}
which weights scatterings by their transverse energy transfer, suppressing both forward and backward directions. In many SIDM applications, constraints and target parameter ranges are therefore expressed in terms of $\sigma_{\rm T}/m$, $\tilde{\sigma}_{\rm T}/m$ or $\sigma_{\rm V}/m$ rather than the total $\sigma/m$, especially for interactions mediated by long-range or light force carriers that generate anisotropic scatterings. For isotropic, elastic, velocity-independent scatterings, one finds $\sigma=\sigma_{\rm T}=(3/2)\sigma_{\rm V}$ and $\tilde{\sigma}_{\rm T}=\sigma/2$, whereas in the highly anisotropic limit $\sigma\gg\sigma_{\rm T},\tilde{\sigma}_{\rm T},\sigma_{\rm V}$.

A useful dimensionless diagnostic of the dynamical regime is the Knudsen number, which compares the local collisional mean free path to a characteristic macroscopic length scale. In the context of the SIDM halo, we write
\begin{equation} 
    \mathrm{Kn} \equiv \lambda/H, \label{eq:knudsen_number}
\end{equation}
where $\lambda \sim (\rho\,(\sigma/m))^{-1}$ is the mean free path, $H\equiv \sqrt{v^2/(4\pi G\rho)}$ the gravitational scale height and $v$ the local one-dimensional velocity dispersion. The Knudsen number cleanly separates two SIDM regimes: when $\mathrm{Kn}\gg 1$ (the long-mean-free-path, LMFP regime), scattering is infrequent on the scale $H$ and heat transport is non-local. When $\mathrm{Kn}\ll 1$ (the short-mean-free-path, SMFP regime), the system approaches a collisional-fluid limit in which transport is well described by local conduction. Different regions of a realistic SIDM halo can lie in different regimes simultaneously.

\subsection{The $N$-body representation of DM}

Solving \cref{eq:CBE} on an Eulerian grid is not computationally feasible for the full cosmological problem: the six-dimensional phase space is enormous and, in cosmological structure-formation simulations, the distribution function is highly inhomogeneous. Because $f$ is conserved along any trajectory $\{\mathbf{x}(t),\mathbf{v}(t)\}$ obtained from \cref{eq:CBE} in the absence of the collision operator (Liouville's theorem), the $N$-body method offers a natural alternative. Rather than evolving the exact microscopic distribution, an $N$-body simulation tracks a coarse-grained sampling of it in phase space~\citep{Hockney1988nbody}. At any given time, the distribution function is approximated as
\begin{equation}
    \hat{f}(\mathbf{x}, \mathbf{v}) =
    \dfrac{1}{\sum_i m_i}
    \sum_i m_i\,W_{\rm soft}(\mathbf{x}-\mathbf{x}_i, \epsilon_i)\,\delta(\mathbf{v}-\mathbf{v}_i),
    \label{eq:nbody}
\end{equation}
where $m_i$, $\mathbf{x}_i$ and $\mathbf{v}_i$ are the mass, position and velocity of the $i$-th simulation particle, $W_{\rm soft}$ is the gravitational-softening kernel and $\epsilon_i$ the associated softening length. Each simulation particle should be interpreted as a phase-space tracer (a ``super-particle'') sampling a finite patch of the underlying distribution rather than as an individual microscopic DM particle. The kernel $W_{\rm soft}$ regularizes the mass distribution and suppresses artificial two-body gravitational scattering~\citep[e.g.,][]{Dehnen2001,Springel2005Gadget2}, allowing the simulation to approximate the collisionless Vlasov limit on resolved scales.

This coarse-grained viewpoint is essential for any SIDM implementation. Microscopic DM scattering takes place between physical particles whose masses lie far below the resolution limit of cosmological simulations. SIDM schemes must construct an effective, resolution-dependent prescription whose stochastic or deterministic updates of the simulation particles reproduce, in the continuum and large-$N$ limit, the desired collision operator in \cref{eq:CBE}. In practice, the collision probabilities, or equivalently effective scattering rates, assigned to a pair of simulation particles must depend on their coarse-grained phase-space kernels, their relative velocity, and the chosen microscopic cross-section model. The velocity-space structure of \cref{eq:nbody} is represented by delta functions, so the unresolved internal velocity distribution of each simulation particle is neglected. Numerical convergence then requires demonstrating that macroscopic observables become insensitive to mass resolution, spatial resolution, and kernel choice in the appropriate limit. These updates, typically stochastic pairwise velocity changes applied at each timestep, must also conserve mass, momentum, and (for elastic scattering) energy of the system. 

\subsection{Numerical schemes for SIDM}
\label{subsec:other_schemes}

A range of numerical methods has been developed to model SIDM across the LMFP and SMFP regimes introduced above. \cref{tab:sidm-methods} summarizes the main families. Below we first introduce the scheme we implement in \arepoTwo: an $N$-body Monte-Carlo scheme, overviewed here with its detailed implementation deferred to \cref{sec:sidm_impl}.  We then review the broader landscape of alternative approaches.

\begin{table*}
    \addtolength{\tabcolsep}{-1.5pt}
    \renewcommand{\arraystretch}{1.05}
    \centering
    \caption{Summary of numerical methods modelling SIDM in the literature.
    $^{\ast}$ Included in the SIDM module of \arepoTwo. \\
    $^{a}$ Unbiased only while the timestep resolves the local collisional time and the scattering kernel remains smaller than the SIDM mean free path. Meeting both requirements deep in the SMFP regime demands prohibitively fine time and mass resolution \citep[e.g.,][]{KodaShapiro2011Gravothermal,Robertson2017SIDM}. \\
    $^{b}$ Currently implemented in the ideal-fluid limit while heat conduction and viscosity are not yet included \citep{Schon2025}. \\
    $^{c}$ Valid with a heat conductivity calibrated against $N$-body simulations in the LMFP regime. \\
    $^{d}$ The core-collapse phase in the SMFP regime is accessible only through an empirical extension \citep{Jiang2023}.}
    \begin{tabular}{lllll}
     \hline
     Algorithm & Description & LMFP? & SMFP? & References \\
     \hline
     $N$-body Monte-Carlo  & stochastic pairwise scatters$^{\ast}$ & yes & yes$^{a}$ & e.g.~\citet{vogelsberger2012subhaloes, rocha2013cosmological} \\
     & & & &  \citet{Robertson2017SIDM,robles2017sidm} \\
     $N$-body drag force & drag + diffusion kick$^{\ast}$ & yes & no & \citet{Kahlhoefer2014,Fischer2021} \\
     \hline
     SIDM--hydro hybrid & blend of Monte-Carlo and SPH & yes & yes$^{b}$ & \citet{Schon2025} \\
     Effective force & coarse-grained from first-principles calculations& no & yes & \citet{Ramos2025} \\
     $N$-body in phase-space  & Monte-Carlo in a reduced-dimensional phase space & yes & yes$^{a}$ & \citet{Kamionkowski2025} \\
     Gravothermal fluid & fast, one-dimensional conducting fluid & yes$^{c}$ & yes & e.g.~\citet{Balberg2002,KodaShapiro2011Gravothermal} \\
     & & & &  \citet{nishikawa2019accelerated-17e,Outmezguine2023} \\
     Isothermal Jeans model & fast, empirical model assuming equilibrium & yes & yes$^{d}$ & e.g.~\citet{KaplinghatTulinYu2016ParticleColliders,Robertson2021} \\
     & & & & \citet{Jiang2023} \\
    \hline
    \end{tabular}
    \label{tab:sidm-methods}
    \renewcommand{\arraystretch}{0.91}
\end{table*}

\subsubsection{Monte-Carlo $N$-body scattering}
\label{subsubsec:mc_overview}

In Monte-Carlo schemes, the elementary operation is a stochastic velocity update of a pair of macroscopic simulation DM particles when the pair is selected to interact. Early implementations estimated a scattering rate for each particle off the local background and selected a partner only after a scatter was triggered \citep[e.g.,][]{Burkert2000,kochanek2000quantitative,Yoshida2000, Colin2002, Randall2008BulletCluster,vogelsberger2012subhaloes}, whereas modern schemes (including ours) assign probabilities to individual pairs \citep[e.g.,][]{Dave2001SIDM,rocha2013cosmological,Robertson2017SIDM}. Explicit conservation of momentum and energy (for elastic scatterings) is required in these operations. We defer a more thorough description of this scheme and the numerical implementation in \arepoTwo to \cref{sec:sidm_impl}.

\subsubsection{$N$-body with drag force}
\label{subsec:dragforce}

For strongly forward-peaked differential cross-sections, the Monte-Carlo approach becomes inefficient: the total cross-section is inflated by the many small-angle deflections (and formally diverges in some models), so resolving every collision individually would demand prohibitively small timesteps. The cumulative effect of these deflections is better described by an effective drag force opposing the relative motion of a particle. This description was derived by \citet{Kahlhoefer2014} in the context of merging clusters, and building on it, \citet{Fischer2021} constructed a momentum- and energy-conserving $N$-body scheme (referred to as ``fSIDM'' therein) in which every pair receives a deterministic drag plus a stochastic, energy-restoring kick. The scheme is implemented in \arepoTwo\ but not used in the present work. We briefly summarize its core idea and defer to Rosenlyst et al. (in prep.) for the full formalism and validation tests. Every neighbouring pair of DM particles receives a drag force controlled by the symmetrized momentum-transfer cross-section $\tilde{\sigma}_{\rm T}$ (see e.g.\ Equation~(9) of \citealt{Fischer2021}). This drag force is applied in the direction opposing the relative velocity between the pair of particles. To ensure energy conservation, one needs to apply an additional stochastic kick to both particles in the plane perpendicular to their relative velocity. The drag-force description relies on a Fokker--Planck expansion about the collisionless solution, and will become inaccurate in the SMFP regime. In that limit, frequent collisions drive the distribution close to local thermodynamic equilibrium while viscous stresses and heat conduction enter at leading order. The correct continuum description is a Chapman--Enskog expansion about a Maxwell--Boltzmann distribution \citep{Ramos2025}.

\subsubsection{Other SIDM schemes}

\noindent \textbf{SIDM--hydro hybrid (SHH) scheme:} \citet{Schon2025} propose an SIDM--hydro hybrid (SHH) scheme that couples the Monte-Carlo solver to smoothed-particle hydrodynamics (SPH), bridging the LMFP regime to the SMFP collisional-fluid limit. Each particle receives both Monte-Carlo scattering kicks and SPH pressure accelerations, blended by a sigmoid function of the local density that serves as a proxy for the local Knudsen number. However, in its present form, the fluid limit is ideal without heat conduction and viscosity, and the bridging function requires additional calibration. 

\vspace{0.1cm}

\noindent \textbf{Effective force scheme:} A conceptually distinct strategy was recently proposed by \citet{Ramos2025}, who revisit the relationship between the microscopic SIDM collisions and the effective dynamics of the macroscopic simulation particles. In the SMFP regime, via a Chapman--Enskog expansion about local equilibrium, they show that the standard assumption that simulation macro-particles obey the same Boltzmann equation as the underlying DM particles, implicit in the common propagation of the microscopic cross-section to the simulation-particle level, is not generically valid. Properly coarse-grained, the interaction between macro-particles becomes a deterministic pairwise effective force at leading-order in the Chapman--Enskog expansion. A numerical implementation of this scheme has not yet been presented.

\vspace{0.1cm}

\noindent \textbf{$N$-body in a reduced phase space:} An alternative to both 3D collisional $N$-body and fluid closures is to exploit the symmetry of the system and evolve the distribution function in phase space with a reduced dimension. \citet{Kamionkowski2025} evolve self-gravitating spherical SIDM haloes with particles in the three-dimensional phase space $(r, v, \cos\theta)$ of a spherical system ($\theta$ is the angle between the particle velocity and the radial direction). Particle orbits are integrated in the self-consistent spherical potential, while self-interactions remain standard pairwise Monte-Carlo scatters, with probabilities estimated from radially coarse-grained neighbour shells and full 3D collision kinematics projected back onto the reduced coordinates. The reduced dimensionality makes the method orders of magnitude cheaper than 3D $N$-body, and it sidesteps many challenges in traditional $N$-body simulation. 

\vspace{0.1cm}

\noindent \textbf{Gravothermal fluid schemes:} A complementary class of methods treats SIDM as a self-gravitating conducting fluid, taking velocity moments of the Boltzmann equation and closing the hierarchy with a prescription for heat conduction~\citep[e.g.,][]{Lynden-Bell1980,Balberg2002}. In spherical symmetry, these ``gravothermal'' models evolve coupled equations for mass conservation, momentum conservation (hydrostatic equilibrium), and energy conservation (with heat conduction due to DM self-interactions). The approach follows the secular evolution of SIDM haloes through core formation and deep into gravothermal collapse at a fraction of the cost of full 3D collisional $N$-body simulations \citep[e.g.,][]{Balberg2002,KodaShapiro2011Gravothermal,Pollack2015,nishikawa2019accelerated-17e,Outmezguine2023}. The heat conductivity in this method depends on both the local halo properties and the microphysical scattering law. In the LMFP regime, energy transport is non-local and limited by the gravitational scale height, while in the SMFP regime, it reduces to a Chapman--Enskog-like local conduction law \citep{Outmezguine2023}. A major strength of these models is that, when expressed in dimensionless variables, the gravothermal solution is self-similar in the LMFP regime, and can therefore be mapped efficiently onto different combinations of halo mass and cross-section. However, a calibration of the heat conductivity, especially in the LMFP regime, against $N$-body simulations is still required.

\vspace{0.1cm}

\noindent \textbf{Empirical isothermal Jeans modelling:} The empirical approach bypasses any explicit solution of the collisional Boltzmann equation and instead assumes that self-interactions thermalize the inner halo, producing an approximately isothermal core embedded in the gravitational potential~\citep[e.g.,][]{KaplinghatTulinYu2016ParticleColliders,Robertson2021}. The halo is split at a characteristic radius $r_{\rm c}$ that separates the inner, scattering-dominated region from an outer region that remains close to a collisionless CDM profile. $r_{\rm c}$ is typically defined as the radius at which an average particle has undergone $\mathcal{O}(1)$ interactions over the halo lifetime. Within $r_{\rm c}$, the SIDM profile is obtained by solving the spherical Jeans equation at constant velocity dispersion together with the Poisson equation. Baryons are included straightforwardly by adding their density to the Poisson equation, which allows the adiabatic contraction induced by baryonic potentials to be explored at minimal cost. \citet{Jiang2023} extended this framework by identifying a new family of Jeans--Poisson solutions that provide a good empirical description of density profiles in the core-collapsed regime. This empirical method is computationally cheap and well-suited to large parameter scans and semi-analytic pipelines. However, there are some ambiguities in the definitions of $r_{\rm c}$ and the halo ``age'' that are left to calibration, and the assumption of a fully isothermal inner region captures only crudely the time-dependent thermalization process. 

\section{The moving-mesh code \arepoTwo}
\label{sec:arepo2}

The massively parallel \arepoTwo code \citep{springel2010pur,pakmor2016improving,weinberger2020arepo} solves the Euler equations on a moving, unstructured Voronoi mesh with a second-order finite-volume method. The Voronoi cells move approximately with the local fluid velocity \citep{vogelsberger2012moving}, yielding a quasi-Lagrangian, Galilean-invariant scheme. An approximately constant baryonic mass resolution is maintained by refining (splitting) cells whose masses become too large and derefining (merging) those whose masses become too small. In addition to the gas, \arepoTwo solves the Poisson equation to evolve collisionless particle species, such as DM and stellar particles, under their mutual gravity. Because the gravity solver and its time integration are central to the SIDM module developed below, we describe them in some detail in \cref{subsec:gravity_solver,subsec:hierarchical_old}.

\citet{pakmor2023millenniumtng} introduced several key algorithmic improvements, including MPI-3 shared memory (data identical across MPI ranks are stored only once per compute node) and a hierarchical domain decomposition that first partitions the simulation volume among compute nodes and then distributes each sub-volume among MPI tasks. These changes enabled \arepo\ to scale, in cosmological-box simulations, to more than $10^{5}$ MPI tasks \citep{pakmor2023millenniumtng, lumina}. \citet{zier2024adapting} subsequently added GPU acceleration for radiative transfer, together with an abstraction layer that will facilitate porting further parts of the code to GPUs. \arepoTwo also features a more modular code base, which allows the SIDM module to be implemented as a self-contained module with minimal impact on the rest of the code.

\subsection{Gravity solver}
\label{subsec:gravity_solver}
\arepoTwo provides several algorithms for computing gravitational forces, including a hierarchical octree \citep[Barnes--Hut;][]{Barnes1986}, a particle--mesh (PM) method \citep{Hockney1988nbody} and their combination in the TreePM scheme \citep{Bagla2002,Bagla2003,Bode2003}, in which short-range forces are evaluated with the tree and long-range forces with the PM solver. The octree groups distant particles into successively larger nodes and approximates their collective force by a low-order multipole expansion, truncated at the monopole by default. Whether a node is accepted or opened is decided either by the geometric Barnes--Hut criterion, based on the node's angular size, or by the relative opening criterion introduced for \textsc{Gadget}, which aims to keep the fractional force error approximately constant by comparing each node's estimated multipole error with the magnitude of the particle's previous acceleration \citep{springel2001gadget}.

By default, the full gravity tree is rebuilt at every timestep through hierarchical subdivision of nodes. This fixed overhead can dominate the runtime when only a sparsely populated time bin is active, the situation that motivates the hierarchical time integration of \cref{subsec:hierarchical_old}.

When a smooth phase-space distribution function is sampled by a finite number of particles, direct interactions between close pairs would introduce spurious two-body scattering and violate the premise of collisionless dynamics. To suppress this, \arepo\ employs a gravitational force softening that replaces the Newtonian $1/r^{2}$ force below a softening scale~$\epsilon_{\mathrm{soft}}$ with a force that smoothly tends to zero at vanishing separation \citep{springel2010pur,weinberger2020arepo}. The softening kernel matches that of the \textsc{Gadget} code family \citep{Springel2005Gadget2}: the potential of a point mass at zero separation equals $-Gm/\epsilon_{\mathrm{soft}}$ and the force becomes fully Newtonian at $2.8\,\epsilon_{\mathrm{soft}}$. The softening length trades spatial resolution against force bias and discreteness noise, and for a given particle number it can be tuned to minimize the mean force error \citep{Dehnen2001}. Collisionless particle types are assigned individual, fixed softening lengths, while for gas cells the softening is set adaptively to $\epsilon_{\mathrm{cell}} = f_{h}\,(3V/4\pi)^{1/3}$, where $V$ is the Voronoi cell volume and $f_{h}$ is a free parameter \citep{weinberger2020arepo}. Interactions between particles with different softening lengths are symmetrized by adopting the larger of the two values.

For the gravitational timestepping, \arepo\ adopts the criterion
\begin{equation}
    \Delta t_{\mathrm{grav}} \leq \sqrt{\frac{2\,\eta\,\epsilon_{\mathrm{soft}}}{|\mathbf{a}|}},
    \label{eq:dt_grav}
\end{equation}
where $|\mathbf{a}|$ is the magnitude of the gravitational acceleration, $\epsilon_{\mathrm{soft}}$ the gravitational softening length and $\eta$ a dimensionless accuracy parameter \citep{weinberger2020arepo}. Tying the step to the softening in this way keeps the integration accurate in regions of high acceleration and is closely related to the convergence criteria established for the inner structure of simulated haloes \citep{Power2003}; alternative criteria based on the local dynamical time have also been proposed \citep{Zemp2007}. For gas cells, additional constraints come from the Courant--Friedrichs--Lewy (CFL) condition, $\Delta t_{\mathrm{CFL}} \leq C_{\mathrm{CFL}}\,r_{\mathrm{cell}}/v_{\mathrm{sig}}$, where $v_{\mathrm{sig}}$ is the signal speed and $C_{\mathrm{CFL}}$ the Courant factor. The timestep of each element is set by the most restrictive of all applicable constraints.

Because the tree uses a truncated multipole expansion, the computed forces are only approximate, and the individual force errors do not in general sum to zero. This produces a small but systematic non-conservation of total momentum and energy, and in non-periodic simulations, the residual net force can cause a spurious drift of the centre of mass over time. \arepo\ mitigates this by randomizing the placement of the tree-domain centre at each tree construction, which decorrelates the force errors in time and greatly improves global momentum conservation \citep{weinberger2020arepo}. Energy conservation is further affected by the time-integration discretization; the corresponding errors shrink with smaller timesteps, i.e.\ smaller $\eta$ in \cref{eq:dt_grav}. A fast multipole method (FMM), in which the multipole expansion is performed about both interacting nodes so that each node--node interaction is evaluated mutually, conserves total momentum by construction up to machine precision \citep{Dehnen2000,Dehnen2002}; such a solver is available in \textsc{Gadget-4} \citep{springel2021simulating}.

\subsection{Hierarchical time integration}
\label{subsec:hierarchical_old}
By default, particles in \arepoTwo are advanced on a power-of-two hierarchy of individual timesteps \citep{pakmor2016improving}; alternatively, a single global timestep can be enforced, in which case all particles are advanced with the smallest step required by any of them. Individual, block-structured timesteps are essential for efficiency when the dynamical times in a simulation span many orders of magnitude \citep{Springel2005Gadget2}, but they complicate the integration, because the leapfrog (kick--drift--kick) scheme is symplectic only for a constant timestep, and adapting the step per particle breaks that property.

A clean way to retain a well-behaved integrator with individual timesteps is to view the evolution as an operator (Hamiltonian) splitting \citep{HairerLubichWanner2006,LeimkuhlerReich2005}. \citet{springel2021simulating} build on this idea to construct a hierarchical time-integration scheme in which, at each level of the hierarchy, the gravitational Hamiltonian is split into a ``slow'' subsystem (particles on long timesteps) and a ``fast'' subsystem (particles on short timesteps). The fast subsystem is integrated on its short step while being coupled symplectically to the slow one. A closely related hierarchical splitting was developed independently by \citet{Pelupessy2012}, and integrators of this type have been shown to be approximately time-reversible, with substantially smaller secular energy drift than adaptive leapfrog schemes \citep{Aguilar2022}. A practical consequence is that the gravity tree is only built for the currently active (fast) particles. The trade-off is that several trees, each containing a different subset of active particles, may have to be constructed per timestep. For a very small set of active particles, the tree walk can be replaced by direct summation. Despite this construction overhead, the hierarchical scheme is typically faster than a global tree rebuild when the timestep hierarchy is deep, and it removes the time-integration contribution to the momentum-conservation error of the standard block-step leapfrog.

The gravity tree in \arepo is also reused as a neighbour-search structure by several physics modules: the live-dust solver \citep{mckinnon2018simulating}, the SIDM implementation of \citet{vogelsberger2012subhaloes}, and the black-hole subgrid model of the TNG framework \citep{weinberger2016simulating,weinberger2018supermassive}. These modules generally require all potential neighbours to be present in the tree, so they are incompatible with the hierarchical gravity scheme whenever they are invoked on non-synchronized timesteps. The SIDM module presented in this work is compatible with the hierarchical scheme. This is achieved by replacing the gravity-tree neighbour search with a dedicated DM-only tree.

\begin{figure*}
    \centering
    \includegraphics[width=\textwidth]{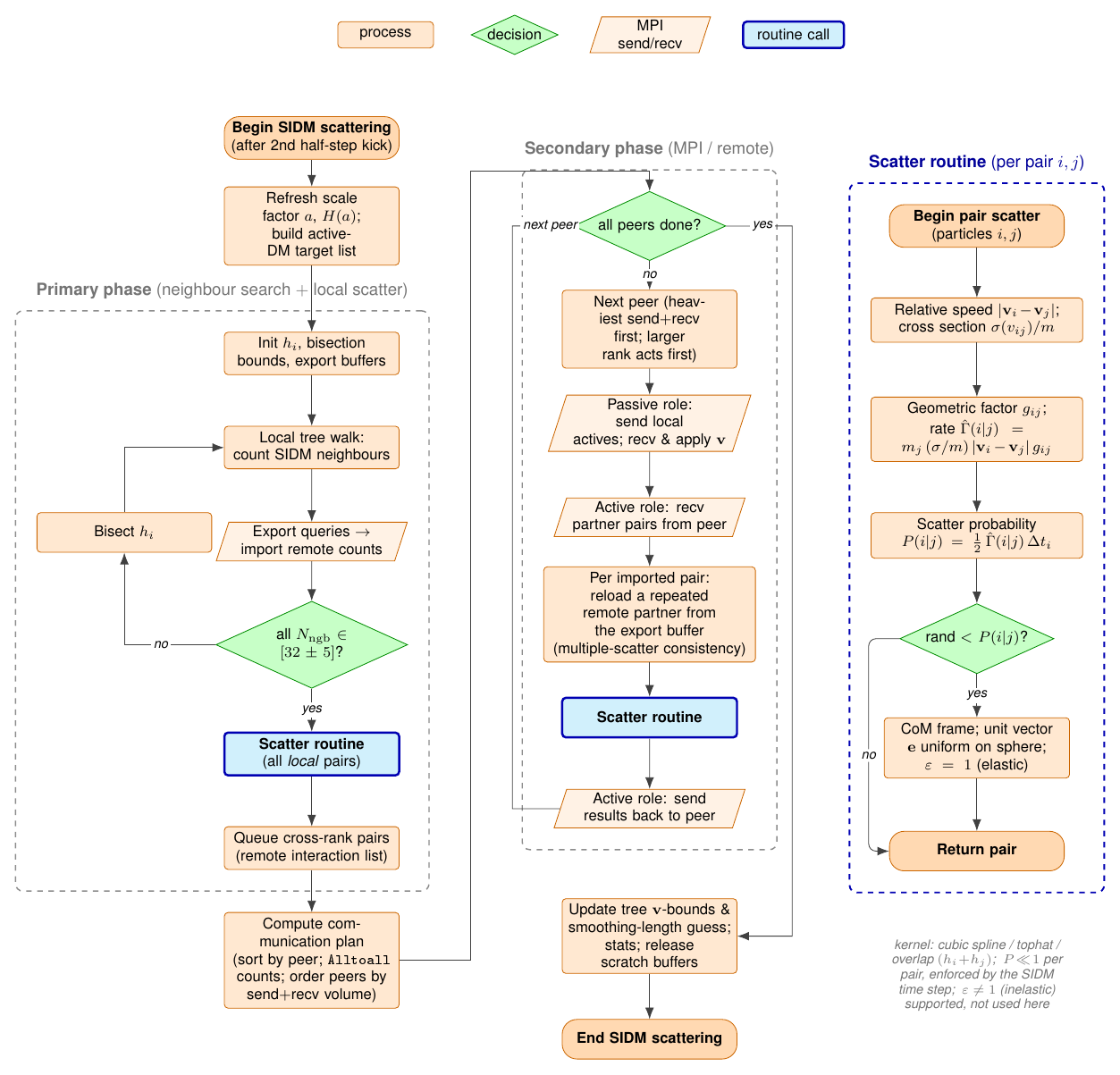}
    \caption{Control flow of the per-timestep SIDM scattering operation in \arepoTwo, read left to right. It traces the steps the module performs each timestep for the dark-matter particles that are \emph{active}, starting just after the gravity solver's second half-step kick. In the legend (top), rounded boxes are local computations, green diamonds are decisions, parallelograms are exchanges between parallel processes, and blue-outlined boxes call the shared per-pair \emph{scatter routine} (right).
    \emph{Primary phase (left; \cref{subsec:sidm_local}).} After refreshing cosmological factors and building the active-target list, each target's neighbours are found with the dedicated DM-only tree: the smoothing length $h_i$ is adjusted by bisection until the neighbour count is $N_{\rm ngb}=32\pm5$. Pairs whose particles share an MPI task are scattered immediately after the search; pairs straddling two tasks are queued, and a communication plan orders the task-to-task exchanges by data volume.
    \emph{Secondary phase (middle; \cref{subsec:sidm_parallel}).} The queued cross-task pairs are resolved by pairing tasks; each pairing exchanges particles with neighbours on the other task. One task is \emph{passive} (it sends its particles and later applies the returned updates) and the other \emph{active} (it receives, scatters, and returns the results), with the higher-rank task taking the active role first. If an imported particle scatters several times in one step, its latest state is reloaded from the export buffer before each scatter (the ``multiple-scatter consistency'' box) so that successive scatters build on one another. Once all pairings finish, the maximum velocity in each tree node is refreshed, and the step ends.
    \emph{Scatter routine (right; \cref{subsec:montecarlo}).} Both phases call this routine for a candidate pair $(i,j)$: it forms the scattering rate $\hat{\Gamma}({i|j})=m_j\,(\sigma/m)\,|\mathbf v_i-\mathbf v_j|\,g_{ij}$ (\cref{eq:rate}) and probability $P({i|j})=\tfrac{1}{2}\hat{\Gamma}({i|j})\,\Delta t_i$ (\cref{eq:probabilityScatter}; the criterion of \cref{subsec:sidm_timestep} keeps $P\ll1$), draws a random number, and only if it falls below $P$ applies an isotropic, momentum- and energy-conserving kick in the centre-of-momentum frame (shown for the elastic case $\epsilon=1$; inelastic, $\epsilon\neq1$, and anisotropic scatterings are also supported).}
    \label{fig:sidm_flow}
\end{figure*}

\begin{figure*}
    \centering
    \includegraphics[width=\linewidth, trim={0 20cm 22cm 0}, clip]{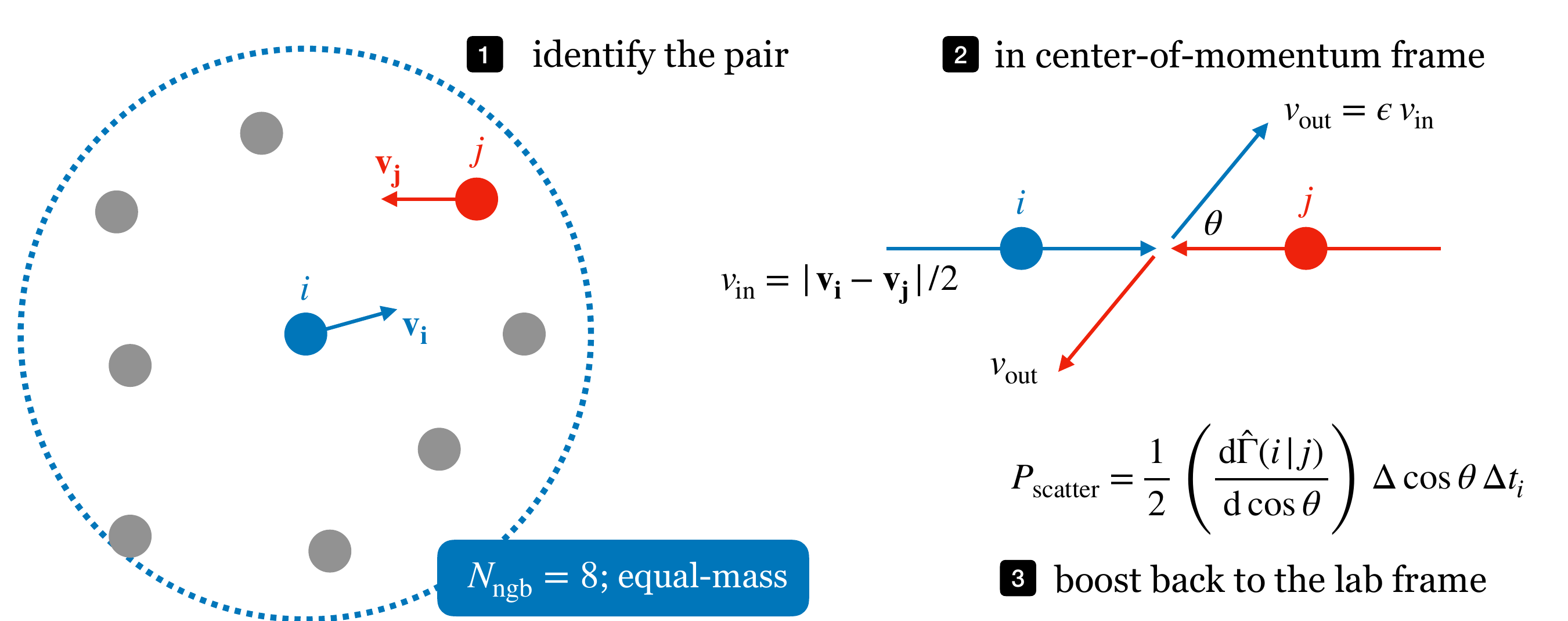}
    \caption{Scattering operation for a pair of DM particles in the Monte-Carlo scheme in \arepoTwo. We assume a target neighbour number $N_{\rm ngb}=8$ and a system of equal-mass particles in this illustration. For an accepted pair of particles $(i,j)$, the scattering deflection angle $\theta$ in the centre-of-momentum frame is drawn randomly according to the differential cross-section. The outgoing centre-of-momentum-frame relative speed is the incoming relative speed multiplied by $\epsilon$ to encapsulate potential inelasticity. The particle velocities are boosted back to the simulation (lab) frame.}
    \label{fig:schematic_methods}
\end{figure*}

\section{The new SIDM module in \arepoTwo}
\label{sec:sidm_impl}

In this section, we describe how SIDM is implemented as a module in \arepoTwo, with flexibility in numerical schemes, kernel choices, and neighbour-finding strategies. We focus on the fiducial Monte-Carlo scheme used throughout this paper that was briefly introduced in \cref{subsec:other_schemes}. We first define the collision operator, scattering probability, and velocity update (\cref{subsec:montecarlo}), together with the kernel and rate-estimator choices (\cref{subsec:kernelchoice}). We then describe the dedicated DM neighbour tree (\cref{subsec:sidm_local}), the per-pair timestep criterion (\cref{subsec:sidm_timestep}), and the consistent parallel scattering protocol (\cref{subsec:sidm_parallel}). Finally, \cref{subsec:sidm_interface} presents the four-function user interface, and \cref{subsec:sidm_arepo1_diff} summarizes the key differences from the \arepoOne\ implementation. \cref{fig:sidm_flow} shows a flow diagram of the SIDM implementation.

\begin{figure*}
    \centering
    \includegraphics[width=\linewidth, trim={0 16cm 22cm 0}, clip]{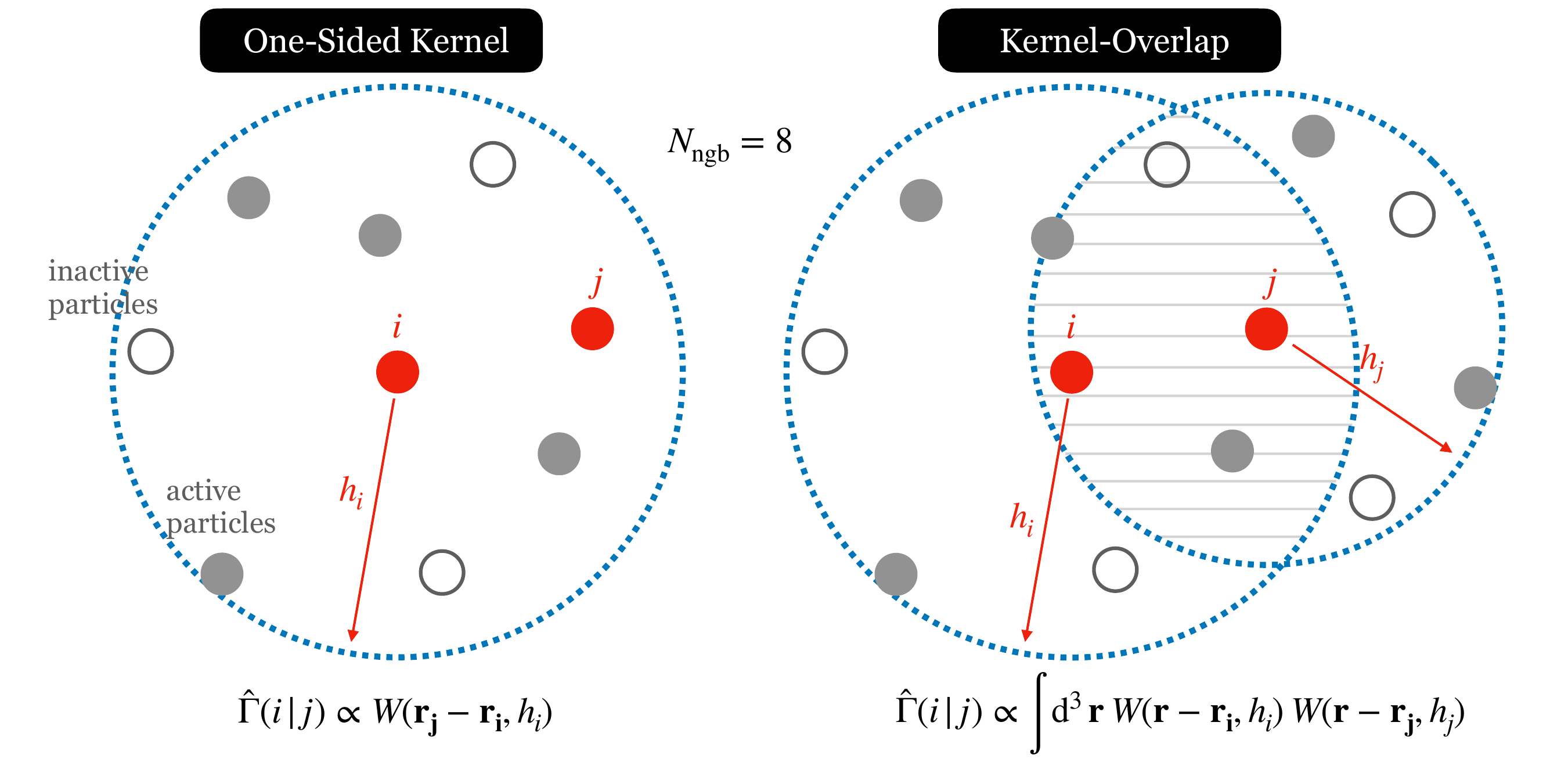}
    \caption{The two approaches to estimate the scattering rate for a candidate pair $(i,j)$ in \arepoTwo, illustrated assuming a target neighbour number $N_{\rm ngb}=8$. \emph{Left:} The fiducial method based on the one-sided kernel (\cref{eq:oneSided}), for which $\hat{\Gamma}({i|j})$ depends only on the smoothing length $h_i$ of particle $i$ and the partner $j$ is weighted by the kernel of $i$ alone. \emph{Right:} The kernel-overlap approach (\cref{eq:kernelOverlap}), for which $\hat{\Gamma}({i|j})$ is built from the spatial overlap of the two particles' kernels and is symmetric under $i\leftrightarrow j$. Red circles highlight the candidate pair, while grey filled (open) circles denote active (inactive) particles in the neighbourhood. Note that the neighbour search and the calculation of $h$ are both based on all particles, including inactive ones.}
    \label{fig:schematic_kernels}
\end{figure*}

\subsection{Monte-Carlo scheme}
\label{subsec:montecarlo}

The collision operator in the Boltzmann equation can be written as
\begin{equation}
    \dfrac{\mathrm{D}\hat{f}(\mathbf{x}, \mathbf{v}, t)}{\mathrm{D}t} = \mathcal{C}[\hat{f}, \sigma] = \sum \hat{\mathcal{C}}_{ij}[\hat{f}, \sigma],
\end{equation}
where $\hat{f}$ is the coarse-grained distribution function in $N$-body simulations, $\hat{\mathcal{C}}_{ij}$ is the numerical collision operator for a pair of DM particles labelled by $(i,j)$, and the sum runs over all DM particle pairs. In practice, this operator is derived by evaluating the scattering rates between a target particle ($i$) and its neighbouring particles ($j$),
\begin{equation}
    \hat{\Gamma}({i|j}) = m_{j}\, (\sigma/m)\,|\mathbf{v}_{i} - \mathbf{v}_{j}|\,g_{ij},
    \label{eq:rate}
\end{equation}
where $g_{ij}$ is a geometric factor that weights the pair according to their spatial separation. Its form is discussed in \cref{subsec:kernelchoice}.

In the Monte-Carlo scheme the probability that particle $i$ scatters with a neighbour $j$ (out of its $N_{\rm ngb}$ neighbours) during its timestep $\Delta t_i$ is
\begin{equation}
    P({i|j}) = \dfrac{1}{2}\, \hat{\Gamma}({i|j})\,\Delta t_i.
    \label{eq:probabilityScatter}
\end{equation}
In our implementation, each interacting pair will be evaluated twice, once in the neighbour search of $i$ and once in that of $j$, so we add the $1/2$ prefactor here. For $P({i|j})\ll1$, this recovers the intended per-pair scattering rate even in the case where particles $i$ and $j$ have different timesteps. 
The velocity-dependent cross-section $\sigma(v)/m$ can be supplied either analytically or as an arbitrary tabulated function of the pair's relative velocity, as in \arepoOne~\citep{vogelsberger2012subhaloes,vogelsberger2019evaporating}, and is evaluated for each interacting pair using linear interpolation in log space for the velocity. In cosmological simulations, the particle coordinates and velocities are stored in comoving form. We evaluate the pair rate of \cref{eq:rate} in physical units at each step, converting the comoving separations, smoothing lengths, and relative velocities with the scale factor $a$, so that the scattering probability tracks the physical collision frequency at every epoch.

During each scattering event, we first compute the centre-of-momentum velocity of the pair
\begin{equation}
\mathbf{v}_{\rm com}=\frac{m_{i}\mathbf{v}_{i} + m_{j}\mathbf{v}_{j}}{m_{i} + m_{j}}.
\end{equation}
The updated velocities of the pair are
\begin{equation}
\mathbf{v}^{\prime}_{i}= \mathbf{v}_{\rm com} + \epsilon\,\frac{m_{j}\,|\mathbf{v}_{i}-\mathbf{v}_{j}|}{m_{i} + m_{j}}\,\mathbf{e},
\qquad
\mathbf{v}^{\prime}_{j}= \mathbf{v}_{\rm com} - \epsilon\,\frac{m_{i}\,|\mathbf{v}_{i}-\mathbf{v}_{j}|}{m_{i} + m_{j}}\,\mathbf{e}.
\end{equation}
The unit vector $\mathbf{e}$ controls the deflection angle. In the isotropic case considered in this paper, it is drawn uniformly on the sphere. For anisotropic scatterings, it should be drawn according to the angle dependence of $\sigma/m$. Here $\epsilon$ is the ratio of the outgoing to incoming relative speed of the pair: $\epsilon=1$ recovers elastic scattering, while $\epsilon\neq 1$ effectively describes inelastic collisions \citep{Essig2019,vogelsberger2019evaporating,Shen2021dSIDM,Shen2022dSIDM} and can have more complicated velocity dependence to mimic arbitrary DM cooling functions. We restrict the present work to the elastic case, $\epsilon=1$. The centre-of-momentum kinetic energy of the pair changes by a factor $\epsilon^{2}$ per scatter, so an inelastic model is fixed by relating $\epsilon$ to the energy released or absorbed in the interaction. The whole operation on a pair of scattering particles is illustrated in \cref{fig:schematic_methods}.

\subsection{Scattering rate estimator and kernel choice}
\label{subsec:kernelchoice}

The geometric factor $g_{ij}$ of \cref{eq:rate} has dimensions of inverse volume and encodes the spatial proximity and overlap of the pair. It is determined by two choices: the scattering rate estimator and the smoothing kernel $W$. We describe both below and state our fiducial selection.

In the fiducial one-sided kernel estimator, the geometric factor is the kernel of particle $i$ evaluated at the partner's position (see \cref{fig:schematic_kernels}, left),
\begin{equation}
    g_{ij} = W(\mathbf{x}_{j} - \mathbf{x}_{i}, h_{i}).
    \label{eq:oneSided}
\end{equation}
The advantage is that the neighbour search depends only on the smoothing length of particle $i$ and does not require any information about particle $j$ other than its coordinates. Therefore, the scattering candidate list can be generated within the same loop as the computation of smoothing lengths. In this estimator, $h_{i}$ is a free parameter and can be chosen adaptively. A caveat is that this operator does not exactly reproduce the collision operator derived from first principles. However, for sufficiently large particle numbers and small smoothing lengths, it converges to the correct result, and the convergence can be established empirically.

An alternative kernel-overlap estimator~\citep{rocha2013cosmological} (\cref{fig:schematic_kernels}, right) is derived directly from the collisional Boltzmann equation by treating each simulation particle as a discrete patch of the phase-space distribution. The resulting geometric factor is given by the overlap integral of the two particles' smoothing kernels,
\begin{equation}
\label{eq:kernelOverlap}
    g_{ij} \;\equiv\; \int \mathrm{d}^3\mathbf{x}\;
    W\!\left(|\mathbf{x}-\mathbf{x}_i|,h_i\right)\,
    W\!\left(|\mathbf{x}-\mathbf{x}_j|,h_j\right).
\end{equation}
For compact kernels, the integral is non-zero only when the particle separation $r_{ij}\equiv|\mathbf{x}_i-\mathbf{x}_j|$ satisfies $r_{ij}<h_i+h_j$. A practical drawback of this estimator is the more complicated neighbour search strategy. Since identifying the scattering partners for particle $i$ requires knowledge about $h_j$, all smoothing lengths must be determined in a first pass before the overlapping partners can be identified based on the overlap condition $r_{ij}<h_i+h_j$. 
We store the maximum smoothing length, $h_{\max}$, in each node of the neighbour tree introduced in the next subsection. This value is used to enlarge the search region relative to $h_i$. Because of the computational overhead associated with this approach, we retain the one-sided cubic spline as our fiducial estimator. The tree-based implementation of the overlap kernel, together with a consistency test against the fiducial method, is presented in \cref{app:OverlapKernel}.

For the one-sided kernel estimator, the choice of the smoothing kernel $W$ is arbitrary. A commonly used one is the cubic spline kernel,
\begin{equation}
W_{\rm CS}(q,h)=\frac{8}{\pi h^{3}}
\begin{cases}
1-6q^{2}+6q^{3}, & 0 \le q \le \frac{1}{2},\\[4pt]
2(1-q)^{3}, & \frac{1}{2} < q \le 1,\\[4pt]
0, & q>1,
\end{cases}
\end{equation}
where $q \equiv r/h$, and the smoothing length $h$ is the radius of compact support. We determine $h$ iteratively by enforcing a target neighbour number of $N_{\rm ngb}=32\pm 5$. Another widely used choice~\citep[e.g.,][]{kochanek2000quantitative,Robertson2017SIDM} is the top-hat kernel,
\begin{equation}
    W_{\mathrm{TH}}(q,h) \;=\;
    \begin{cases}
        \dfrac{3}{4\pi h^3}, & q \le 1, \\[6pt]
        0, & q > 1,
    \end{cases}
\end{equation}
for which all neighbours within the fixed search radius contribute equally to the scattering probability, simplifying the algorithm at the cost of a discontinuity at the kernel edge and the loss of the central weighting provided by smooth kernels.

\subsection{Dedicated DM neighbour tree}
\label{subsec:sidm_local}

Standard $N$-body codes typically use a hierarchical octree to accelerate gravitational force evaluations. In principle, the SIDM algorithm of the previous section could reuse this gravity tree for neighbour searches. However, as discussed in \cref{sec:arepo2}, the gravity tree in the hierarchical time-integration scheme of \citet{springel2021simulating} is not fully rebuilt at every timestep, so using it for SIDM would require a full tree construction at every synchronization point. In hydrodynamical simulations, the gravity tree also contains baryonic particles, so additional type checks during the tree walk would be needed to filter out non-DM partners, adding further overhead. We therefore implement a dedicated, dynamically updated neighbour tree containing only DM particles. Its structure closely follows the existing gas-cell neighbour tree in \arepoTwo.

\vspace{0.1cm}

\noindent\textbf{Tree structure and dynamic updates.} Immediately after each domain decomposition, we construct a distributed DM-only octree covering all DM particles. To allow the tree to track the system between domain decompositions, we store additional kinematic information in every node: (i) the minimum and maximum particle coordinates along each axis, and (ii) the minimum and maximum particle velocity components along each axis. Each node's axis-aligned bounding box is then drifted between decompositions by advancing its lower and upper faces along each axis with the corresponding minimum and maximum velocity components over the elapsed time, giving a conservative, directional bound. During neighbour searches for a particle, we test whether the SIDM kernel overlaps with these time-expanded boxes. If there is no overlap, the node can be discarded; otherwise, the node is opened.
Whenever a particle's velocity changes, either through SIDM scattering or gravitational acceleration, we walk from that particle to the root and update the per-node velocity extrema. Because the bounding boxes grow monotonically between decompositions, node overlaps gradually increase, and the neighbour search becomes less efficient. However, as shown below, this degradation is mild and has no significant impact on the overall performance.

\vspace{0.1cm}

\noindent\textbf{Tree-walk optimizations.} The implementation also includes two architecture-aware optimizations. (i) \emph{Grouped traversal.} We walk the tree for groups of four particles in parallel. Because particles are ordered along a Peano--Hilbert curve, consecutive particles are spatially correlated and tend to require opening the same nodes. If any particle in a group needs a node to be opened, we open it for all four. Some particles thus pay for unnecessary work, but the gain in cache locality more than compensates, yielding a typical speed-up of $\sim 2.5$ in our tests over independent walks. (ii) \emph{Shared memory.} The top-level tree can use intra-node shared memory, reducing the overall memory footprint. A natural future improvement is to store multiple particles directly in the tree leaves, accelerating the construction phase.

\subsection{Timestep criterion}
\label{subsec:sidm_timestep}
A key feature of our algorithm is that, even when a particle undergoes multiple scatters within a single timestep, its state is updated self-consistently immediately after each scattering event. Every accepted scatter conserves momentum and kinetic energy by construction, so for elastic scatterings, the SIDM scattering substep does so to floating-point round-off. The scheme becomes inaccurate only if the scattering probability for an individual pair approaches unity. We therefore impose a per-pair timestep criterion,
\begin{equation}
\Delta t_i < C_{\rm SIDM}\,\frac{2}{ m_{\rm DM}\, W(0,h_i)\, \left(v\, \sigma/m\right)_{\max}},
\label{eq:sidm_timestep}
\end{equation}
with the denominator evaluating the upper bound of the per-pair scattering rate involving particle $i$. $\left(v\,\sigma/m\right)_{\max}$ is the maximum of the product of the relative velocity (between particle $i$ and its neighbours from the previous timestep) and cross-section at this velocity. $m_{\rm DM}$ is the (constant) DM particle mass. The factor of two in the numerator compensates for the per-encounter factor of $1/2$ of \cref{eq:probabilityScatter}: each pair is processed once from $i$'s neighbour list and once from $j$'s. 
This guarantees that the pairwise scattering probability $P({i|j})$ between particle $i$ and any single neighbour $j$ remains $\ll 1$. This condition is less restrictive than the criterion adopted by \citet{vogelsberger2012subhaloes}, which required the total scattering probability of particle $i$ during the timestep, $P_i^{\rm tot}=\sum_j P({i|j})$, to remain $\ll 1$. In this work, we adopt the fiducial value $C_{\rm SIDM}=0.1$.

It is instructive to compare this criterion with the gravitational one of \cref{eq:dt_grav}. For the cubic-spline kernel $W(0,h_i)=8/(\pi h_i^{3})$ and a fixed neighbour number, $\rho\simeq N_{\rm ngb}\,m_{\rm DM}/(\tfrac{4}{3}\pi h_i^{3})$, so \cref{eq:sidm_timestep} reduces to $\Delta t_{\rm SIDM}\simeq (3 N_{\rm ngb}\,C_{\rm SIDM}/16)\,[\rho\,(\sigma/m)\,v]^{-1}$, a fixed fraction of the local collision time. Estimating the local acceleration as $|\mathbf{a}|\simeq v\,\sqrt{4\pi G\rho}$, the ratio of the two criteria is
\begin{equation}
\frac{\Delta t_{\rm SIDM}}{\Delta t_{\rm grav}} \simeq \frac{3 N_{\rm ngb}\,C_{\rm SIDM}}{16}\,\frac{(4\pi G)^{1/4}}{(\sigma/m)\,\sqrt{2\eta\,\epsilon_{\mathrm{soft}}}}\;\rho^{-3/4}\,v^{-1/2}.
\label{eq:dt_ratio}
\end{equation}
The SIDM timestep criterion, therefore, dominates in dense regions during the core collapse.

\subsection{Consistent parallel scattering}
\label{subsec:sidm_parallel}

\begin{figure*}
    \centering
    \includegraphics[width=\linewidth, trim={0 2cm 1cm 8.5cm}, clip]{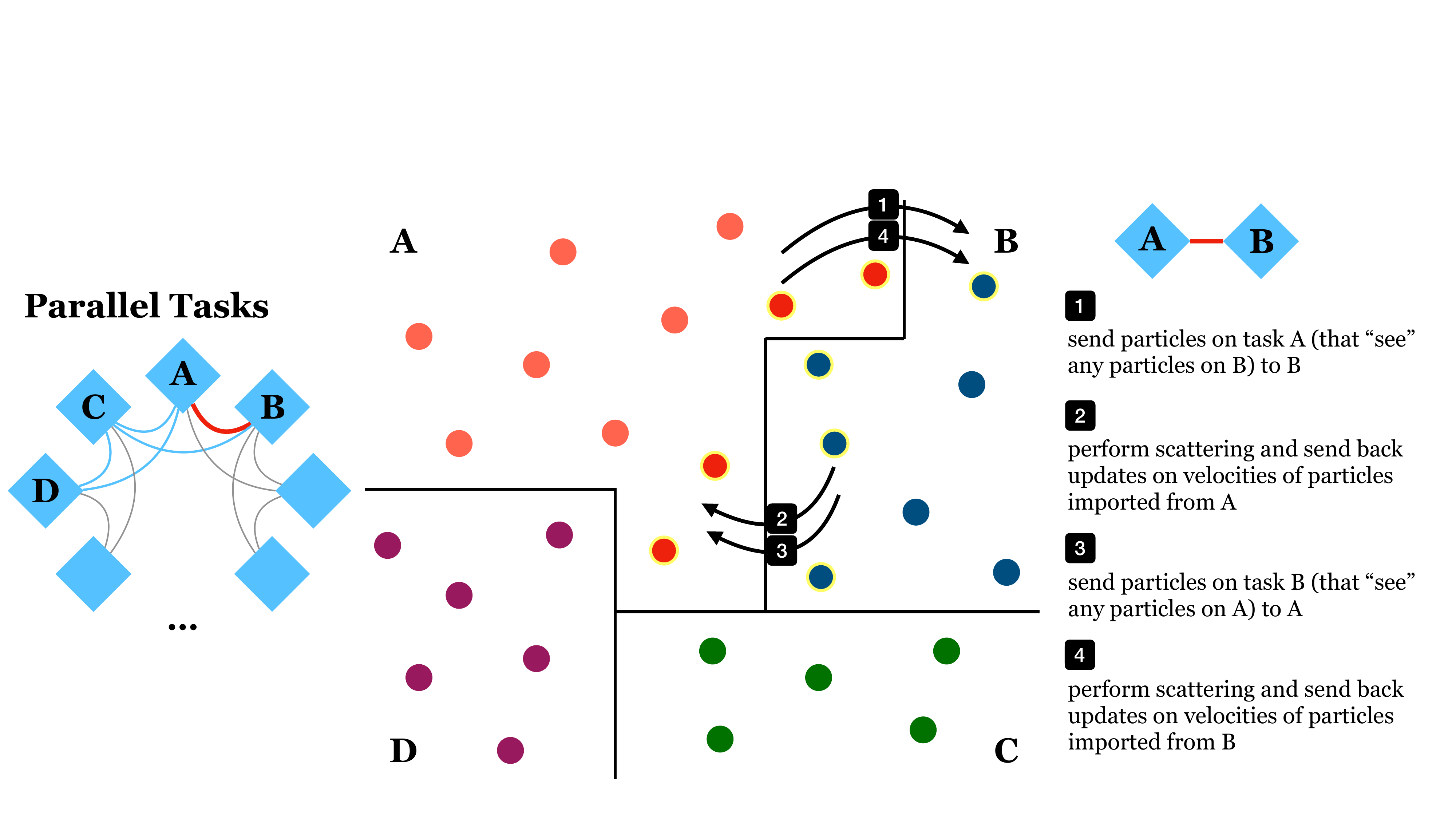}
    \caption{The consistent pairwise SIDM parallelization adopted in this work. MPI tasks are paired over a sequence of communication sub-steps sorted by their communication costs (\emph{left}). We highlight four MPI tasks (A--D) that are next to each other in the domain decomposition.
    For a given pair of tasks (A,B) that share a domain boundary, the boundary particles of A that can ``see'' particles on B are first sent to B (step~1); B then performs the scattering and returns the resulting velocity updates (step~2); the symmetric exchange is subsequently carried out in the opposite direction (steps~3--4). Because only one side of each task pair performs the update at any time, no particle is modified concurrently on two tasks, and momentum and energy are conserved to machine precision.}
    \label{fig:schematic_parallel}
\end{figure*}

The default domain decomposition in \arepoTwo distributes particles across MPI tasks, so scattering events between local and remote partners must be handled carefully to preserve conservation. For every active DM particle, we first perform an iterative neighbour search with the dedicated tree. The iteration follows a standard SPH approach and terminates when $32\pm 5$ neighbours are found within a smoothing length $h_i$. We store the local index of each neighbour and the rank of the MPI task that hosts it. Interactions are then processed in two stages, local and remote, which are done in the primary and secondary phases shown in \cref{fig:sidm_flow}. For each pair $(i,j)$ where both particles reside on the same MPI task, we compute the scattering probability according to \cref{eq:probabilityScatter}. If a scatter is accepted, the velocities of both particles are updated immediately, so that any subsequent interaction involving them within the same timestep uses their updated state. The remaining pairs, which involve at least one remote particle, are handled with an ordered pairwise communication scheme similar to that of \citet{Fischer2021}: the MPI tasks are iteratively grouped into pairs, each pair directly exchanges the data of its interacting particles, and the order in which a task visits its partners is chosen according to the expected communication volume. An illustration of this scheme is shown in \cref{fig:schematic_parallel}, with key steps summarized below:

\begin{enumerate}[leftmargin=*]
\item \textbf{Sorting and exchange.} On each task, the candidate remote interactions are sorted first by the rank of the remote task and then by the local particle index. An all-to-all exchange then communicates the number of export interactions per task, from which each task also learns how much data it will exchange with every other task.

\item \textbf{Pairwise iteration.} The code pairs MPI tasks iteratively, each communicating with one partner at a time; the order in which a task visits its partners is set by the load-balancing criterion described at the end of this section. Consider a pair with task ranks $A$ and $B$ with $A<B$.

\item \textbf{Export ($A\to B$).} Task $A$ identifies all interactions involving its local particles and remote partners on $B$. For each, it packages the necessary data (indices, position, velocity, $h_i$, timestep) and sends the list to $B$.

\item \textbf{Processing ($B$).} Task $B$ iterates through the list, loads the corresponding local particles, and computes the scattering probability from \cref{eq:probabilityScatter}. Updates to the \emph{local} particle (on $B$) are applied immediately, while updates to the \emph{remote} particle (from $A$) are accumulated in an export buffer. To remain self-consistent when a particle undergoes multiple scatters, $B$ checks whether the current remote particle is identical to the previous one processed. If so, it reloads the particle's properties from the export buffer (which holds the most recent update); otherwise, it initializes from the imported list. This relies on the prior sort by particle index.

\item \textbf{Return update.} Task $B$ sends the accumulated updates back to $A$, which applies them to its local particle array.
\end{enumerate}

\noindent The same pattern is then repeated with the roles reversed ($B$ exports, $A$ processes). Once both directions are complete, each task selects its next partner. The partner order is chosen for load balance: using the interaction counts exchanged in step~(i), each task estimates the volume of data to be exchanged with every potential partner and processes its partners in order of decreasing volume, communicating with the most expensive partners first. Front-loading the heaviest exchanges in this way keeps the busiest tasks occupied throughout the communication phase and improves the overall load balance.

\subsection{Module interface}
\label{subsec:sidm_interface}

A central design goal is to make the framework easy to extend to new pairwise interactions, such as angle-dependent cross-sections, state-dependent SIDM, and drag-like models. We achieve this by treating each candidate interaction independently, while the algorithm in \cref{subsec:sidm_parallel} guarantees self-consistency when a particle participates in multiple non-local interactions within a single timestep.

A user can change the scattering physics by implementing only the following four functions, without any knowledge of the neighbour search and parallelization scheme:
\begin{itemize}[leftmargin=*]
    \item \textbf{\texttt{scatter(state\_i, state\_j) -> (state\_i', state\_j')}}: Perform the pairwise interaction. In our case, this draws a random number and decides whether the two particles scatter.
    \item \textbf{\texttt{get\_state(particle) -> state}}: Extract the model state from the code's particle structure.
    \item \textbf{\texttt{set\_state(particle, state)}}: Write the updated state back into the particle structure.
    \item \textbf{\texttt{overwrite\_state(state\_in, state\_out) -> state}}: Replace the current input state with an updated output state, used during non-local scattering when a remote particle appears in multiple interactions.
\end{itemize}

\subsection{Key differences from the SIDM implementation in \arepoOne}
\label{subsec:sidm_arepo1_diff}

Although our model is mathematically equivalent to that of \citet{vogelsberger2012subhaloes}, the implementation differs in several important respects. In the previous code, the initial smoothing-length iteration also accumulated, for each particle, the sum of pairwise scattering probabilities over all neighbours. A local decision was then made as to whether that particle would scatter during the current timestep. For particles selected to scatter, a second tree walk retrieved the individual neighbour-wise probabilities. Each MPI task chose a partner by sampling from them. Finally, a third tree walk exported the particle, and the MPI task owning the selected neighbour executed the scatter.

This algorithm cannot consistently handle multiple scatters. If, for example, particle $a$ is selected to scatter with $b$ while $c$ is independently selected to scatter with $a$, both $b$ and $c$ work from the same initial state of $a$ and both try to update it, violating momentum and energy conservation. To mitigate this, after partners had been chosen, the code performed a global check for particles participating in more than one event and cancelled all scatters involving any such particle. In regimes where scattering is frequent, as for example during gravothermal collapse, this can significantly undercount the true number of scatters. To suppress conflicts, the previous implementation imposed a stringent timestep limit ensuring that the \emph{total} probability of a particle scattering with any neighbour satisfies $P_i\ll 1$.

Concretely, the \arepoOne\ scheme bounds the timestep so that the total probability for particle $i$ to scatter with any of its $N_{\rm ngb}$ neighbours during $\Delta t_i$,
\begin{equation}
    P_i = \Delta t_i \sum_{j=1}^{N_{\rm ngb}} \hat{\Gamma}({i|j}),
    \label{eq:total_prob_arepo1}
\end{equation}
remains below a small threshold, $P_i \le \kappa$. This corresponds to the timestep criterion
\begin{equation}
    \Delta t_i \le \frac{\kappa}{\sum_{j=1}^{N_{\rm ngb}} \hat{\Gamma}({i|j})},
    \label{eq:sidm_timestep_arepo1}
\end{equation}
with $\kappa = 2.5\times 10^{-3}$ \citep{vogelsberger2019evaporating,Mace2024}. In contrast to our per-pair criterion in \cref{eq:sidm_timestep}, which only requires the probability of \emph{each individual} pair to be small, the sum over all neighbours in \cref{eq:sidm_timestep_arepo1}, together with the much smaller value of $\kappa$ compared with our $C_{\rm SIDM}=0.1$, makes the \arepoOne\ timestep substantially more restrictive, and is the main reason for its higher cost in high-density regions such as the core-collapse phase.

The two criteria can be compared directly. Writing the bound in the denominator of \cref{eq:sidm_timestep} as the maximal single-pair rate $\hat{\Gamma}_{\mathrm{max},i}\equiv m_{\rm DM}\,W(0,h_i)\,(v\,\sigma/m)_{\max}$, our per-pair criterion reads $\Delta t_{\rm new}=2C_{\rm SIDM}/\hat{\Gamma}_{\mathrm{max},i}$, whereas \cref{eq:sidm_timestep_arepo1} reads $\Delta t_{\rm old}=\kappa/\sum_j\hat{\Gamma}({i|j})$, so that
\begin{equation}
\frac{\Delta t_{\rm old}}{\Delta t_{\rm new}}=\frac{\kappa}{2\,C_{\rm SIDM}}\,\frac{\hat{\Gamma}_{\mathrm{max},i}}{\sum_{j=1}^{N_{\rm ngb}}\hat{\Gamma}({i|j})}\simeq\frac{\kappa}{2\,C_{\rm SIDM}}\,\frac{32}{3\,N_{\rm ngb}}\,\frac{v_{\max}}{\langle v\rangle}.
\label{eq:dt_ratio_old_new}
\end{equation}
The second step uses the cubic-spline central weight $W(0,h_i)=8/(\pi h_i^{3})$ together with $\rho\simeq N_{\rm ngb}\,m_{\rm DM}/(\tfrac{4}{3}\pi h_i^{3})$ to write $m_{\rm DM}\,W(0,h_i)=(32/3)\,\rho/N_{\rm ngb}$, and $\sum_j\hat{\Gamma}({i|j})=\rho\,(\sigma/m)\,\langle v\rangle$ with $\langle v\rangle$ the kernel-weighted mean relative speed and $v_{\max}$ the maximum relative speed over the neighbours. For the fiducial $\kappa=2.5\times10^{-3}$, $C_{\rm SIDM}=0.1$ and $N_{\rm ngb}=32$, and with $v_{\max}/\langle v\rangle\sim 2$--$3$, this gives $\Delta t_{\rm old}/\Delta t_{\rm new}\sim10^{-2}$, i.e.\ the \arepoOne\ criterion forces timesteps roughly two orders of magnitude shorter. Because both timesteps scale as $[\rho\,(\sigma/m)\,v]^{-1}$, this factor is essentially density-independent, so the advantage persists from the long heating phase into the deep-collapse regime.

In addition, all tree walks in the previous code used the gravity tree, making the scheme incompatible with hierarchical gravity time integration and significantly increasing the cost of tree walks in simulations that include gas cells.

\section{Code verification}
\label{sec:codeVerification}

We now validate the new implementation through a suite of idealized tests of increasing complexity. Where an analytic solution is available, we compare against it directly. Otherwise, we benchmark the new module against the original \arepoOne\ implementation in \citet{vogelsberger2012subhaloes,vogelsberger2019evaporating}.

\subsection{Deflection by a uniform lattice}

Following e.g. \citet{rocha2013cosmological,Robertson2017SIDM,Fischer2021}, we first consider a beam-deflection test in which a beam of test particles with the same velocity propagates through a uniform, static target medium. We compare the simulated distribution of deflection angles (measured relative to the initial beam axis) against the analytic expectation. We assume isotropic scattering in this test.

The initial conditions consist of $80^3=512\,000$ target particles placed on a uniform Cartesian grid and initially at rest, together with $512\,000$ beam particles that are randomly distributed within the same volume and all move along a common axis with speed $v=2.5\,\kms$. Placing the target particles on a regular lattice enforces a genuinely constant target density and avoids artefacts (e.g.\ spurious low-density regions) that would arise from a random Poisson sampling of the target medium. Each particle, in both the beam and the target population, has the same mass $10^{5}\,\Msun$. Both populations uniformly fill a cubic volume of side $25\,\kpc$, so the target particles define a uniform target density $\rho_{\rm target}=3.28\times10^{6}\,\Msun\,\kpc^{-3}$. Beam particles scatter off target particles with a cross-section per unit mass $\sigma/m=10\,\cpm$, while beam--beam and target--target scatterings are disabled and gravity is switched off. 

For the angular-deflection validation, we adopt the fixed-target, or no-recoil, limit. In this limit, the target velocities are not updated during a beam--target scattering, corresponding formally to infinite target inertia in the scattering kinematics. This approximation affects only the recoil update: the scattering rate is still set by the finite target mass density $\rho_{\rm target}$. Consequently, the magnitude of the beam-particle velocity remains constant and only its direction changes, allowing a direct comparison to the analytic fixed-target solution for the angular distribution. We evolve the system until every beam particle has scattered at least once.

\begin{figure}
    \centering
    \includegraphics[width= 1.02 \linewidth]{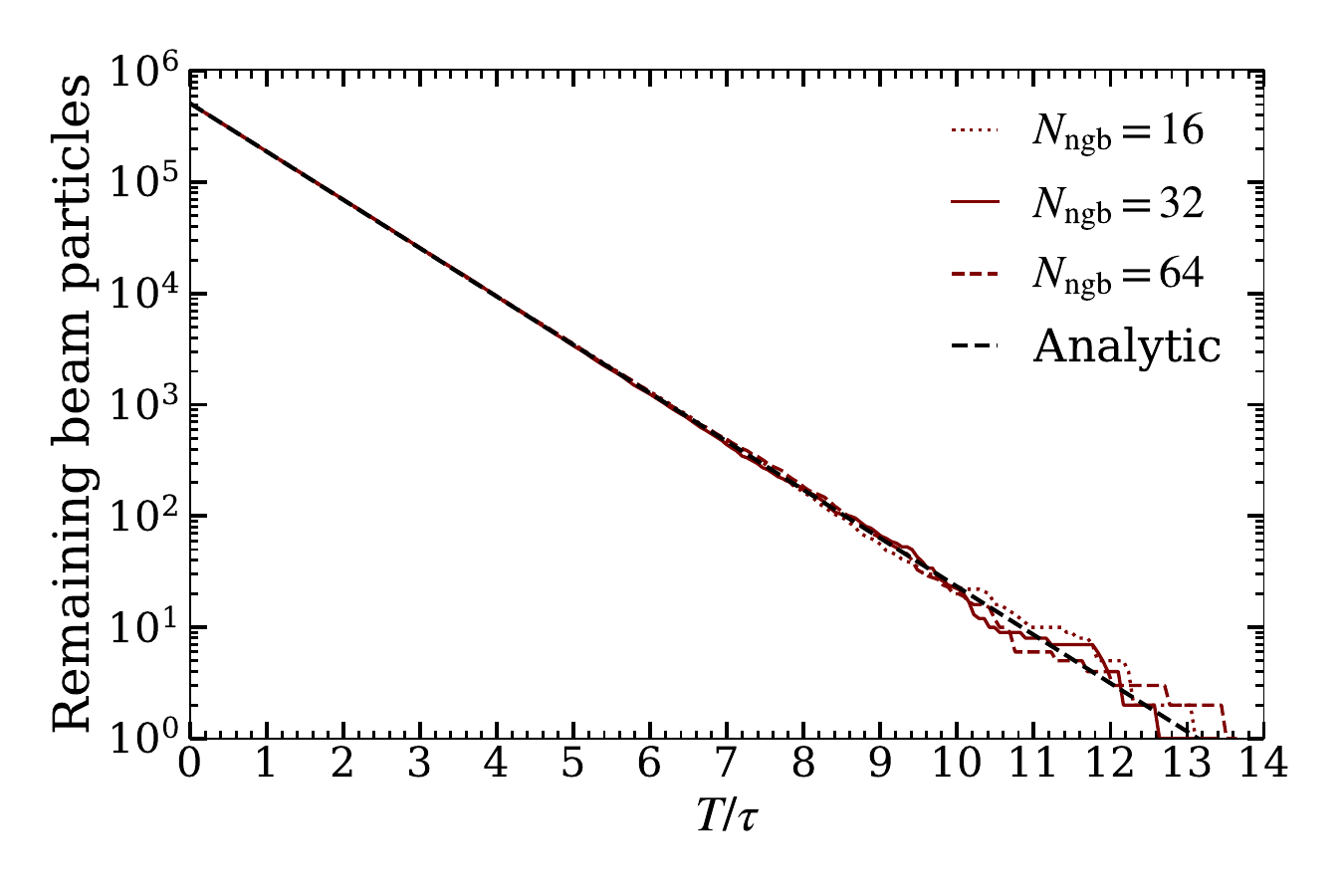}
    \includegraphics[width= 0.95\linewidth]{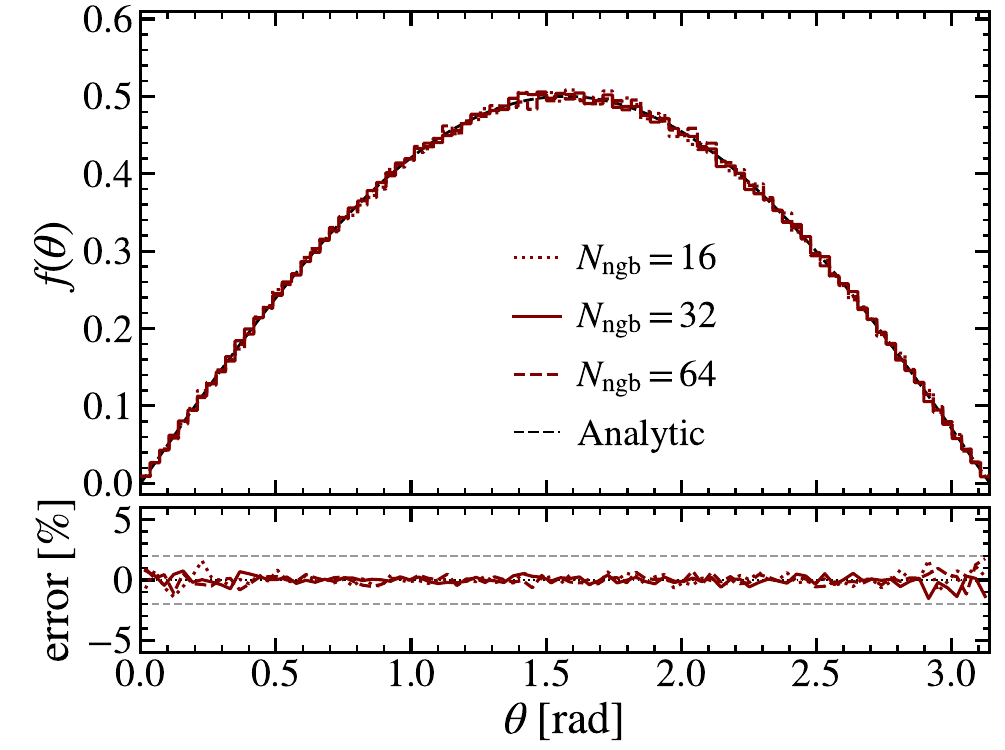}
    \caption{Uniform-lattice beam tests. Top: Time evolution of the number of unscattered beam particles as a function of scaled time $T/\tau$ with mean collision time $\tau\simeq 57.1~\Gyr$. The red curves correspond to neighbour numbers $N_{\rm ngb}=16$ (dotted), $32$ (solid) and $64$ (dashed); the black dashed curve is the analytic expectation $N(T)=N_0\exp(-T/\tau)$ with $\tau^{-1} = \rho_{\rm target}\,(\sigma/m)\,v$. The numerical results follow the predicted exponential decay and are converged with respect to $N_{\rm ngb}$. At late times, when only a small number of unscattered beam particles remain, the curves naturally become noisier due to the reduced counting statistics. Bottom: Probability distribution of beam-particle deflection angles, where $\theta$ is the deflection angle relative to the initial beam axis. The black dashed curve is the exact isotropic solution $f(\theta)=(1/2)\sin\theta$. The test uses $\sigma/m =10\,\cpm$. All neighbour choices reproduce the expected angular dependence, confirming that the deflection distribution is insensitive to $N_{\rm ngb}$. The lower sub-panel shows the percentage error between the numerical and analytic distributions.}
    \label{fig:deflection_tests}
\end{figure}

The upper panel in \cref{fig:deflection_tests} shows the time evolution of the number of unscattered beam particles for the three neighbour numbers $N_{\rm ngb}=16,\,32,\,64$. Because beam--beam interactions are disabled, each beam particle scatters independently at a constant rate $\Gamma=\rho_{\rm target}\,(\sigma/m)\,v$, so the number of survivors decays exponentially, $N(T)=N_0\exp(-\Gamma\, T)$, with $N_0=512\,000$ and a mean interaction time $\tau\equiv 1/\Gamma\simeq 57.1~\Gyr$. The simulations reproduce this exponential decay (black dashed line) and converge with $N_{\rm ngb}$. The lower panel shows the corresponding distribution of the deflection angles. For isotropic scattering, the outgoing directions are uniform in $\cos\theta$, so the expected probability distribution is $f(\theta)=(1/2)\sin\theta$. Defining $\theta$ as the angle between each particle's final velocity and the initial beam axis, we find that the simulation results for $N_{\rm ngb}=16,\,32,\,64$ closely follow the analytic curve over the full range $0\le\theta\le\pi$, with only minor bin-to-bin noise. The implementation thus recovers the expected isotropic angular statistics, independently of the neighbour number.

\subsection{Thermalization in a uniform box}
\label{subsec:Thermalisation}
\begin{figure}
    \centering
    \includegraphics[width= 1\linewidth]{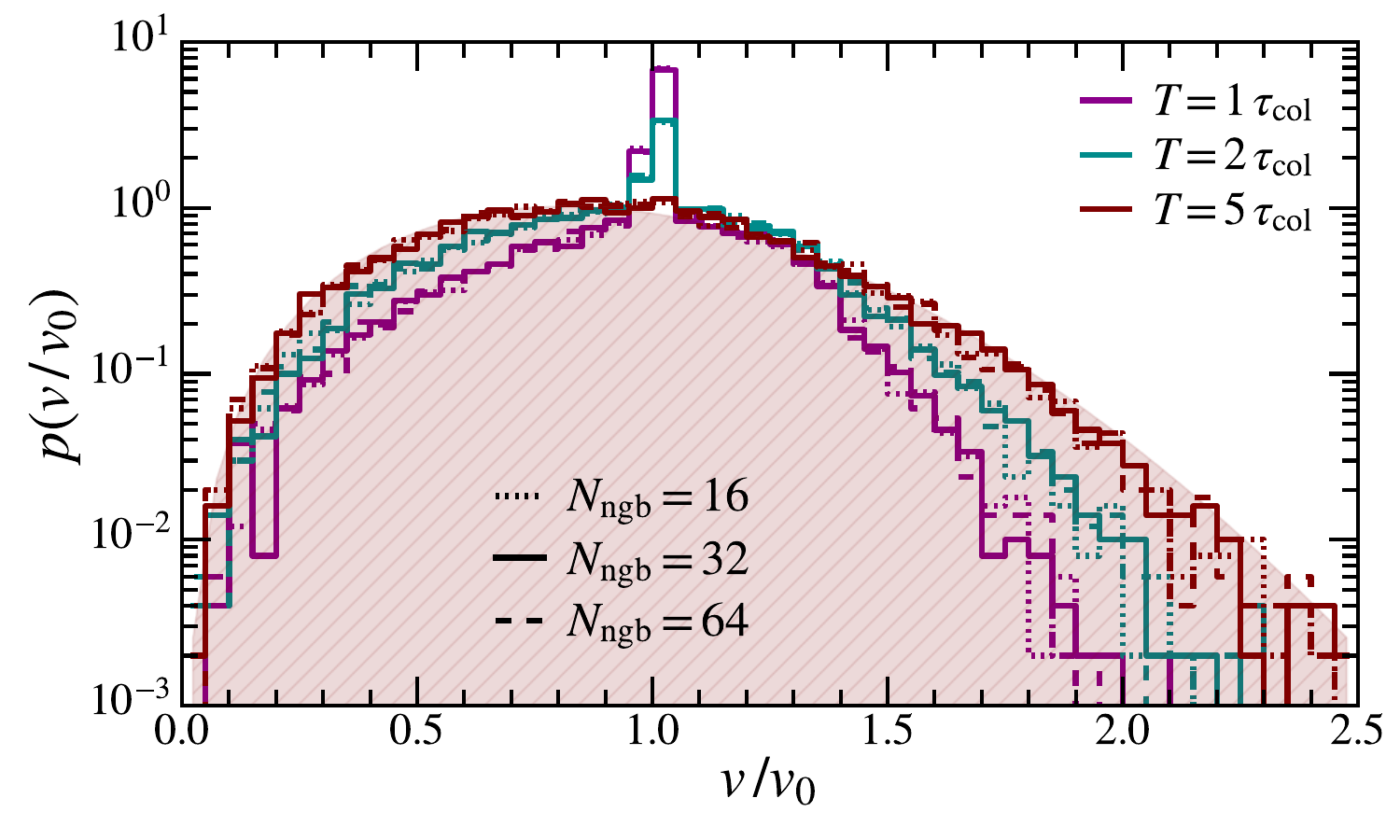}
    \includegraphics[width= 1\linewidth]{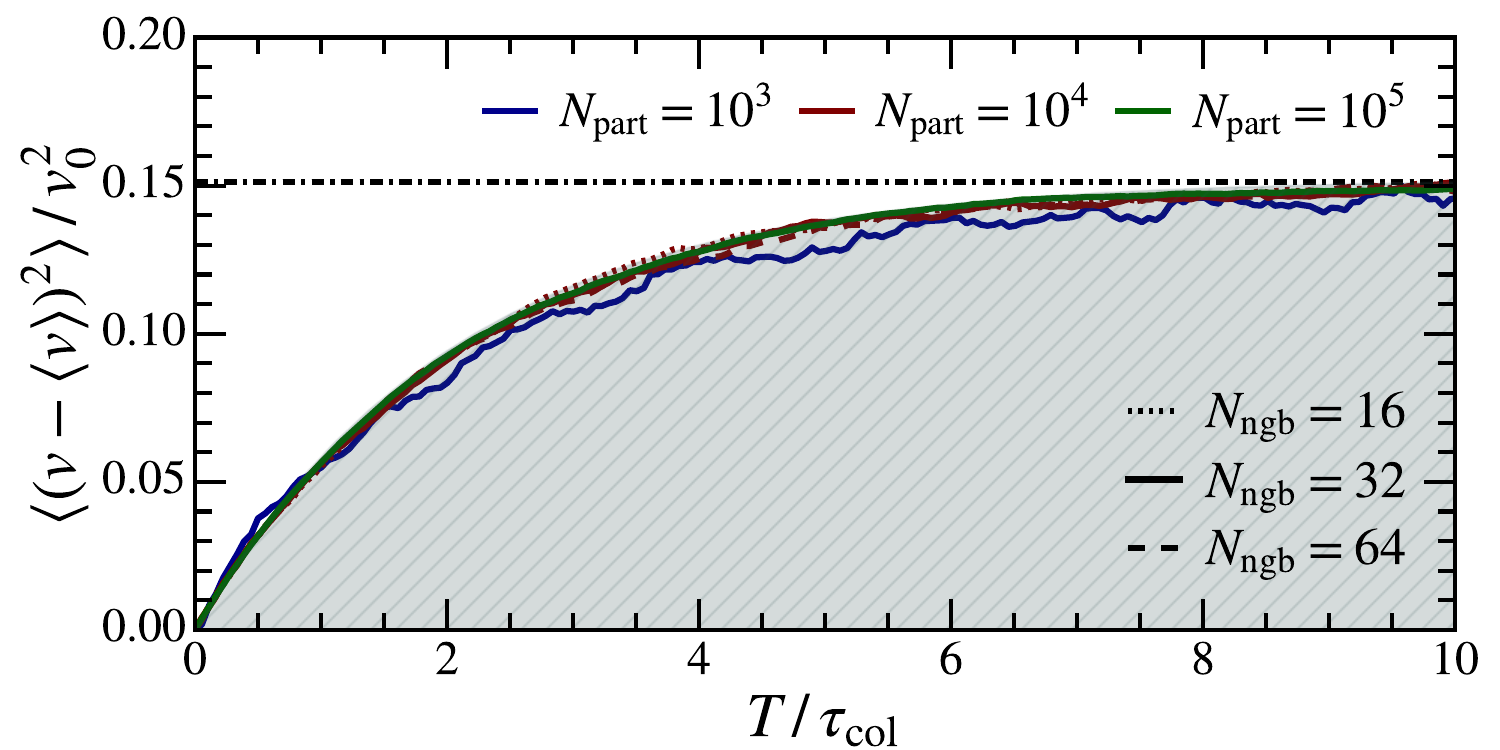}
    \caption{Top: speed distribution in the thermalization test with $N_{\rm{part}} = 10^4$ particles, evolved with $N_{\rm ngb} = 16,\,32,\,64$ (dotted, solid, and dashed). Snapshots at $T = 1\tau_{\rm col} , \, 2\tau_{\rm col} , \, 5\tau_{\rm col}$ are shown in purple, cyan, and red. The shaded region shows the fully thermalized MB distribution of \cref{eq:MB_distribution}. Bottom: evolution of the squared speed dispersion for these runs (red), together with higher- (green) and lower-resolution (blue) runs; the black dash-dot line marks the expected MB value and the shaded band the semi-analytic result from the Boltzmann solver of \cref{apd:MB_solver}. Thermalization proceeds as expected, independently of the numerical choices, and at a rate consistent with the analytic prediction.}
    \label{fig:thermalize_test}
\end{figure}

\begin{figure}
    \centering
    \includegraphics[width= 1\linewidth]{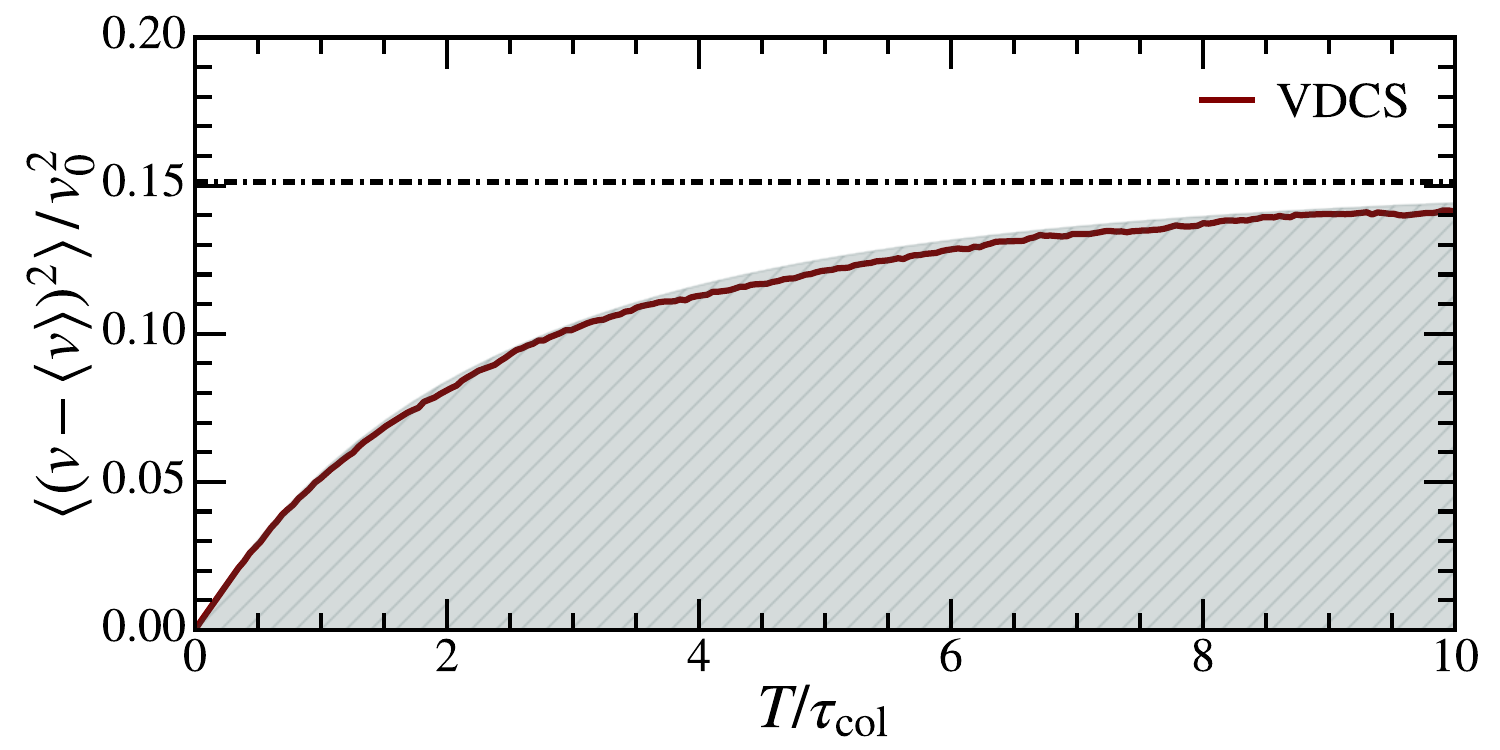}
    \caption{Similar to the bottom panel of \cref{fig:thermalize_test}, but for the Yukawa-type velocity-dependent cross-section with $\sigma_0/m=60\,\cpm$ and $w=2\,\kms$. Once again, the system thermalizes as expected, closely following the evolution predicted by the kinetic Boltzmann solver.}
    \label{fig:thermalize_test_VDCS}
\end{figure}

In the next step, we perform a thermalization experiment to test the SIDM solver in the absence of self-gravity. Although the problem is formally scale-free, we adopt a periodic box of side length $10\,\kpc$ containing $N_{\rm{part}}=10^4$ particles of total mass $M=10^{10}\,\Msun$ (density $\rho=10^7\,\Msun\,\kpc^{-3}$), similar to \citet{Fischer2021}. All particles are initialized with velocities of fixed magnitude $v_0=2\,\kms$ and isotropic directions. Elastic self-interactions should then drive the probability distribution of particle velocities towards a Maxwell--Boltzmann (MB) distribution with the same mean energy,
\begin{equation}
    \label{eq:MB_distribution}
    f_{\rm{MB}}(v) = 4\pi v^2 \left( \frac{3}{2\pi v_0^2} \right)^{3/2} \exp\left( -\frac{3v^2}{2v_0^2} \right).
\end{equation}
We adopt a cross-section $\sigma/m=10\,\cpm$, which corresponds to an asymptotic collision timescale
\begin{equation}
    \label{eq:thermalize_box_collision_time}
    \tau_{\rm col}
    = \frac{\sqrt{3 \pi}}{4 v_0 \left(\sigma/m\right) \rho} = 17.97 \, \Gyr.
\end{equation}
The factor of $4 v_0 / \sqrt{3 \pi}$ corresponds to the average relative velocity of an MB distribution with an average specific energy of $v_0^2 / 2$. This value is close to the average relative velocity $4 v_0 / 3$ of an isotropic mono-speed distribution with velocity $v_0$ as in the initial conditions. We therefore treat the collision timescale as constant throughout the evolution of the system. For velocity-dependent cross-sections (VDCSs), the cross-section per unit mass $\sigma/m$ is taken as the value at the aforementioned average relative velocity.

The runs are evolved to $T_{\rm max}=200\,\Gyr \simeq 11 \tau_{\rm col}$ using $N_{\rm ngb}=16,\,32,\,64$. Additionally, we probe mass-resolution convergence using runs with $N_{\rm{part}}=10^3$ and $10^5$. \cref{fig:thermalize_test} summarizes the time evolution of the velocity distribution and the velocity dispersion. In all cases, the system relaxes towards the MB configuration, converging after $\sim 5\tau_{\rm col}$, independently of neighbour number and resolution, and follows the Boltzmann-equation solution of \cref{apd:MB_solver} (also shown). The velocity dispersion of the system likewise converges to the MB value (the horizontal dash-dot line in the bottom panel) after ${\sim}10\tau_{\rm col}$. The total energy and momentum of the system are also conserved to machine precision. The observed relaxation is about a factor of two slower than reported by \citet{Fischer2021}. The difference is expected: their frequent-scattering (drag) rate is built on the symmetrized momentum-transfer cross-section $\tilde{\sigma}_{\rm T}$, whereas our isotropic rare-scattering rate uses the total cross-section $\sigma$; for identical particles $\tilde{\sigma}_{\rm T}=\sigma/2$, which accounts for the factor of two.

We repeat the thermalization test with an isotropic but velocity-dependent (Yukawa-type) cross-section, $\sigma(v)/m = (\sigma_0/m)\,[1+(v/w)^2]^{-2}$, where we adopt $\sigma_0/m = 60\,\cpm$ and $w= 2\,\kms = v_0$, so that the cross-section varies by more than an order of magnitude across the populated relative velocities. Because elastic collisions conserve energy, the equilibrium is again the MB distribution of \cref{eq:MB_distribution} and only the relaxation rate changes. The corresponding reference solution is obtained from the same kinetic Boltzmann solver (\cref{apd:MB_solver}), now evaluated with the VDCS. \cref{fig:thermalize_test_VDCS} shows the evolution of the squared velocity dispersion for the thermalization test, together with the corresponding reference solution from the kinetic Boltzmann solver. Once again, the simulation closely follows the reference solution, while both the total energy and momentum of the system are conserved to machine precision.

We also use this test to examine the convergence against the SIDM timestep parameter $C_{\rm SIDM}$ in \cref{eq:sidm_timestep}. Keeping $N_{\rm{part}}=10^4$ and $N_{\rm ngb}=32$ but removing the externally imposed maximum timestep, so that \cref{eq:sidm_timestep} alone sets $\Delta t$, we scan $C_{\rm SIDM}$ from $0.1$ to $5$. This yields timesteps $\Delta T=C_{\rm SIDM}\times 15.625\,\Gyr$, i.e.\ roughly $C_{\rm SIDM}$ collision times. Even for $C_{\rm SIDM}=5$, the system still relaxes correctly to the MB distribution, because the scattering probability for each particle pair remains below unity and the expected number of scattering events is therefore recovered. The total energy and momentum are conserved to machine precision despite the large individual timesteps.

\subsection{Isolated halo core collapse}
\label{subsec:isolatedCoreCollapse}
\begin{figure}
    \centering
    \includegraphics[width= 1 \linewidth]{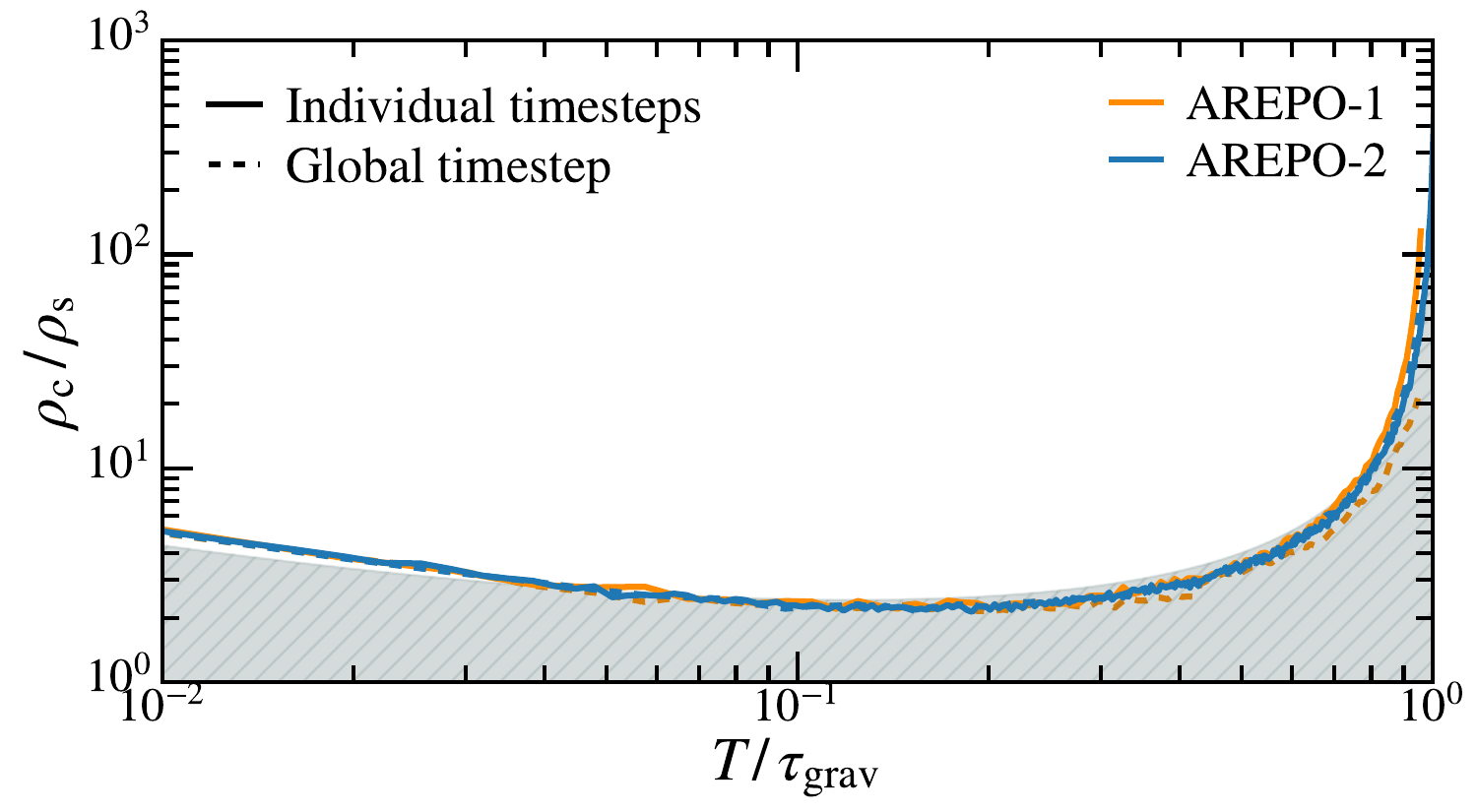}
    \includegraphics[width= 1 \linewidth]{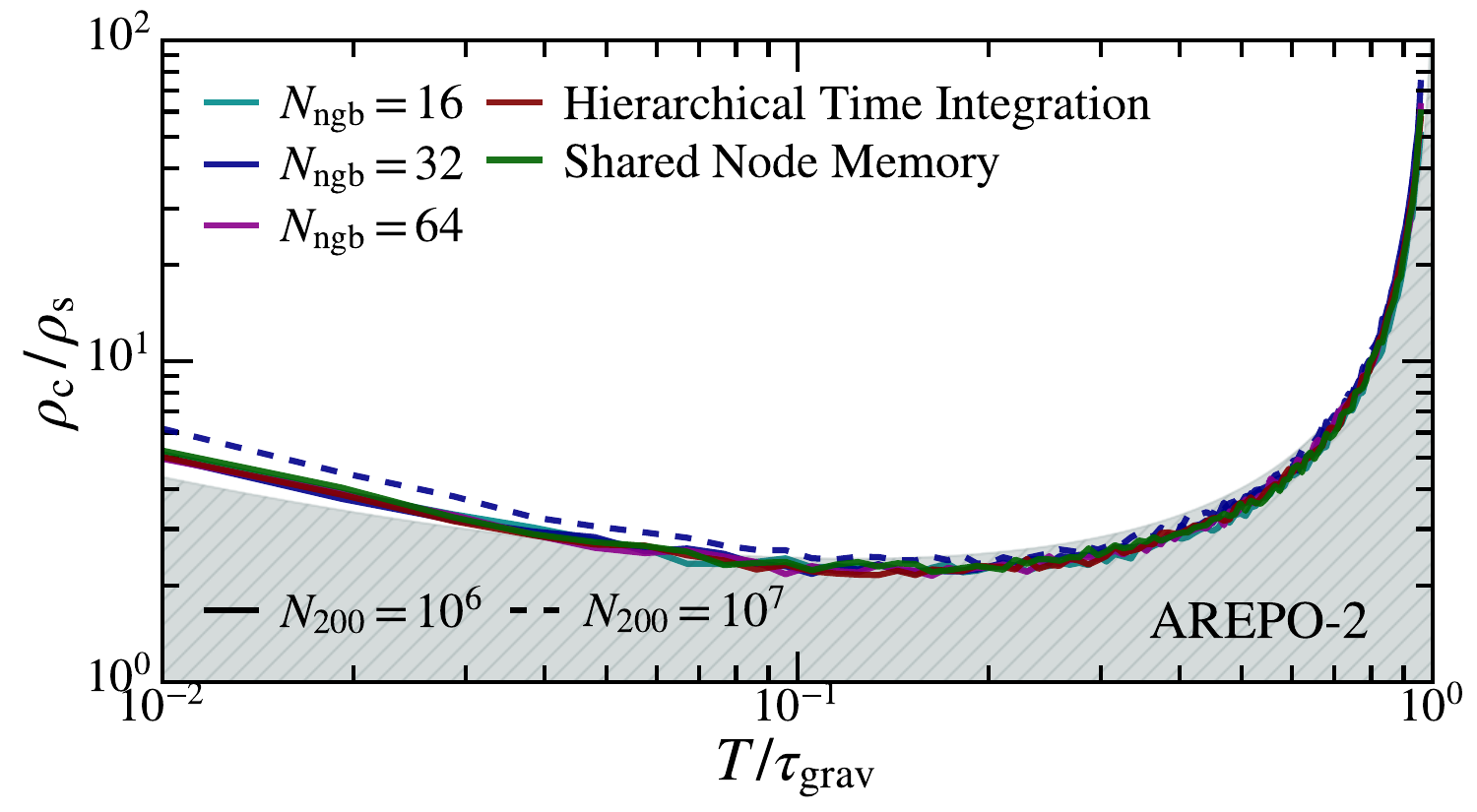}
    \caption{Core-density evolution of an isolated halo with mass $M_{200} = 10^{10} \, \Msun$ and concentration $c = 13.08$. The core density $\rho_{\rm c}$ is normalized to the NFW scale density $\rho_{\rm s}$, while the physical time $T$ is normalized to the collapse timescale $\tau_{\rm grav}$ of \cref{eq:collapse_timescale}. $N_{200}$ indicates the number of particles within the virial radius $r_{200}$ for each run. The top panel compares the new and old codes with both adaptive individual timesteps (solid) and a global timestep (dashed); the bottom panel compares different numerical choices within the new code. For reference, the shaded bands show the gravothermal-fluid solution (see \cref{apd:fluid_code} for details) assuming $C_{\kappa}=0.8$. The collapse is essentially independent of the numerical choices, agrees closely with the fluid model, and is somewhat more stable in the new code than in the old one.}
    \label{fig:collapse_test}
\end{figure}

To test the coupling between the SIDM module and self-gravity, we now follow the gravothermal collapse of an isolated halo. The halo is initialized as a stable CDM configuration following \citet{Tran2024_1}, with the modifications of \citet{Tran2025_3} that ensure stability in the presence of finite gravitational softening. The corresponding initial-condition (IC) generator, \textsc{SoftIsoICs}, is publicly available\footnote{\url{https://github.com/vinh-qtran/SoftIsoICs}}. Within the virial radius $r_{200}$, i.e. the radius within which the average density is $200$ times the critical density of the Universe, the halo follows a classical NFW profile \citep{navarro1997universal}, and it transitions to a modified exponential cut-off \citep{Springel1999} beyond $r_{200}$,
\begin{align}
    \label{eq:nfw_inner_density}
    \rho_{r < r_{200}} (r) &= \frac{\rho_{\rm{s}}}{\left(r/r_{\rm s}\right) \left(1 + \left(r/r_{\rm s}\right)\right)^2}, \\
    \label{eq:nfw_outer_density}
    \rho_{r > r_{200}} (r) &= \frac{\rho_{\rm s}}{c \left(1 + c\right)^2} \left(\frac{r}{r_{200}}\right)^{\epsilon_{\rm d}} \exp\left(-\frac{r-r_{200}}{r_{\rm d}}\right).
\end{align}
We adopt a scale density $\rho_{\rm{s}}=1.10\times 10^{7}\,\Msun\,\kpc^{-3}$ and scale radius $r_{\rm{s}}=3.48\,\kpc$, a virial mass (the mass enclosed within $r_{200}$) of $M_{200}=10^{10}\,\Msun$, and a concentration $c=13.08$; for simplicity we set the decay scale $r_{\rm d}=r_{200}$. Continuity of the logarithmic density slope at $r_{200}$ then fixes the decay index $\epsilon_{\rm d}$,
\begin{equation}
    \label{eq:nfw_decay_exponential}
    \epsilon_{\rm d} = \frac{r_{200}}{r_{\rm d}} - \frac{1+3c}{1+c}.
\end{equation}
We sample $N_{200}=10^6$ particles within $r_{200}$ and extend the sampling out to $r_{\rm max}=5\,r_{200}$. The resulting Plummer-equivalent gravitational softening, $\epsilon=0.04\,\kpc$, satisfies the convergence criteria of \citet{Mace2024}. We measure the core density following \citet{Tran2025_2}. At each snapshot, we determine the halo centre using the shrinking-sphere method \citep{Power2003} and partition the halo into shells with log-linearly spaced radial bins $r_i$, merging adjacent shells as needed to ensure each contains a minimum bin count of $400$ particles. Beginning from the innermost radius $r_i$ at which the enclosed particle count reaches $N_i \ge 2500$, i.e.\ a minimum core count of $2500$, we compute the mean enclosed density $\bar\rho_i \pm \Delta\bar\rho_i$ under Poisson statistics. We identify $\bar\rho_i$ as the core density $\rho_{\rm c}$ at the first radius satisfying
\begin{equation}
    \label{eq:core_density}
    \lvert \bar\rho_i - \rho_{i+1} \rvert \, \geq 2 \left[\left(\Delta{\bar\rho_i}\right)^2+\left(\Delta{\rho_{i+1}}\right)^2\right]^{1/2},
\end{equation}
where $\rho_{i+1}$ and $\Delta{\rho_{i+1}}$ are the mean density of the shell between $r_i$ and $r_{i+1}$ and its associated Poisson uncertainty. If this criterion is not met, we advance to $r_{i+1}$ and repeat. A profile fit \citep[e.g.,][]{Yang2023,Fischer2024,Tran2025_1} yields very similar results. We quote times in units of the characteristic gravothermal collapse time \citep{Essig2019,Yang2024},
\begin{equation}
    \label{eq:collapse_timescale}
    \tau_{\rm grav} = \frac{150}{C_{\kappa}} \frac{1}{\sigma/m} \frac{1}{\rho_{\rm{s}}} \left(\frac{1}{4 \pi \, G \, \rho_{\rm{s}} \, r_{\rm{s}}^2}\right)^{1/2}.
\end{equation}
For $\sigma/m=30\,\cpm$ and $C_{\kappa}=0.8$ (calibrated against the fluid model in \cref{apd:fluid_code}), this gives $\tau_{\rm grav} = 31.33\,\Gyr$. Guided by this collapse time and the results of \citet{Mace2024}, we set the gravity-timestep accuracy parameter to $\eta=0.005$ in \cref{eq:dt_grav}; \cref{app:eta} verifies this choice explicitly.

We evolve the halo with both the new implementation in \arepoTwo\ and the old one in \arepoOne\ on a single node, using either adaptive individual timesteps or a single global timestep. To stress-test the new code further, we add runs with $16\pm 5$ and $64\pm 5$ neighbours, one run with hierarchical time integration enabled, and one using shared intra-node memory for the DM neighbour tree on $8$ computing nodes. We also evolve a higher-resolution halo with $N_{200}=10^7$ particles within $r_{200}$ and softening length $\epsilon=0.016\,\kpc$. \cref{fig:collapse_test} compares the resulting core-density evolution with the gravothermal-fluid solution \citep{Outmezguine2023,Gad-Nasr2024} of \cref{apd:fluid_code}. \arepoOne and \arepoTwo agree well, with \arepoTwo slightly more stable across timestep choices. As expected, the evolution is essentially independent of the numerical settings and closely tracks the gravothermal fluid prediction. In addition to numerical settings, we investigate the convergence of core collapse with respect to resolution, random seed, and timestep, with results presented in \cref{app:convergence_all}.

\subsection{Cosmological zoom-in simulations}
\label{subsec:zoominCoreCollapse}

\begin{figure}
    \centering
    \includegraphics[width=1\linewidth]{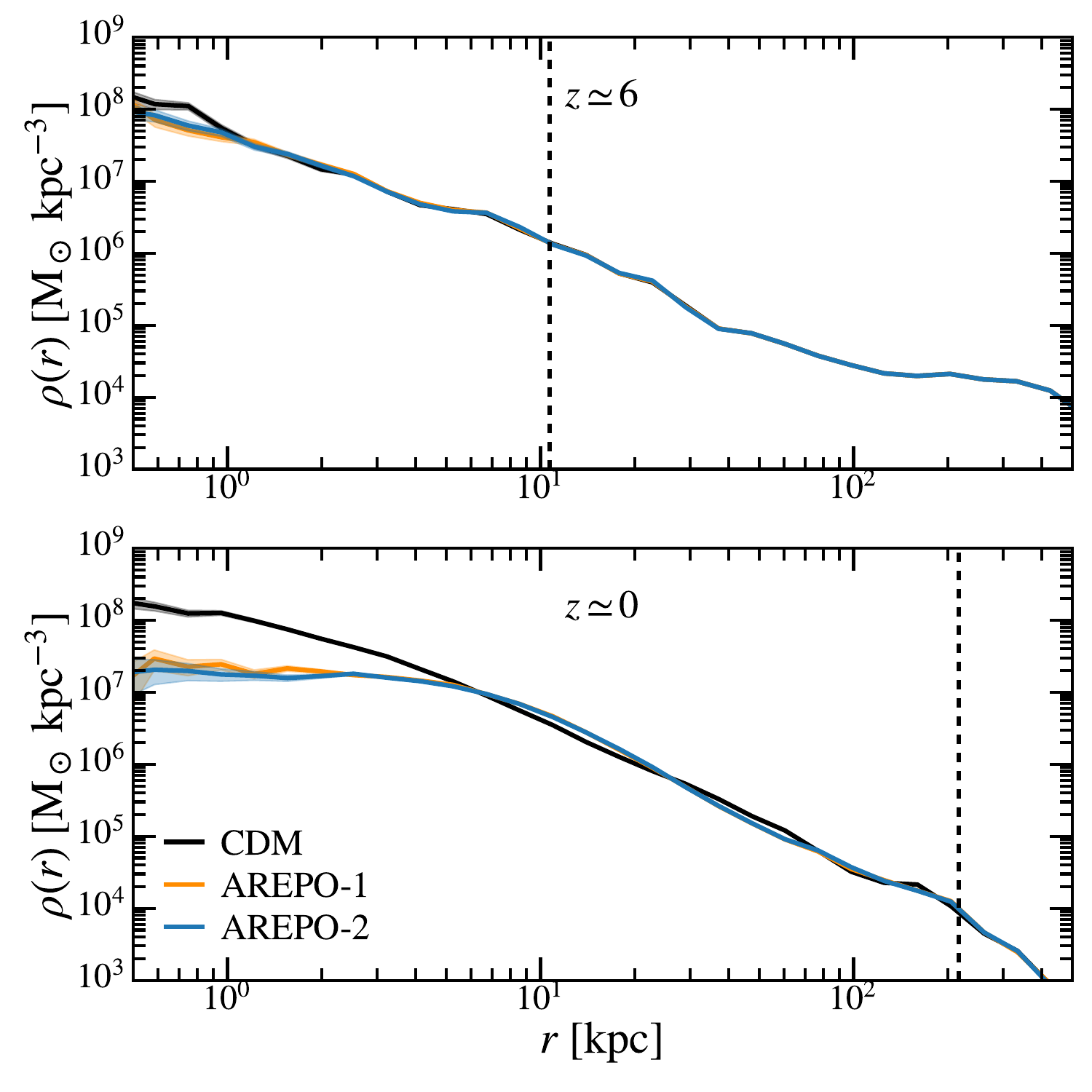}
    \caption{Density profiles of a Milky Way-mass halo at $z=6$ (top) and $z=0$ (bottom), for zoom-in simulations in CDM (black) and SIDM (orange: \arepoOne; blue: \arepoTwo). Shaded bands show the $1\sigma$ uncertainties and the dashed vertical line marks the virial radius $r_{200}$. The two SIDM implementations are consistent at both redshifts.}
    \label{fig:zoomin_main_halo}
\end{figure}

\begin{figure}
    \centering
    \includegraphics[width=1\linewidth]{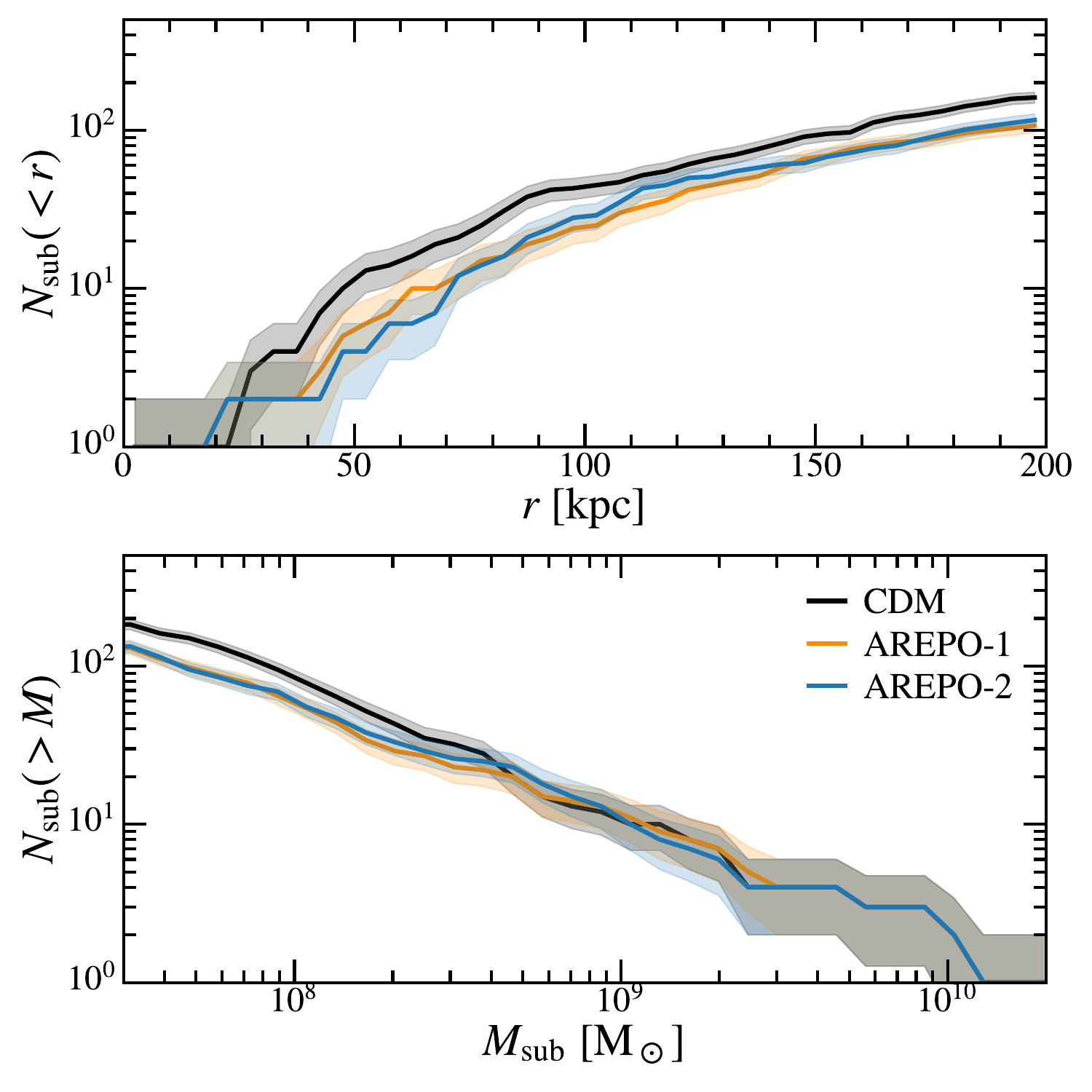}
    \caption{Top: cumulative number of subhaloes with $M_{\rm sub} > 10^7 \msun$ within radius $r$ at $z=0$ for CDM (black) and SIDM (orange: \arepoOne; blue: \arepoTwo). Bottom: cumulative subhalo mass function (number with mass greater than $M$) at $z=0$ for the same runs. Shaded bands show the $1\sigma$ Poisson uncertainties. As expected, the subhalo spatial distribution and mass function agree between the old and new SIDM implementations, within the statistical uncertainties.}
    \label{fig:zoomin_sub_halo}
\end{figure}

In the next step, we test the SIDM module in cosmological settings. We start with DM-only zoom-in simulations of Milky Way-mass haloes selected at $z=0$. The initial conditions are generated with \textsc{music} \citep{HahnAbel2011} at $z=127$, embedded in a comoving box of side $100\,h^{-1}\,{\rm cMpc}$, and target a main halo of mass $M_{200}\simeq 1\times 10^{12}\,\Msun$ at $z=0$. We adopt $\Omega_{\rm m}=0.302$, $\Omega_\Lambda=0.698$ and $H_0=100\,h\,\kms\,\Mpc^{-1}$ with $h=0.691$. Each run resolves the target halo with $7\,159\,600$ high-resolution particles of mass $1.764\times 10^6\,\Msun$, with a comoving Plummer-equivalent softening of $0.61\,{\rm c}\kpc$ at $z\geq 1$ and a maximum physical softening of $0.305\,\kpc$ at lower redshifts. Self-interactions are only activated for the high-resolution particles, so the background low-resolution particles do not scatter. The setup is a lower-resolution version of the one studied in \citet{ONeil2023} in the context of exothermic and endothermic SIDM. We use a FoF group finder and the SUBFIND algorithm \citep{springel2001populating} to identify substructures.

\cref{fig:zoomin_main_halo} shows the density profiles of this halo at $z=6$ and $z=0$ in simulations of CDM and SIDM with $\sigma/m=10\,\cpm$, in both the \arepoOne and \arepoTwo implementations. We observe that even at $z=6$, when deviations between the CDM and SIDM models first begin to emerge, the halo evolution is nearly identical between \arepoOne and \arepoTwo. By $z=0$, both implementations show clear core formation, with core properties in close agreement. \cref{fig:zoomin_sub_halo} shows the subhalo spatial distribution and mass function at $z=0$ for the same runs. Both implementations produce equivalent suppression of the subhalo abundance, particularly in the low-mass regime, and are consistent across all spatial scales.

\subsection{Large-volume cosmological simulations}
\label{subsec:cosmo_profiles}

As a final cosmological check, we conduct a suite of DM-only simulations in periodic boxes in \arepoOne\ and \arepoTwo. This is the same suite later used for the performance study of \cref{sec:performance}, with parameters collected in \cref{tab:simulation_overview}.

\begin{figure*}
    \centering
    \includegraphics[width=1\linewidth]{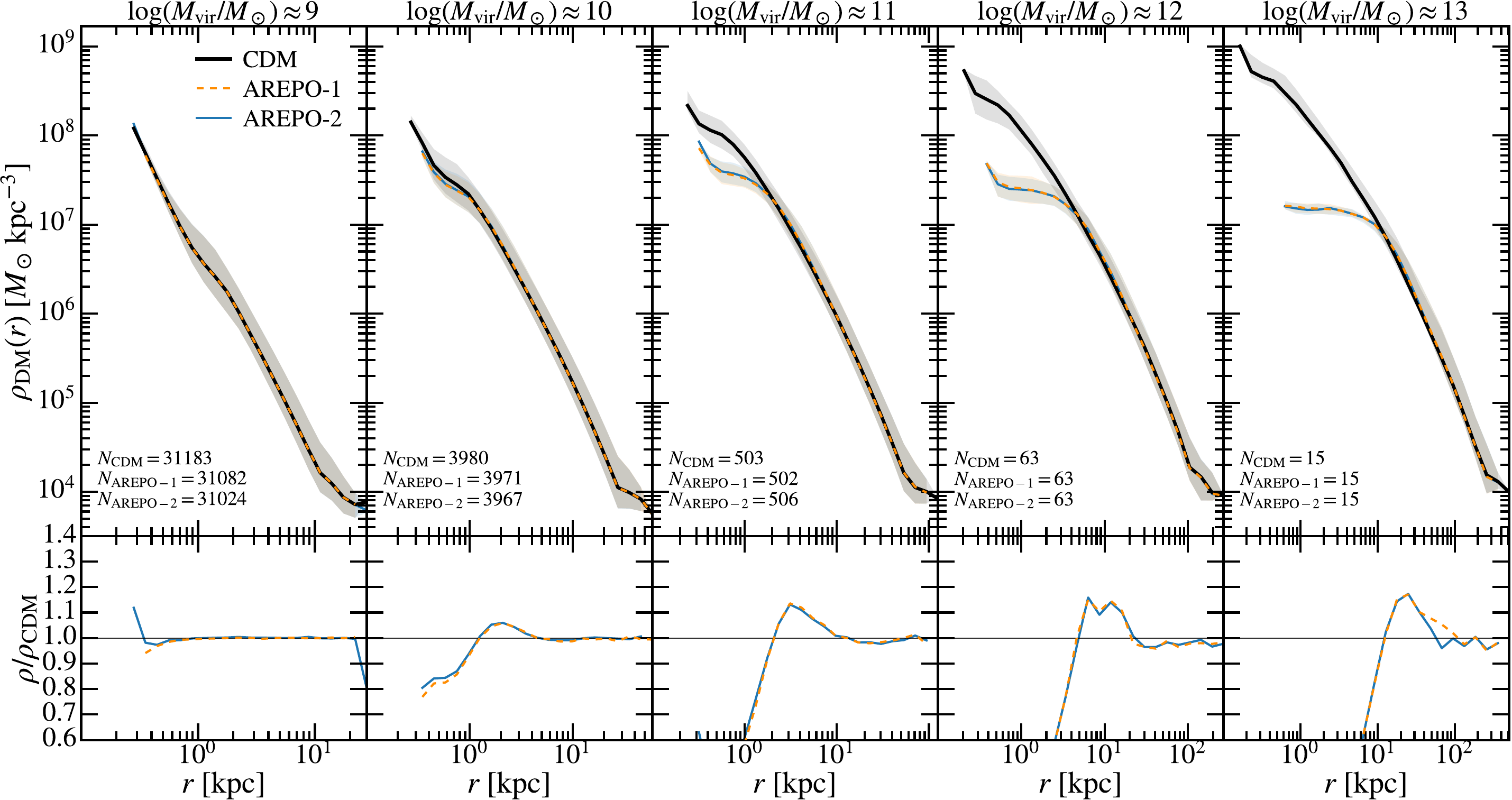}
    \caption{Median DM density profiles at $z=0$ in five virial-mass bins of half-width $\Delta \log\left(M_{\rm vir} / \Msun\right) = 0.5$ (left to right), from the DM-only runs with $1024^3$ particles.
    Each bin is labelled with the numbers of haloes in the CDM and in the \arepoOne\ (SIDM) and \arepoTwo\ (SIDM) runs.
    We compare the density profiles in CDM (black), the \arepoOne\ SIDM implementation (orange), and the \arepoTwo\ SIDM implementation (blue). The upper panels show the median profiles with $16$th--$84$th percentile bands, and the lower panels show the ratio of densities with respect to the CDM profile.
    Massive haloes develop a central core due to the larger SIDM scattering rates (for a constant cross-section), and the two SIDM implementations agree well.}
    \label{fig:dm_profiles}
\end{figure*}

As a basic verification, we compute the median DM density profile of haloes at $z=0$ in five bins of halo mass, each of half-width $\Delta\log(M_{\rm vir}/\Msun)=0.5$. \cref{fig:dm_profiles} shows the results for the simulations with $1024^3$ particles, assuming a constant cross-section of $1\,\cpm$.
The three most massive bins develop a central core, while the inner profiles of lower-mass haloes are only weakly affected---a consequence of the velocity-independent cross-section, since more massive haloes have larger characteristic velocities and hence higher scattering rates.
At our resolution, the SIDM implementations in \arepoOne and \arepoTwo produce profiles that agree within the halo-to-halo scatter.

\begin{figure}
    \centering
    \includegraphics[width=\linewidth]{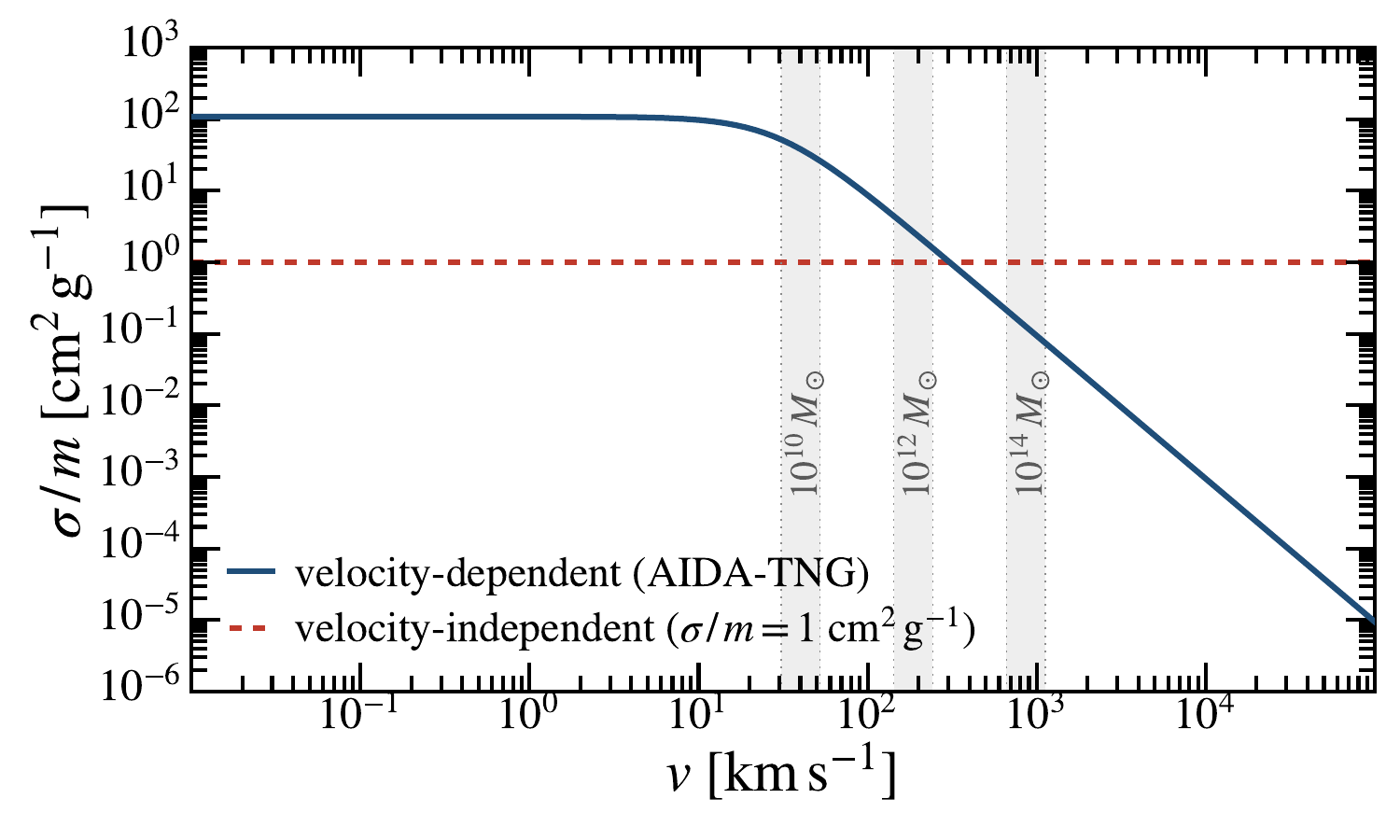}
    \caption{The two SIDM cross-section models used for our cosmological DM-only simulations (\cref{subsec:cosmo_profiles}): the constant, velocity-independent cross-section $\sigma/m=1\,\cpm$ and the velocity-dependent \textsc{AIDA}-TNG model \citep{Correa2021,Despali2025AIDATNG}, supplied to the module as a tabulated function of the pair relative velocity. Vertical bands mark the characteristic velocities for haloes of mass $10^{10}$, $10^{12}$, and $10^{14}\,\Msun$, spanning dwarf to cluster scales. The velocity bands extend from the virial circular velocity $v_{200}$ at $z=0$ to the heat conductivity-weighted most probable relative velocity $v_{\kappa} = \sqrt{14} \sigma_{\rm 1D}$ \citep{Tran2025_2}. Here, the effective one-dimensional velocity dispersion is taken as $\sigma_{\rm 1D} = 1.1 \sqrt{G \rho_{\rm s} r_{\rm s}^2}$). The velocity-dependent cross-section is largest at the low velocities of dwarf haloes and decreases towards the higher velocities of more massive haloes, dropping below the constant case at group and cluster scales.}
    \label{fig:cross_section}
\end{figure}

To test the velocity-dependent path in a cosmological setting, we additionally rerun the DM-only box with the velocity-dependent (vSIDM) cross-section adopted by the \textsc{AIDA}-TNG project \citep{Despali2025AIDATNG}, which follows \citet{Correa2021}. This isotropic, velocity-dependent cross-section is shown together with the constant case in \cref{fig:cross_section}. It is enhanced at the low relative velocities of dwarf haloes and falls off towards the high velocities of more massive haloes. We have verified that, for these simulations, the two SIDM implementations agree with each other.

\section{Performance and scalability}
\label{sec:performance}

Having verified the accuracy of the module, we now turn to its computational performance and scalability. Throughout this section we compare three simulation setups on identical hardware, Intel Sapphire Rapids nodes ($112$ cores each) of the Harvard Cannon cluster. Collisionless CDM and our new SIDM module are both run with the current \arepoTwo\ code base and its hierarchical gravity time integration. As a reference, we additionally run the legacy SIDM implementation of \citet{vogelsberger2012subhaloes} from the \texttt{master} branch of \arepoOne. Because it performs its neighbour search on the gravity tree, that scheme is incompatible with hierarchical gravity, so the \arepoOne\ runs instead use the standard, non-hierarchical gravity integration. Unless stated otherwise, the new SIDM runs adopt $C_{\rm SIDM}=0.1$ (see \cref{eq:sidm_timestep}) and the old ones $\kappa = 2.5\times 10^{-3}$ (see \cref{eq:sidm_timestep_arepo1}).

\subsection{Isolated core-collapse benchmark}
\label{sec:IsolatedCoreCollapse}
\begin{figure}
    \centering
    \includegraphics[width=1\linewidth]{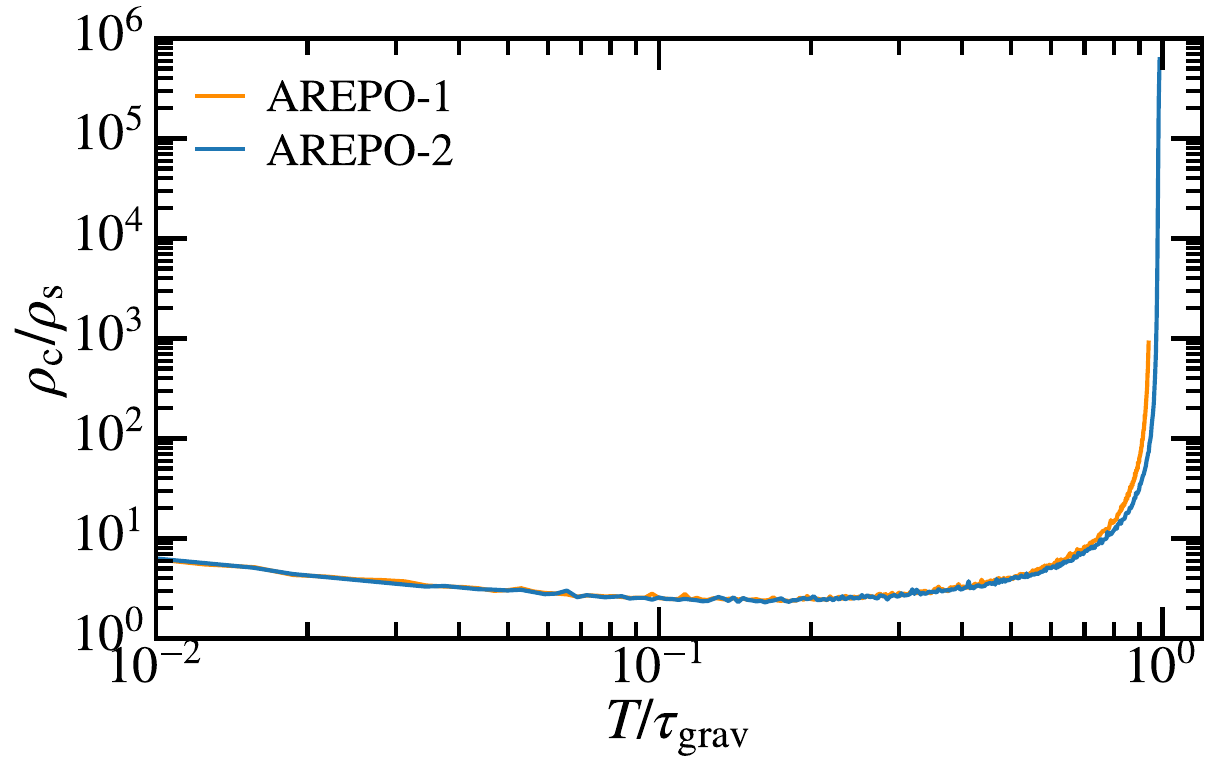}
    \includegraphics[width=1\linewidth]{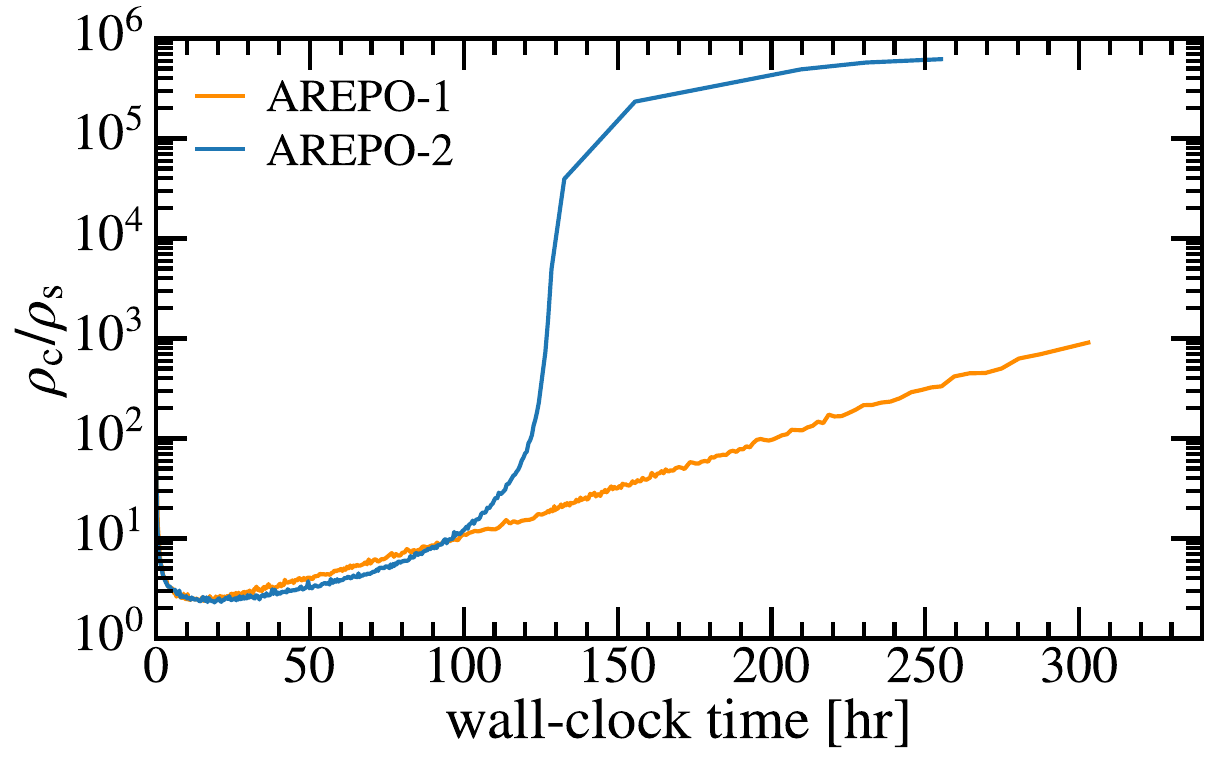}
    \caption{Gravothermal core collapse of the isolated halo sampled with $N_{200}=10^{7}$ particles (see \cref{subsec:isolatedCoreCollapse}), evolved on a single node with the \arepoOne\ (orange) and \arepoTwo\ (blue) SIDM modules. \emph{Top:} core density $\rho_{\rm c}$, in units of the NFW scale density $\rho_{\rm s}$, versus physical time in units of the collapse time. The two codes trace the same gravothermal evolution throughout the long core-formation phase and the onset of collapse. We stop the \arepoOne\ run before it reaches the highest core densities, because its more restrictive timestep criterion (\cref{eq:sidm_timestep_arepo1}) makes the deep-collapse phase prohibitively expensive. \emph{Bottom:} the same core density versus cumulative wall-clock time. The two modules cost essentially the same up to the onset of collapse, but during the runaway \arepoTwo\ reaches $\rho_{\rm c}/\rho_{\rm s}\sim 5\times10^{5}$, whereas at comparable cost \arepoOne\ advances only to $\sim10^{3}$, nearly three orders of magnitude lower.}
    \label{fig:CoreCollapseBenchmark}
\end{figure}

As a first controlled benchmark, we return to the isolated core-collapse problem described in \cref{subsec:isolatedCoreCollapse}, rerunning it with $N_{200}=10^7$ particles, a Plummer-equivalent softening of $\epsilon=16\,{\rm pc}$, and adaptive individual timesteps with $\eta=0.005$ on a single node. \cref{fig:CoreCollapseBenchmark} shows the resulting evolution of the core density as a function of both physical time (top) and cumulative wall-clock time (bottom).

Against physical time, the two implementations agree closely through core formation and into the onset of collapse, confirming that they follow the same gravothermal evolution. They differ only in how far the core-collapse phase can be traced: \arepoOne\ reaches a central density some three orders of magnitude below \arepoTwo\ before the run becomes impractical, because its far stricter timestep criterion (\cref{eq:sidm_timestep_arepo1}) makes the deep-collapse phase prohibitively expensive. The lower panel makes this cost explicit: the two modules track each other in wall-clock time up to the onset of collapse, but during the runaway the new module advances to $\rho_{\rm c}/\rho_{\rm s}\sim 5\times10^{5}$ at a cost for which the old module has reached only $\sim10^{3}$. This is precisely the regime that the multi-scattering scheme and the less restrictive per-pair timestep are designed to make tractable.

\subsection{Cosmological boxes}

To further quantify the performance and scalability of the new SIDM module relative to collisionless CDM and the old SIDM implementation, we run a suite of cosmological simulations: DM-only runs---the same ones whose $z=0$ density profiles we validated in \cref{subsec:cosmo_profiles}---and full-physics runs using the \illustrisTNG\ galaxy-formation model, in uniformly resolved periodic volumes.
As a compromise between cost and volume, we fix the comoving box size to $L=25\,{\rm cMpc}$ and evolve every run to $z=0$.
Initial conditions are generated at $z=49$ using \textsc{N-GenIC} (distributed with \textsc{Gadget-4}; \citealt{springel2021simulating}) for a \textit{Planck}\,2018 cosmology \citep{Planck2018Parameters}: $\Omega_\Lambda=0.69035856$, $\Omega_{\rm m}=0.30964144$, $\Omega_{\rm b}=0.04897468$, $h=0.6766$, $\sigma_8=0.8102$ and $n_{\rm s}=0.9655$. We generate the initial conditions using second-order Lagrangian perturbation theory and the fixed-amplitude technique of \citet{Angulo2016} to reduce sample variance. All runs use the same input linear power spectrum and random seed, so that the Fourier phases are identical for modes common to all resolutions; this isolates the effect of mass and spatial resolution. We use the same linear transfer function for the SIDM models, assuming that the adopted self-interactions do not modify the linear matter power spectrum at the initial redshift.
For the scalability study, we vary the DM resolution from $256^3$ to $1024^3$ particles, targeting $\approx 1.2\times 10^{6}$ particles per MPI task and adjusting the task count accordingly to keep the load per task roughly fixed.
For the SIDM runs, we use both a velocity-independent cross-section $\sigma/m=1\,\cpm$ and the velocity-dependent (vSIDM) cross-section of \cref{subsec:cosmo_profiles} following \citet{Correa2021}, as adopted by the \textsc{AIDA}-TNG project \citep{Despali2025AIDATNG}.
\cref{tab:simulation_overview} summarizes the full set of runs.

\begin{table*}
    \centering
    \caption{Parameters of the cosmological simulations performed in this work, all using a fixed comoving box size $L=25\,{\rm cMpc}$. We run CDM, our new SIDM implementation, and---for comparison---the legacy SIDM implementation of \protect\citet{vogelsberger2012subhaloes,vogelsberger2019evaporating}, both with and without baryons. For each run we list the numbers of gas and DM resolution elements ($N_{\rm gas}$, $N_{\rm dm}$), the mean baryonic and DM mass resolutions ($m_{\rm b}$, $m_{\rm dm}$), the comoving Plummer-equivalent DM softening ($\epsilon$), and the minimum gas softening $\epsilon_{\rm gas,min}$ (gas softenings are adaptive and tied to the cell size). The last three columns give the number of CPU cores $N_{\rm cores}$ ($112$ per node), the DM model, and the total wall-clock time in hours.
    We also include two simulations with a velocity-dependent cross-section (vSIDM).}
    \label{tab:simulation_overview}
\setlength{\tabcolsep}{12pt}
\begin{tabular}{@{}llccccc c c cc@{}}
\toprule
Run  & $N_{\rm gas}$ & $N_{\rm dm}$ & $m_{\rm b}$ & $m_{\rm dm}$ & $\epsilon$ & $\epsilon_{\rm gas, min}$ &$N_{\rm cores}$ & Model & Time\\
 &    &  &
$({\rm M}_\odot)$ & $({\rm M}_\odot)$ & $[\mathrm{cpc}]$ & $[\mathrm{cpc}]$ & & & [hr] \\
\midrule
CDM-256 & 0 & $256^3$ & - & $3.6\times 10^7$  &1200 & -&14 & CDM & 20.3\\
CDM-512 & 0 & $512^3$ & - & $4.5\times 10^6$  &600 & -&112 & CDM & 57.9\\
CDM-1024 & 0 & $1024^3$ & - &$5.7\times 10^5$  &300 & -&896 & CDM & 170\\
SIDM-256 & 0 & $256^3$ & - & $3.6\times 10^7$  &1200& - &14 & new SIDM & 28.5\\
SIDM-512 & 0 & $512^3$ & - & $4.5\times 10^6$  &600& - &112 & new SIDM & 74.3\\
SIDM-1024 & 0 & $1024^3$ & - &$5.7\times 10^5$   &300& - &896 & new SIDM & 210\\
SIDM-256-O & 0 & $256^3$ & - & $3.6\times 10^7$  &1200& - &14 & old SIDM & 35.4\\
SIDM-512-O & 0 & $512^3$ & - & $4.5\times 10^6$  &600 & -&112 & old SIDM & 93.7\\
SIDM-1024-O & 0 & $1024^3$ & - & $5.7\times 10^5$  &300 & -&896 & old SIDM & 255\\
vSIDM-512 & 0 & $512^3$ & - & $4.5\times 10^6$  &600& - &112 & new vSIDM & 79.4 \\
vSIDM-512-O & 0 & $512^3$ & - & $4.5\times 10^6$  &600 & -&112 & old vSIDM & 305 \\
\hline
B-CDM-256 & $256^3$ & $256^3$ & $5.8\times 10^6$ & $3.0\times 10^7$  &1200 & 300&112 & CDM &  44.7\\
B-SIDM-256 & $256^3$ & $256^3$ & $5.8\times 10^6$ & $3.0\times 10^7$ &1200 & 300&112 & new SIDM & 57.1\\
B-SIDM-256-O & $256^3$ & $256^3$ & $5.8\times 10^6$ & $3.0\times 10^7$  &1200& 300&112 & old SIDM & 530\\
\hline
\end{tabular}
\end{table*}

\subsubsection{Dark-matter-only box: constant cross-section}
\label{sec:DM_only_box}

\begin{figure}
    \centering
    \includegraphics[width=1\linewidth]{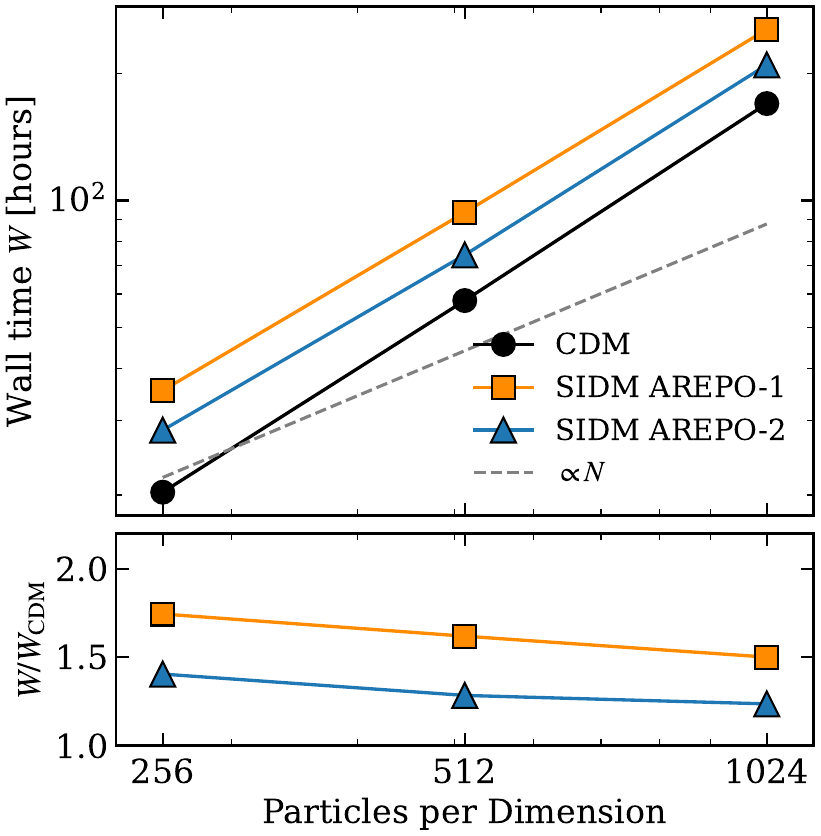}
    \caption{Upper panel: total wall-clock time of the DM-only runs (evolved from $z=49$ to $z=0$ in a $25\,{\rm cMpc}$ box) at three resolutions, for CDM (black), the old SIDM implementation in \arepoOne (orange), and the new SIDM implementation in \arepoTwo (blue). The dashed grey line indicates linear growth with the particle number per dimension as a guide to the eye. The measured runtimes rise somewhat faster ($\propto N_{\rm 1D}^{1.5}$), owing to the shorter timesteps required at higher resolution.
    Both SIDM simulation suites use constant cross-sections.
    Lower panel: runtime of the two SIDM implementations relative to CDM, which decreases with resolution for both.}
    \label{fig:walltime_vs_resolution_dm_only}
\end{figure}

To compare the efficiency of the different SIDM implementations, we measure the total wall-clock time to $z=0$ as a function of resolution (\cref{fig:walltime_vs_resolution_dm_only}). 
The new SIDM implementation adds an overhead that decreases with resolution, from $\approx 40$ per cent at $256^3$ to $\approx 24$ per cent at $1024^3$ (\cref{tab:simulation_overview}). The overhead is dominated by the neighbour search, with the relative SIDM cost shrinking as the particle number grows. The \arepoOne\ implementation carries about twice this overhead ($\approx 74$, $62$ and $50$ per cent at $256^3$, $512^3$ and $1024^3$).
Because all haloes here lie outside the core-collapse regime, the capability of handling multiple scatterings in \arepoTwo\ yields only a marginal gain in accuracy. The lower cost despite the more complex parallelization comes mainly from the tree-walk optimizations, in particular the grouped four-particle neighbour search.

\subsubsection{Dark-matter-only box: velocity-dependent cross-section}

We also benchmark the velocity-dependent (vSIDM) path with the \textsc{AIDA}-TNG cross-section of \cref{subsec:cosmo_profiles}. To limit the cost, we run only the $512^3$ configuration. Because this cross-section exceeds the constant $1\,\cpm$ case at the low velocities typical of low-mass haloes, it raises the local scattering rate and tightens the SIDM timestep. In \arepoOne, this drives a deeper timestep hierarchy, which, combined with its non-hierarchical gravity time integration, raises the runtime to 427 per cent above CDM (see \cref{tab:simulation_overview}). In \arepoTwo, the per-pair timestep criterion (\cref{eq:sidm_timestep}) is not triggered, so the overhead is approximately 37 per cent.
The small additional cost relative to the constant-cross-section run is due to the extra cross-section lookup performed for each particle pair.

\subsubsection{Full-physics cosmological box}
\label{subsec:baryon_box}

Baryons can cool and condense into denser structures than collisionless CDM, so hydrodynamical simulations typically develop a deeper timestep hierarchy and are correspondingly more expensive. To test the module in this regime, we repeat a subset of the cosmological runs with baryons added to the initial conditions, using the \illustrisTNG\ galaxy-formation model \citep{weinberger2016simulating, Pillepich2018TNGModel} at a reduced particle load of $2\times 256^3$ particles on one node in order to keep the total wall-clock time tractable.

With a constant cross-section, the new SIDM code adds an overhead of around $28$ per cent relative to CDM. By contrast, the old SIDM implementation is about an order of magnitude slower than the new SIDM run and more than an order of magnitude slower than CDM. The main reason is that the legacy module is not compatible with \arepoTwo's hierarchical time integration: it reuses the gravity tree for its neighbour search and therefore forces a full reconstruction of the gravity tree at every timestep, which becomes prohibitively expensive once a deep timestep hierarchy develops, so that the cost is dominated by the small number of particles in the lowest time bins. The new module avoids this precisely because its dedicated DM neighbour tree (\cref{subsec:sidm_local}) decouples scattering from gravity, so that the optimized hierarchical gravity solver can be used together with SIDM at all. 

These costs depend on the CPU architecture and the local network, so the quantitative factor of ten is not necessarily portable; nevertheless, similar slowdowns were found in high-resolution simulations of the \textsc{AIDA}-TNG project when baryons were added (Despali, private communication).

\section{Conclusions}
\label{sec:conclusions}

In this paper, we present a novel implementation of SIDM in the multi-physics (cosmological) simulation code \arepoTwo, which offers greater flexibility and extensibility along with improved numerical accuracy and efficiency, relative to the original SIDM implementation in \arepo\ \citep{vogelsberger2012subhaloes}. This new module features several key numerical improvements:

\begin{itemize}[leftmargin=*]
    \item \textbf{Flexibility in the scattering rate estimator.} We implement two SIDM scattering rate estimators based on either one-sided kernel searches or kernel overlap. 

    \item \textbf{Parallelization scheme that can handle multiple scatterings.} We introduce a new communication pattern between parallel processes that handles multiple scattering events per particle within a single timestep. This allows an SIDM timestep criterion that is orders of magnitude less restrictive than those of canonical methods, while maintaining energy and momentum conservation to machine precision in the SIDM module.
    
    \item \textbf{Decoupled neighbour search.} We replace the canonical neighbour search method based on gravity trees with a DM-only neighbour tree. 
    This allows us to use the hierarchical time integration for gravity from \citet{springel2021simulating}, which can significantly decrease end-to-end runtime in configurations with deep timestep hierarchies.
    
    \item \textbf{Batched tree walks.} The neighbour tree is walked for a batch of particles at once, substantially reducing the cost of the neighbour search.

    \item \textbf{Extensible SIDM interface.} We focus on isotropic, elastic scattering in this work, but the framework is straightforward to extend to more complicated SIDM models. These interactions can be implemented through four user-supplied functions, without any knowledge of the underlying parallelization and neighbour search schemes.
\end{itemize}

To validate and stress-test this implementation, we performed a series of numerical tests, including
(i) the scattering of a beam of SIDM particles through a static lattice, (ii) thermalization to a Maxwellian velocity distribution in a uniform box, (iii) the gravothermal collapse of an isolated halo, (iv) DM-only cosmological zoom-in simulations, and (v) large-volume DM-only cosmological simulations. Across all of these tests, predictions derived using the new SIDM module agree exceptionally well with analytic expectations wherever available. The module demonstrates excellent conservation of energy and momentum and is robust to variations in numerical parameters. We have further measured the performance and scalability of the module during an isolated core collapse and in large-volume cosmological simulations with both DM-only runs and runs including baryons. This new SIDM module provides a solid foundation for simulations of SIDM haloes deep in the core-collapse phase, where a highly efficient and accurate numerical implementation is required.

\section*{Acknowledgements}
We thank Stephanie O'Neil for providing the initial conditions used in \cref{subsec:zoominCoreCollapse}, Giulia Despali for the velocity-dependent cross-section file, and Zihan Wang for validating the gravothermal fluid model. OZ acknowledges support from Harvard University through the Institute for Theory and Computation Fellowship. MR acknowledges support from the Carlsberg Foundation, grant CF24-1996.

\section*{Data Availability}
The initial-condition generator and the Boltzmann reference solver used for the validation tests in this work are publicly available. The SIDM implementation and the data underlying this article can be shared on reasonable request to the corresponding author (OZ).



\bibliographystyle{mnras}
\bibliography{main} 




\appendix

\section{Comparison to overlap kernel}
\label{app:OverlapKernel}

\begin{figure}
    \centering
    \includegraphics[width= 1 \linewidth]{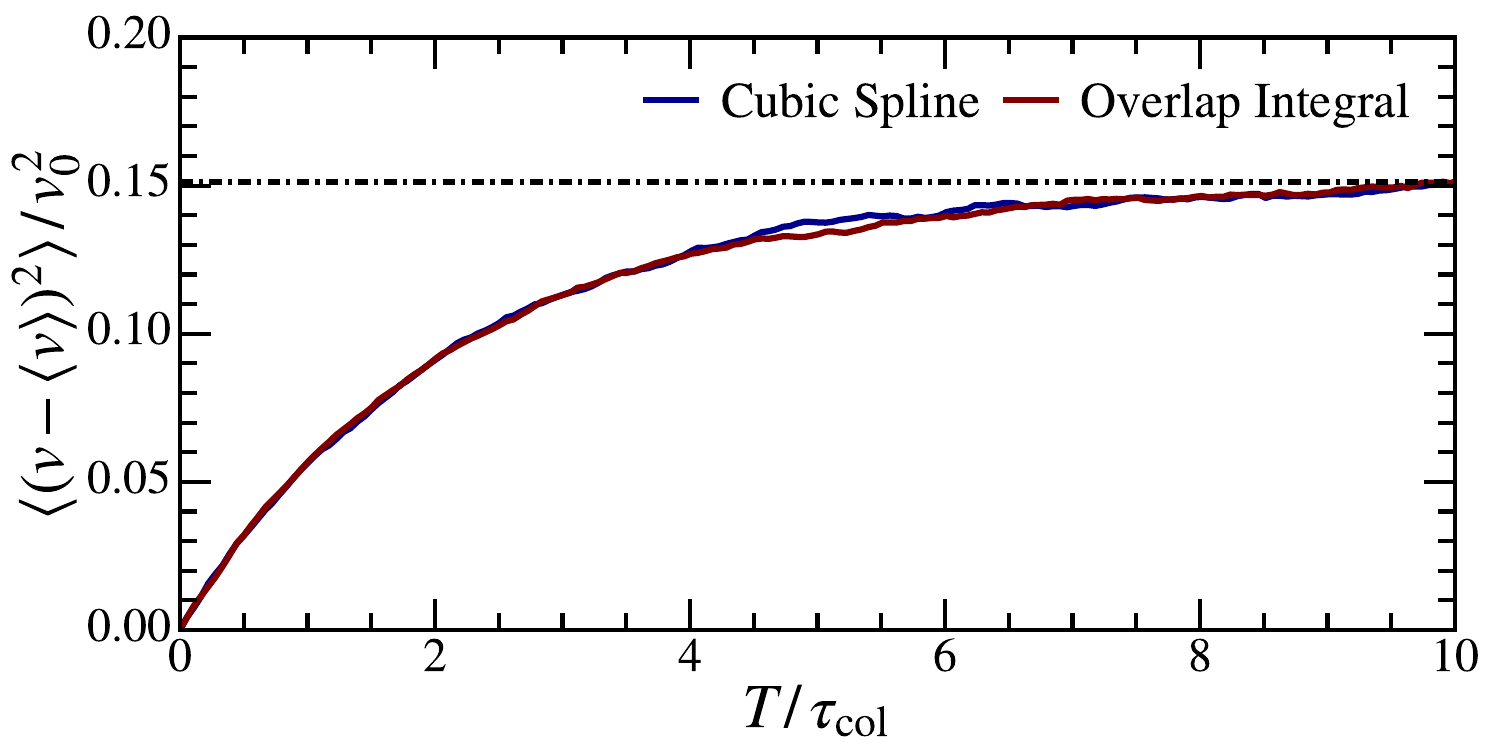}
    \includegraphics[width= 1 \linewidth]{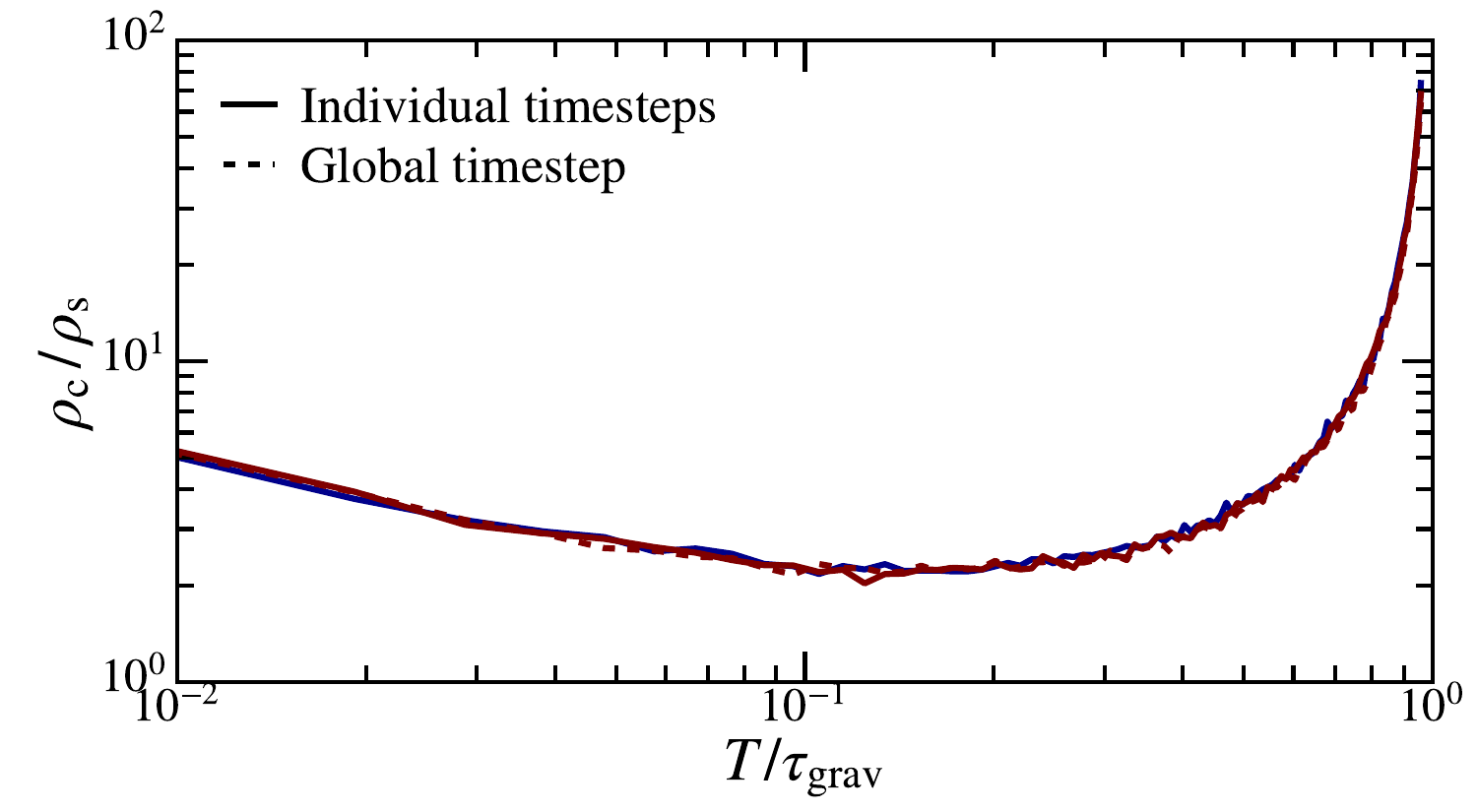}
    \caption{Top: The evolution of the squared speed dispersion for the thermalization test using the cubic-spline kernel (blue) and the kernel-overlap integral (red); the black dash-dot line marks the expected Maxwell--Boltzmann value. Bottom: The time evolution of the core density for the $M_{200} = 10^{10} \, \Msun$, $c = 13.08$ isolated halo evolved using the aforementioned self-interaction methods (colours as in the top panel: cubic spline blue, kernel overlap red), employing both adaptive individual timesteps (solid) and a global timestep (dashed). The results from all setups are consistent.}
    \label{fig:kernel_test}
\end{figure}

To implement the kernel-overlap SIDM method, we replace the one-sided spline weight by the symmetric overlap factor in \cref{eq:kernelOverlap}, which is non-zero only when $r_{ij}<h_i+h_j$, where $r_{ij}\equiv|\mathbf{x}_i-\mathbf{x}_j|$. Relative to the fiducial one-sided pipeline of \cref{fig:sidm_flow}, only the neighbour-search front end changes; after this, control returns to the unchanged per-pair scatter routine.
To support this efficiently in our tree-based neighbour search, we store the maximum smoothing length $h_{\max}$ in every tree node.
We then perform two tree walks: (i) the standard fixed-$N_{\rm ngb}$ smoothing-length search to determine $h_i$, and (ii) an overlap search that returns all neighbours with $r_{ij}<h_i+h_j$.
Between the two walks the tree is updated recursively with the newly determined smoothing lengths.
The overlap search does not bound the neighbour count by construction, so we use a larger neighbour buffer for the second walk; the subsequent scattering logic is unchanged.
For efficiency, we evaluate $g_{ij}$ from a tabulation of the dimensionless overlap $\hat g\!\left(h_i/(h_i+h_j),\,r_{ij}/(h_i+h_j)\right)$, exploiting the symmetry $g_{ij}=g_{ji}$ to reduce the lookup domain.

Keeping the neighbour number fixed at $32\pm 5$, we repeat the thermalization test of \cref{subsec:Thermalisation} with kernel overlap at $C_{\rm SIDM}=0.1$, and the isolated core-collapse test of \cref{subsec:isolatedCoreCollapse} with hierarchical gravity, and compare with the cubic-spline results in \cref{fig:kernel_test}. We utilize the medium resolutions of $N_{\rm part} = 10^4$ and $N_{200} = 10^6$ for the two tests, respectively. The two methods agree within the statistical scatter.

\section{Reference solutions for the verification tests}
\label{app:reference_solutions}

The two reference solutions described here are independent, non-$N$-body benchmarks used to validate the module: a kinetic Boltzmann solver for the thermalization test of \cref{subsec:Thermalisation}, and a gravothermal fluid model for the core-collapse test of \cref{subsec:isolatedCoreCollapse}.

\subsection{Kinetic Boltzmann solver for the thermalization test}
\label{apd:MB_solver}

To obtain the reference solution for the thermalization test of \cref{subsec:Thermalisation}, we numerically solve the collisional Boltzmann equation,
\begin{equation}
    \label{eq:full_Boltzmann_eqn}
    \frac{\partial f}{\partial t} + \mathbf{v}\!\cdot\!\nabla_{\mathbf{x}} f - \nabla_{\mathbf{x}}\Phi \!\cdot\! \nabla_{\mathbf{v}} f = \left(\frac{\partial f}{\partial t}\right)_{\mathrm{coll}} .
\end{equation}
Here $f\equiv f(\mathbf{x},\mathbf{v},t)$ is the phase-space distribution function and $\Phi$ the gravitational potential, which plays no role in this test. The right-hand side is the collision term, discussed in more detail below. Because the system is homogeneous and isotropic, $f$ depends only on the particle speed and time, $f=f(v,t)$, so only the first term on the left-hand side survives. For isotropic scattering it is more convenient to use the particle energy as the independent variable, $f=f(E,t)$, in which case the Boltzmann equation reads
\begin{multline}
    \label{eq:reduced_Boltzmann_eqn}
    \frac{\partial f (E,t)}{\partial t} = \left(\frac{\partial f(E,t)}{\partial t}\right)_{\mathrm{coll}} = - \Gamma(E) \, f(E,t) \\
    + \int_0^\infty \Gamma(E' \rightarrow E) \, f(E',t) \, \mathrm{d}E' ,
\end{multline}
where $\Gamma(E)$ is the total scattering rate for particles with initial energy $E$ and $\Gamma(E'\to E)$ is the differential rate from initial energy $E'$ to final energy $E$. These scattering rates are functionals of $f$, and the discretized form below evaluates this dependence explicitly. The first term on the right of \cref{eq:reduced_Boltzmann_eqn} is the out-scattering loss from energy $E$, while the second is the in-scattering contribution from other energies.

We solve \cref{eq:reduced_Boltzmann_eqn} numerically by binning $f(E,t)$ into energy bins $E_i$. For each timestep $\Delta T\ll\tau$, the procedure is:
\begin{enumerate}[leftmargin=*]
    \item For each unordered pair of energy bins $i \leq j$ and incoming angle $\delta$, we calculate the centre-of-momentum velocity
    \begin{equation}
        \label{eq:V_cm_Boltzmann}
        V_{i,j} (\delta) = \frac{1}{2} \sqrt{v_i^2 + v_j^2 + 2 v_i v_j \cos{\delta}},
    \end{equation}
    the relative velocity
    \begin{equation}
        \label{eq:g_Boltzmann}
        u_{i,j} (\delta) = \sqrt{v_i^2 + v_j^2 - 2 v_i v_j \cos{\delta}},
    \end{equation}
    and the angle $\gamma_{i,j} (\delta)$ between them
    \begin{equation}
        \label{eq:gamma_Boltzmann}
        \cos{\gamma_{i,j}} (\delta) = \frac{| v_i^2 - v_j^2 |}{2 V_{i,j} (\delta) u_{i,j} (\delta)}.
    \end{equation}
    Here $v_i$ and $v_j$ are the speeds of bins $i$ and $j$, with the angle between the two incoming velocity vectors $\delta$ isotropically distributed, i.e. $p(\cos\delta)=1/2$. 
    \item We then compute the post-scattering energy distribution $p_{i,j}(E',\delta)$ for the forward-scattered particle,
    \begin{align}
        \label{eq:Eprime_Boltzmann}
        E'(\theta,\phi,\delta) = \frac{1}{2} \Bigg( &V_{i,j}(\delta)^2 + \frac{u_{i,j}(\delta)^2}{4} \\
        &+ V_{i,j}(\delta) u_{i,j}(\delta) \cos{\gamma_{i,j}(\delta)} \cos{\theta} \notag \\
        &+ V_{i,j}(\delta) u_{i,j}(\delta) \sin{\gamma_{i,j}(\delta)} \sin{\theta} \cos{\phi} \Bigg). \notag
    \end{align}
    Here $\theta$ is the scattering angle and $\phi$ is the azimuthal angle, uniformly distributed on $[0,2\pi]$. For isotropic scattering,
    $E'$ is uniformly distributed on $[E_{i,j,\rm{min}}(\delta),E_{i,j,\rm{max}}(\delta)]$ with $E_{i,j,\rm{min}}(\delta)=\tfrac{1}{2}(V_{i,j}(\delta)-u_{i,j}(\delta)/2)^2$ and $E_{i,j,\rm{max}}(\delta)=\tfrac{1}{2}(V_{i,j}(\delta)+u_{i,j}(\delta)/2)^2$. In the anisotropic case, $p_{i,j}(E',\delta)$ must be obtained analytically or numerically (for example, by sampling $\theta$ and $\phi$). For later use, we also define the cumulative distribution function (CDF),
    \begin{equation}
        \label{eq:cdf_Boltzmann}
        P_{i,j}(E',\delta) = \int_0^{E'} p_{i,j}(\tilde{E}',\delta) \, {\rm d}\tilde{E}',
    \end{equation}
    and its anti-derivative 
    \begin{equation}
        \label{eq:I_Boltzmann}
        I_{i,j}(E',\delta) = \int_0^{E'} P_{i,j}(\tilde{E}',\delta) \, {\rm d}\tilde{E}'.
    \end{equation}
    \item The system's energy distribution is then updated using $I_{i,j}(E',\delta)$ and the scattering count
    \begin{multline}
        \label{eq:Gamma_Boltzmann}
        \Delta N_{i,j} (\delta) = \rho \left(\sigma_{\rm tot} (u_{i,j}) / m\right) u_{i,j} (\delta) \, n_{i,j} \\
        \times f(E_i) f(E_j) \Delta E^2 \, \Delta (\cos\delta)/2 \, \Delta T.
    \end{multline}
    Here $\rho$ is the mass density, $\sigma_{\rm tot}/m$ the total cross-section per unit mass, $n_{i,j}=1/2$ for $i=j$ and $n_{i,j}=1$ otherwise, and $\Delta E$ and $\Delta(\cos\delta)$ are the bin widths in energy and in $\cos\delta$ (the incoming angle is sampled uniformly in $\cos\delta$). The appearance of $I_{i,j}(E',\delta)$ reflects how we populate the energy bins: rather than assigning a ``particle'' with energy $E'$ to a single bin, we split it between the two nearest bins so as to conserve energy, namely
    \begin{equation}
        \label{eq:lever_update_Boltzmann}
        \Delta f_{i,k,\delta}(E_k) = \Delta N_{i,j} (\delta) / \Delta E \int_{E_{k-1}}^{E_{k+1}} p_{i,j}(E',\delta) \, \Lambda_k(E') \, {\rm d}E',
    \end{equation}
    where $\Lambda_k(E')=\max\!\left(0,1-|E'-E_k|/\Delta E\right)$ is the lever function. Integration by parts gives
    \begin{multline}
        \label{eq:lever_update_clean_Boltzmann}
        \Delta f_{i,k,\delta}(E_k) = \Delta N_{i,j} (\delta) / \Delta E \\
        \times \frac{I_{i,j}(E_{k+1},\delta) - 2\,I_{i,j}(E_{k},\delta) + I_{i,j}(E_{k-1},\delta)}{\Delta E}.
    \end{multline}
    At the lower and upper energy boundaries, we introduce virtual energy bins to help with conservation of normalization and energy. For $E < E_{\rm min}$ we set $I = 0$, while for $E > E_{\rm max}$ we set $I = I_{\rm max} + \left(E - E_{\rm max}\right)$. Note that the expression of \cref{eq:lever_update_clean_Boltzmann} accounts only for the forward-scattered particle. The contribution of the second particle is included by adding an analogous term with the substitution $E\to(E_i+E_j)-E$ (from energy conservation). One must also subtract $\Delta N_{i,j}(\delta)/\Delta E$ from bins $i$ and $j$ to account for the corresponding out-scattering. Looping over all pairs $(i,j)$ and incoming angles $\delta$ advances the system by one timestep.
\end{enumerate}

To test the Boltzmann solver, we evolve the system from a Maxwell--Boltzmann distribution,
\begin{equation}
    \label{eq:MB_energy}
    f(E) = \frac{2}{\sqrt{\pi}} \frac{E^{1/2}}{\left(k_{\rm{B}} \Theta\right)^{3/2}} \exp\left(- \frac{E}{k_{\rm{B}} \Theta}\right),
\end{equation}
where $k_{\rm B}\Theta=2E_0/3$ and $E_0$ is the mean energy. As expected, the energy distribution, total energy and normalization $\int f(E)\,{\rm d}E$ remain stable in time, with deviations at the level of $\lesssim 10^{-3}$. We then apply the same procedure to the initial condition $f(E)=\delta(E-v_0^2/2)$ and obtain the curves shown in \cref{fig:thermalize_test}. The solver is publicly available\footnote{\url{https://github.com/vinh-qtran/ThermBoxSIDM}}.

\subsection{Gravothermal fluid model for core collapse}
\label{apd:fluid_code}

\begin{figure}
    \centering
    \includegraphics[width= 1 \linewidth]{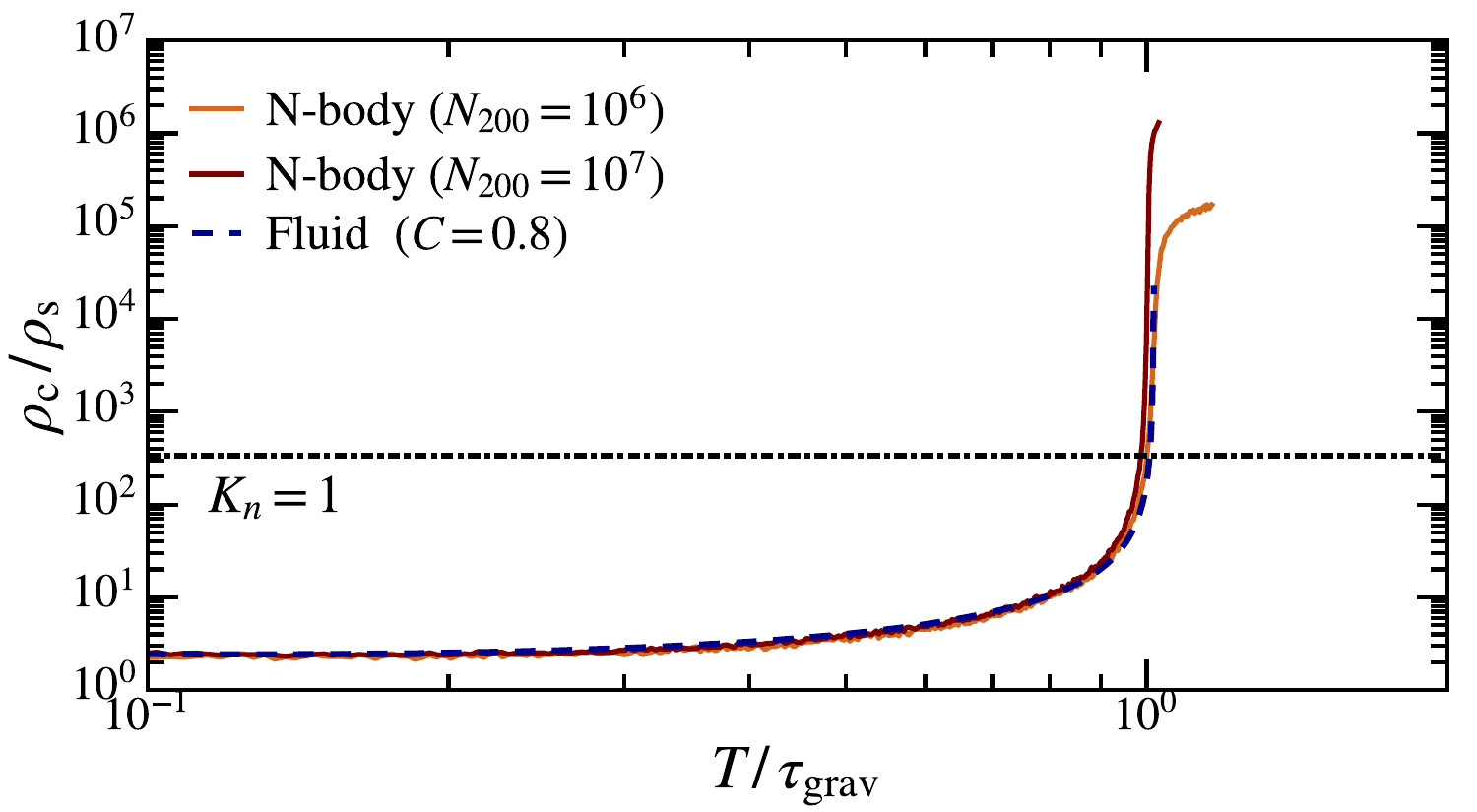}
    \caption{Comparison of the late-time evolution of an isolated halo with $M_{200} = 10^{10} \, \Msun$ and $c = 13.08$, evolved using our new $N$-body implementation with $N_{200} = 10^6$ (orange) and $N_{200} = 10^7$ (red), as well as using the fluid code (blue). The horizontal dash-dot line marks the core density at which the Knudsen number (\cref{eq:Knudsen}) crosses unity. Late-time agreement is obtained with $C_{\kappa} = 0.8$. Note that the core density flattening observed in the late stages of the core-collapse regime is not a physical result, but rather an artefact of requiring the core to contain a minimum number of particles. This condition can no longer be satisfied deep into the collapse, when the core mass shrinks exponentially, and the particle count falls below the minimum threshold required by the measurement, artificially flattening the density profile.}
    \label{fig:fluid_test}
\end{figure}

To validate the $N$-body heat-transfer implementation in the core-collapse regime, we compare our simulations with a semi-analytical gravothermal fluid model \citep{Outmezguine2023,Gad-Nasr2024}. The model numerically solves the gravothermal fluid equations \citep[e.g.,][]{Lynden-Bell1980,Balberg2002}
\begin{align}
    \label{eq:gravo_fluid_hydrostatic}
    \frac{\partial M}{\partial r} &= 4\pi r^2 \rho; \hspace{0.5cm} \frac{\partial (\rho v^2)}{\partial r} = -\frac{GM\rho}{r^2} \\
    \label{eq:gravo_fluid_heat}
    \frac{L}{4 \pi r^2} &= - \kappa \frac{\partial T}{\partial r}; \hspace{0.5cm} \frac{\partial L}{\partial r} = -4\pi r^2 \rho v^2 \left( \frac{\partial}{\partial t} \right)_M \ln \left( \frac{v^3}{\rho} \right),
\end{align}
by alternating heat conduction with a rebalancing of hydrostatic equilibrium \citep{Pollack2015}. Here $v$ is the one-dimensional velocity dispersion, the temperature is identified as $T=mv^2$, and $(\partial/\partial t)_M$ denotes the Lagrangian derivative at fixed enclosed mass.

Heat conduction is modelled with an effective conductivity $\kappa_{\rm eff}$ that interpolates between the long-mean-free-path (LMFP) and short-mean-free-path (SMFP) regimes,
\begin{equation}
    \label{eq:kappa_eff}
    \frac{1}{\kappa_{\rm{eff}}} = \frac{1}{\kappa_{\rm{LMFP}}} + \frac{1}{\kappa_{\rm{SMFP}}}.
\end{equation}
In the LMFP regime, heat transport is non-local. The fluid approximation requires a calibration parameter $C_{\kappa}$, fitted so that the heat flux matches that measured in $N$-body simulations \citep[e.g.,][]{Mace2026},
\begin{equation}
    \label{eq:kappa_LMFP}
    \kappa_{\rm{LMFP}} = \frac{3 a C_{\kappa}}{8 \pi G} \left(\sigma/m\right) \frac{\rho v^3}{m},
\end{equation}
with $a=4/\sqrt{\pi}$ and $C_{\kappa}$ an order-of-unity calibration factor whose value is discussed below. In the SMFP regime the system approaches local thermodynamic equilibrium and the conductivity follows rigorously from a Chapman--Enskog expansion of the Boltzmann equation, giving the parameter-free expression
\begin{equation}
    \kappa_{\text{SMFP}} = \frac{3}{2} \frac{b v}{\sigma},
    \label{eq:kappa_smfp}
\end{equation}
with $b=25\sqrt{\pi}/32$. The fluid prediction for the SMFP density-evolution rate therefore provides both a benchmark for the $N$-body solver and a reference for calibrating $C_{\kappa}$. We define the LMFP-to-SMFP transition in the $N$-body simulations by the Knudsen number $\mathrm{Kn}$, the ratio of the collisional mean free path to the gravitational scale height \citep{Balberg2002}, crossing unity,
\begin{equation}
    \label{eq:Knudsen}
    \mathrm{Kn} = \frac{1}{\rho_{\rm c} \left(\sigma/m\right)} \left(\frac{v_{\rm c}^2}{4 \pi G \rho_{\rm c}}\right)^{-1/2} = 1,
\end{equation}
where $\rho_{\rm c}$ and $v_{\rm c}$ are the core density and the one-dimensional velocity dispersion, respectively. We find that $C_{\kappa}=0.8$ gives the best agreement with the $N$-body benchmark in both collapse time and late-time behaviour (\cref{fig:fluid_test}); the corresponding core-density evolution is also shown in \cref{fig:collapse_test}.
To ensure that the fluid model is an accurate benchmark, we have checked its spatial-resolution convergence. The fluid code uses a logarithmic radial grid spanning from $r_{\rm min}=0.01\,r_{\rm s}$ to $r_{\rm max}=100\,r_{\rm s}$, in which the fractional shell spacing $s\equiv\Delta r/r$ controls the finite-difference truncation error. The default resolution of $\approx 100$ shells gives a coarse $\sim10$ per cent step that, while adequate for the early LMFP evolution, blunts the steep $\rho \propto r^{-2.2}$ gradients that drive the gravothermal catastrophe and thereby biases the collapse time. We therefore adopt a finer spacing $s\approx 0.03$ ($3$ per cent steps, $N_{\rm shells}=310$) for our benchmark comparisons, for which the collapse times and runaway slopes are converged.

\section{Convergence of the core-collapse test}
\label{app:convergence_all}

We verify the robustness of the isolated core-collapse test of \cref{subsec:isolatedCoreCollapse} with two complementary convergence checks: the first varies the mass resolution and initial-condition realizations (and random seeds), and the second varies the gravitational-timestep accuracy parameter $\eta$.

\subsection{Resolution and initial-condition convergence}
\label{app:Convergence}

\begin{figure}
    \centering
    \includegraphics[width=1\linewidth]{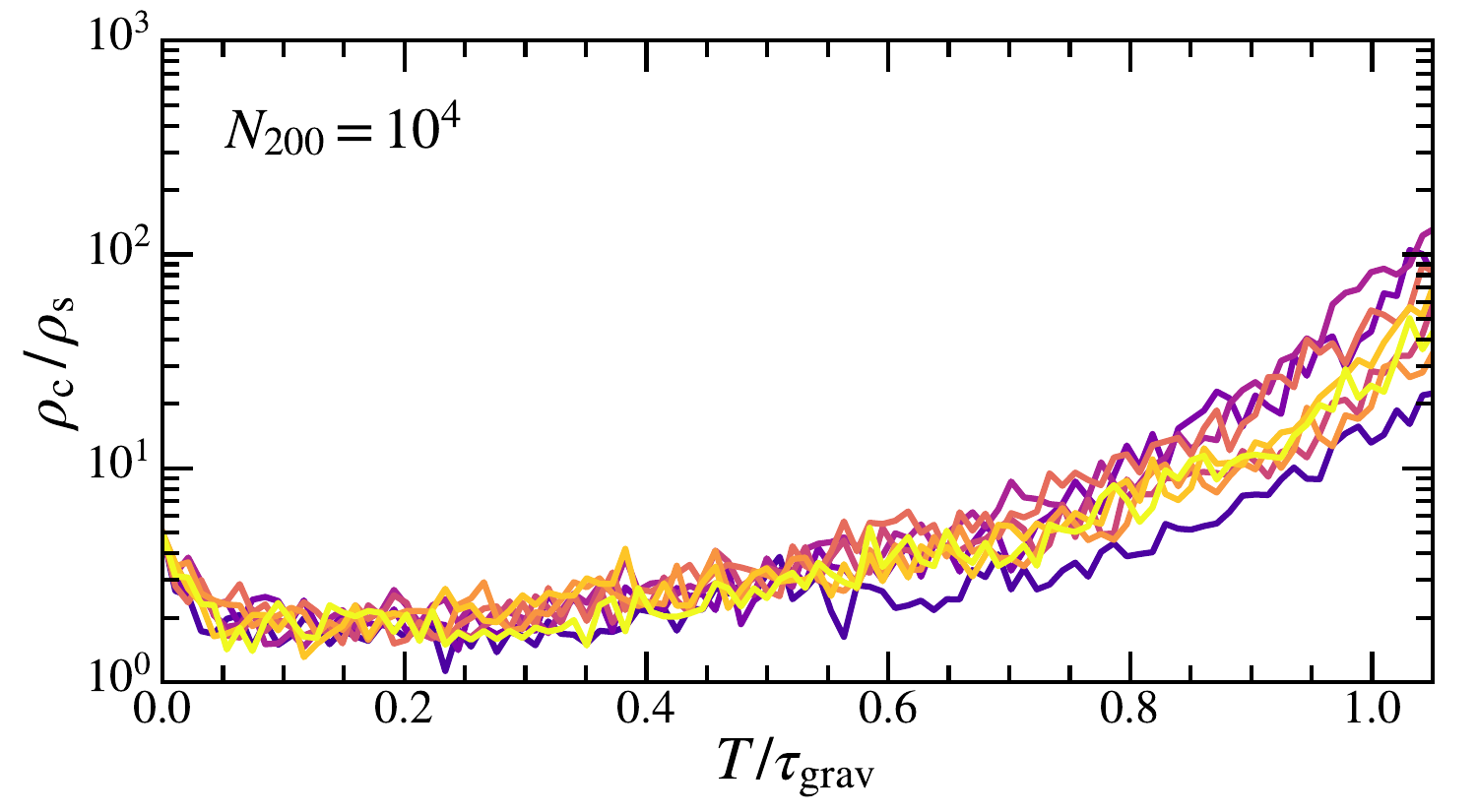}
    \includegraphics[width=1\linewidth]{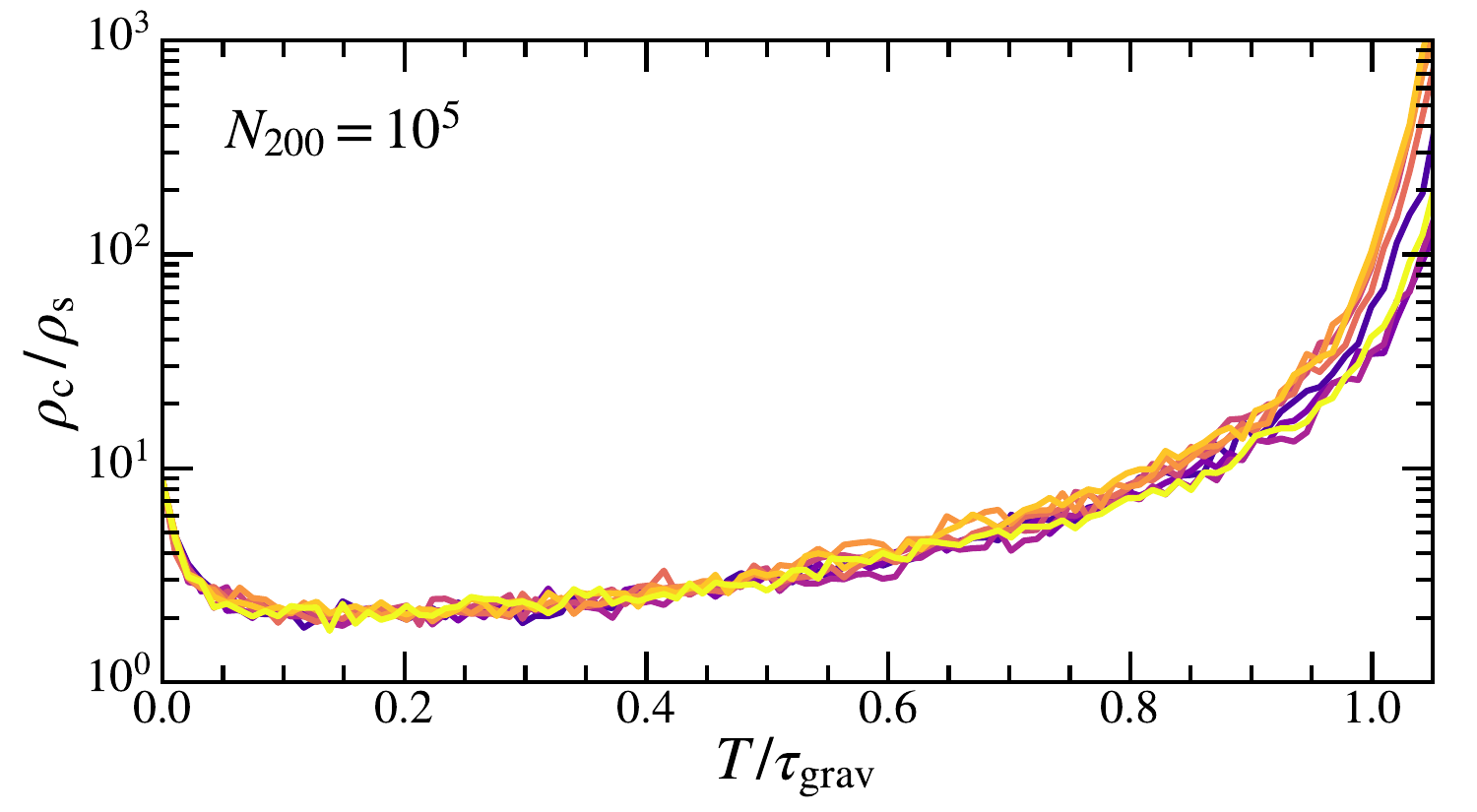}
    \includegraphics[width=1\linewidth]{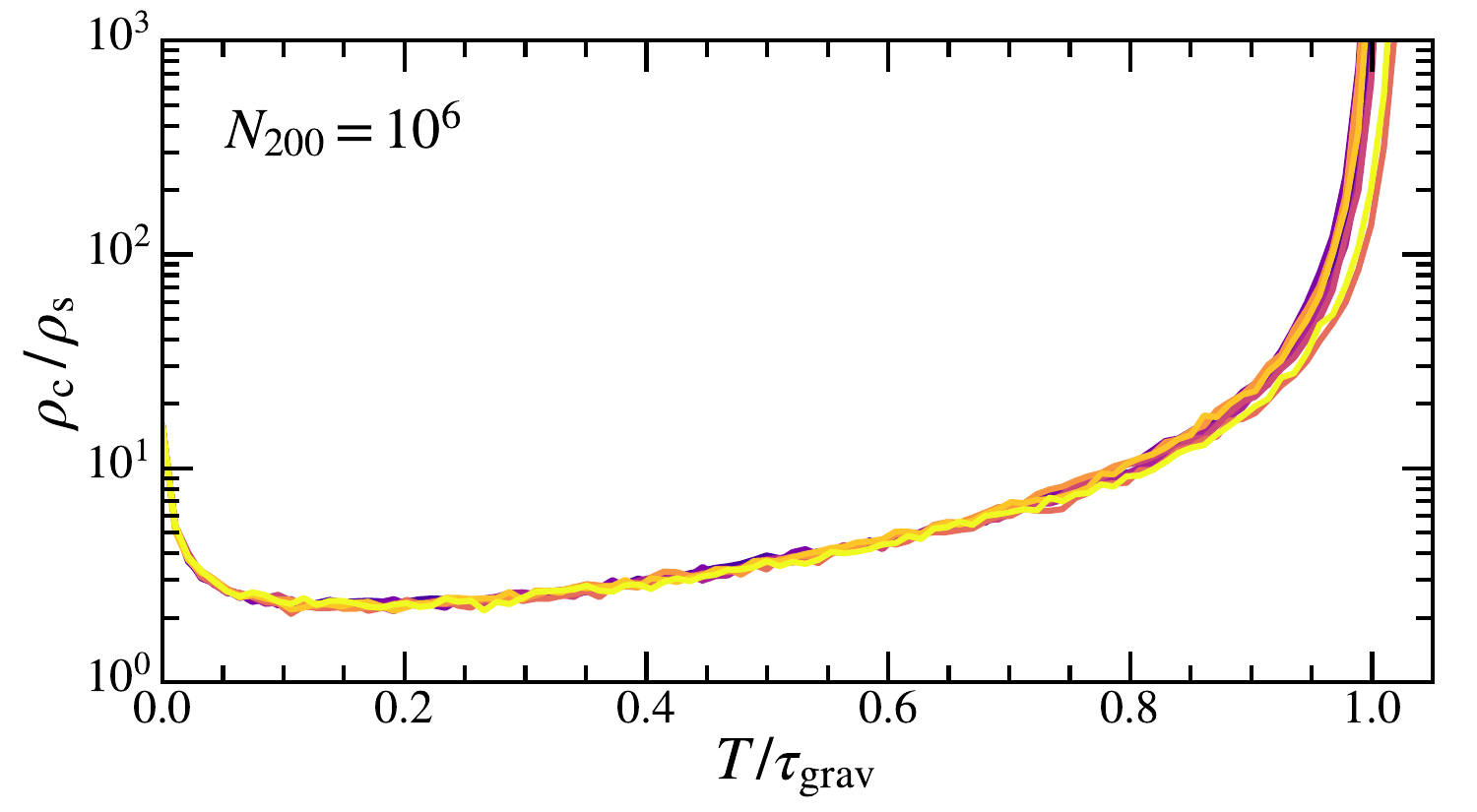}
    \caption{Convergence of the core-density evolution for the $M_{200} = 10^{10} \Msun$, $c = 13.08$ isolated halo evolved from identical ICs with different random-number-generator seeds. The top, middle, and bottom panels show the results for the $N_{200} = 10^4$, $N_{200} = 10^5$, and $N_{200} = 10^6$ runs, respectively. The evolution appears well converged by $N_{200} = 10^5$, despite a slight delay in collapse time.}
    \label{fig:convergence}
\end{figure}

\begin{figure}
    \centering
    \includegraphics[width=1\linewidth]{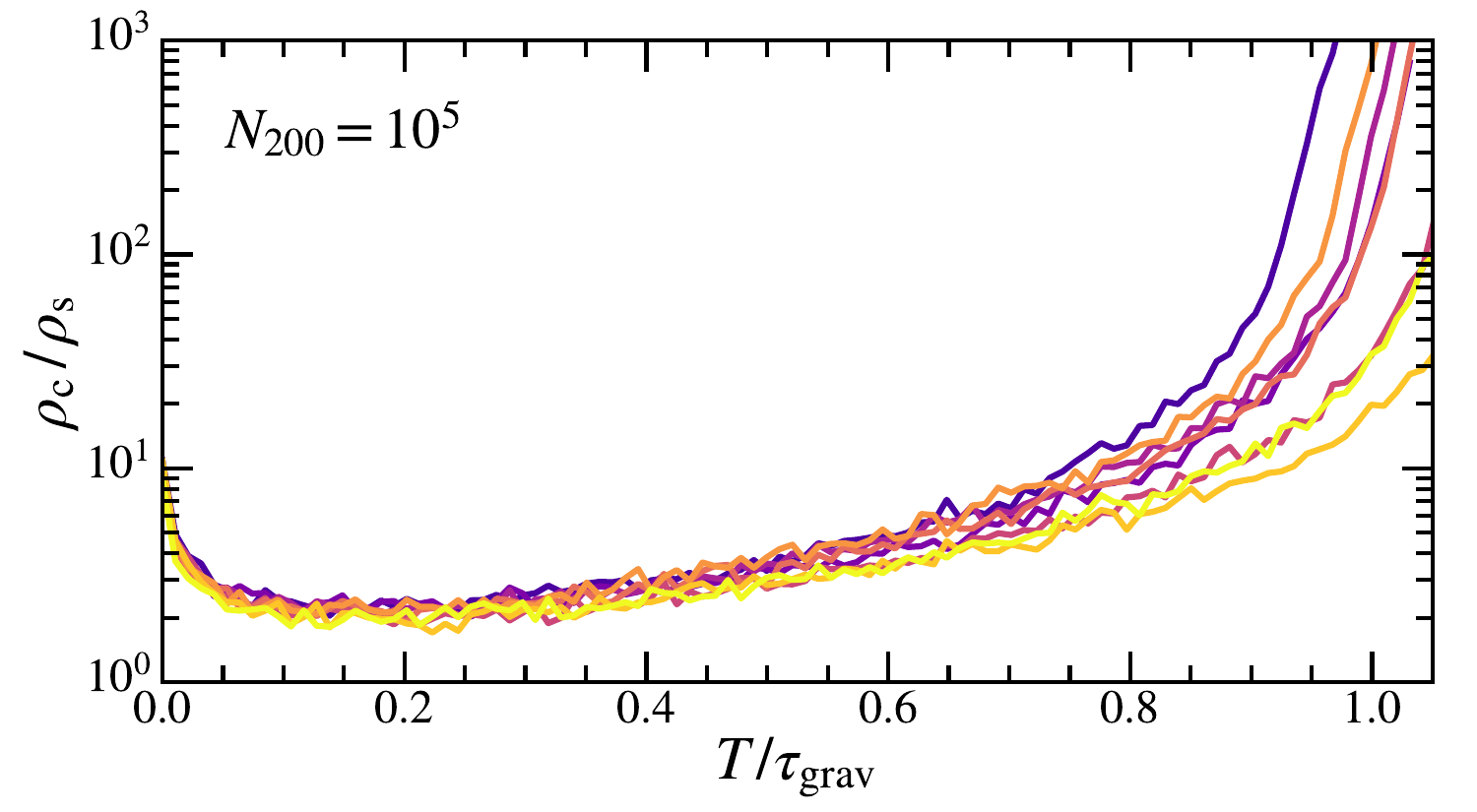}
    \caption{Convergence of the core-density evolution for the $M_{200} = 10^{10} \Msun$, $c = 13.08$ isolated halo evolved with $N_{200} = 10^5$ from different IC realizations and random number generator seeds. Deviations between haloes are much more significant than in the case of evolutions from the same ICs. This is expected, as different realizations result in different total energies, which in turn lead to different collapse times.}
    \label{fig:convergence_ICs}
\end{figure}

To assess the numerical stability of the code, we perform a convergence study using the isolated halo of \cref{subsec:isolatedCoreCollapse} at several resolutions. At each resolution we evolve eight haloes from identical ICs but with different internal random-number seeds. For the $N_{200}=10^4$ and $N_{200}=10^5$ runs the core density is computed as in \cref{subsec:isolatedCoreCollapse}, with a minimum core count of $900$ and a minimum bin count of $100$. The results are shown in \cref{fig:convergence}. At low resolution the evolution is visibly delayed, most likely because too few collisions are available to represent the heat transfer accurately. At $N_{200}=10^5$ the collapse time is already converged at the $\sim 5$ per cent level, and at $N_{200}=10^6$ the differences between seeds shrink to $1$--$2$ per cent. We also perform a convergence test using different IC realizations of the same halo at $N_{200}=10^5$ (\cref{fig:convergence_ICs}); convergence is significantly weaker, which we attribute to differences in the realized halo energy that arise from the limited resolution \citep[see also][]{Mace2024}.

\subsection{Gravitational-timestep convergence}
\label{app:eta}

\begin{figure}
    \centering
    \includegraphics[width=1\linewidth]{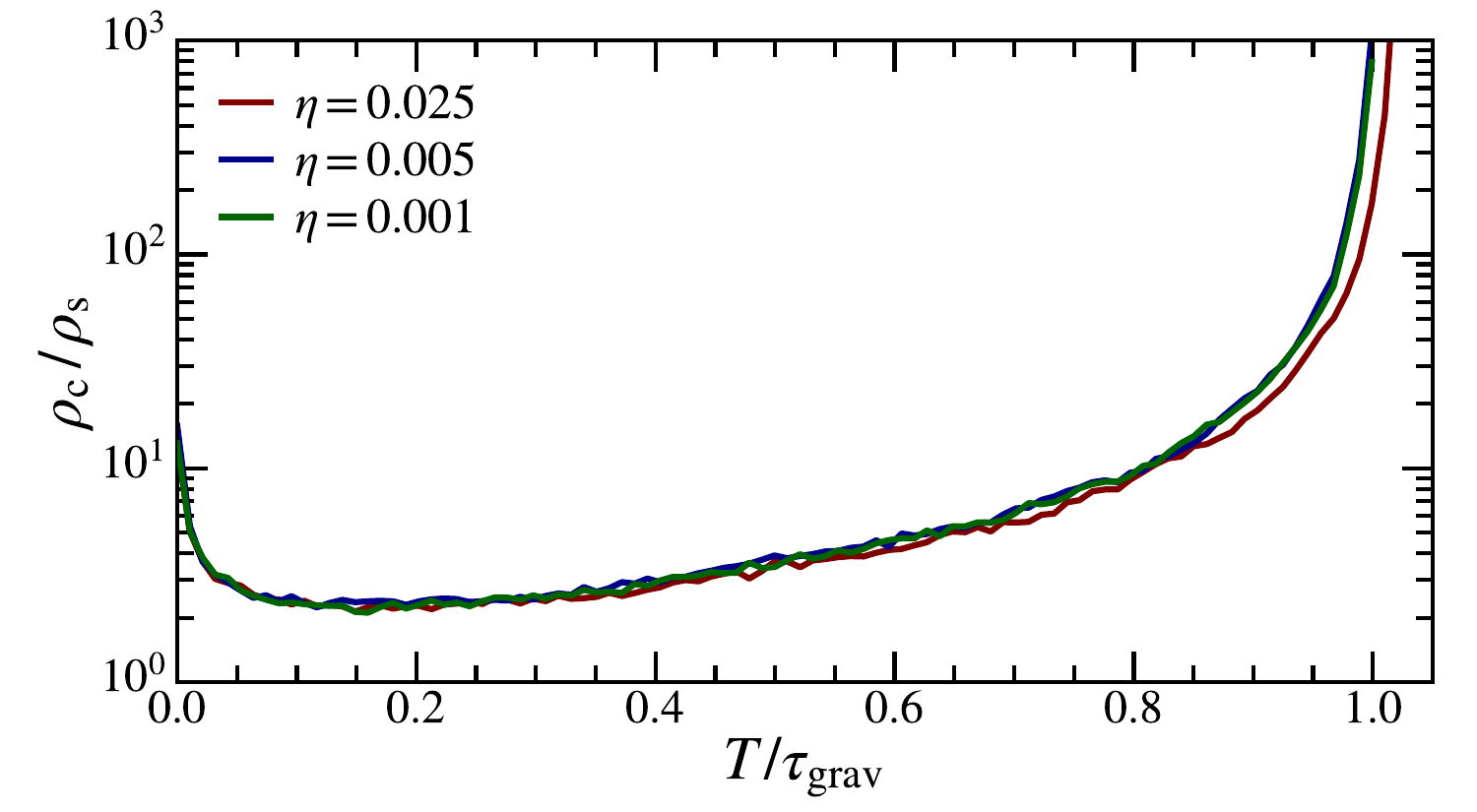}
    \caption{The time evolution of the core density for the $M_{200} = 10^{10} \, \Msun$, $c = 13.08$ isolated halo evolved with $N_{200} = 10^6$ under different choices of the gravitational-timestep accuracy parameter $\eta$ (\cref{eq:dt_grav}). The core collapse time appears to converge for $\eta \lesssim 0.005$, consistent with the findings of \protect\citet{Mace2024}, given our collapse timescale of $\tau_{\rm grav} = 31.33 \, \Gyr$.}
    \label{fig:convergence_eta}
\end{figure}

\citet{Mace2024} have shown that, for simulations evolved well beyond the age of the Universe, the standard gravitational timestep criterion with $\eta=0.025$ is insufficient to converge the core-collapse time, and that the lack of convergence persists at $\eta=0.005$ even in high-resolution runs with $N_{200}=10^6$. Tightening to $\eta=0.001$ does converge the result but is computationally prohibitive. Given our much shorter collapse timescale of $\tau_{\rm grav}=31.33\,\Gyr$ in comparison with ${\sim}225\,\Gyr$ in the long runs of \citet{Mace2024}, we expect $\eta=0.005$ to be sufficient. \cref{fig:convergence_eta} shows the core-density evolution of the halo of \cref{subsec:isolatedCoreCollapse} for $\eta=0.025$, $0.005$ and $0.001$. As anticipated, $\eta\lesssim 0.005$ is sufficient for convergence of the collapse time.


\bsp	
\label{lastpage}
\end{document}